\documentclass[a4,10pt]{article}
\usepackage{verbatim}
\usepackage{graphicx} 
\usepackage{placeins} 
\usepackage{fullpage}
\usepackage{subfigure} 
\usepackage{amsmath}
\usepackage{amssymb} 
\usepackage{nomencl}
\usepackage{setspace}
\usepackage{multicol}
\usepackage{enumitem}
\usepackage[normalem]{ulem}
\usepackage{soul}
\usepackage{afterpage}
\usepackage{epsfig}
\usepackage{eepic}
\usepackage{csquotes}

\usepackage{natbib}
\usepackage[abs]{overpic}
\usepackage{pinlabel}
\usepackage{xcolor}
\usepackage{anyfontsize}
\usepackage{undertilde}
\usepackage{accents}

\usepackage[export]{adjustbox}
\usepackage{multirow}
\usepackage{booktabs}
\usepackage{fixmath}
\usepackage[normalem]{ulem}
\usepackage{soul}
\usepackage{bm}

\begin{document}


\begin{center}
\textbf{\Large{Turbulent Jet: A DNS Study}}
\end{center}

\vspace{5 mm}

\begin{center}

\textbf{Sachin Y. Shinde$^{1,2,a)}$, Prasanth Prabhakaran$^{1,3,b)}$, Roddam Narasimha$^{1,c)}$}

\end{center}

\vspace{5 mm}

\begin{flushleft}

$^{1)}$Engineering Mechanics Unit, Jawaharlal Nehru Centre for Advanced Scientific Research, Bangalore 560064, India

$^{2)}$Department of Mechanical Engineering, Indian Institute of Technology, Kanpur 208016, India

$^{3)}$Department of Physics, Michigan Technological University, Houghton 49931, USA

\vspace{3 mm}

$^{a)}$Electronic mail: sachin@iitk.ac.in

$^{b)}$Electronic mail: prasantp@mtu.edu

$^{c)}$Electronic mail: roddam@jncasr.ac.in

\end{flushleft}

\vspace{10 mm}

\begin{abstract}

The entrainment of ambient fluid into a turbulent shear flow has been a topic of wide interest for several decades. To estimate the entrainment of ambient fluid into turbulent jet, it is essential to define the boundary of the jet. The question arises as to what should be the appropriate criterion to determine the edge of the turbulent jet. From the present DNS simulations, we observe that there is a need to define two boundaries for the turbulent jet, namely, the Rotational/Irrotational (R/IR) interface termed here as ``Outer boundary" and the Turbulent/Non-Turbulent (T/NT) interface as ``Inner boundary". The main aim of the present paper is to identify the need for defining the two boundaries for the jet. We also present some interesting observations of the jet flow, essentially on self similarity and self preservation.

\end{abstract}

\vspace{5 mm}

\section{Introduction}

The entrainment of ambient fluid into a turbulent shear flow has been
a topic of wide interest for several decades. The study on entrainment
was inspired by a striking instantaneous shadowgraph of the turbulent
wake of a bullet traveling at supersonic speeds carried out by
\cite{Corrsin_NACA_1955}. To understand the several facets of the
entrainment mechanism, the jet has been the most widely studied
turbulent shear flow (experiments on round jets reported by
\cite{Westerweel_ExpFluids_2002}, \cite{Wolf_PoF_2012}; DNS results on
a temporal round jet by \cite{Mathew_PoF_2002}, on a temporally
evolving plane jet by \cite{Reeuwijk_JFM_2014} and on a spatially
developing plane jet by \cite{Silva_PoF_2010}).

To estimate the entrainment of ambient fluid into turbulent jet, it
is essential to define the boundary of the jet. The question arises as
to what should be the appropriate criterion to determine the edge of
the turbulent jet. From the present DNS simulations, we observe that
there is a need to define two boundaries for the turbulent jet,
namely, the Rotational/Irrotational (R/IR) interface termed here as
``Outer boundary'' and the Turbulent/Non-Turbulent (T/NT) interface as
``Inner boundary''.

The main aim of the present paper is to identify the need for
defining the two boundaries for the jet. We begin with providing the
details of the direct numerical simulation (DNS) and its
validation. We also present some interesting observations of the jet
flow, essentially on self similarity and self preservation. This is
followed by a detailed discussion on the inner (T/NT) and outer (R/IR)
boundaries. We conclude by presenting a summary of the findings.

\section{DNS details}

\begin{figure}
\begin{center}
{\includegraphics[width = 8.0 cm]{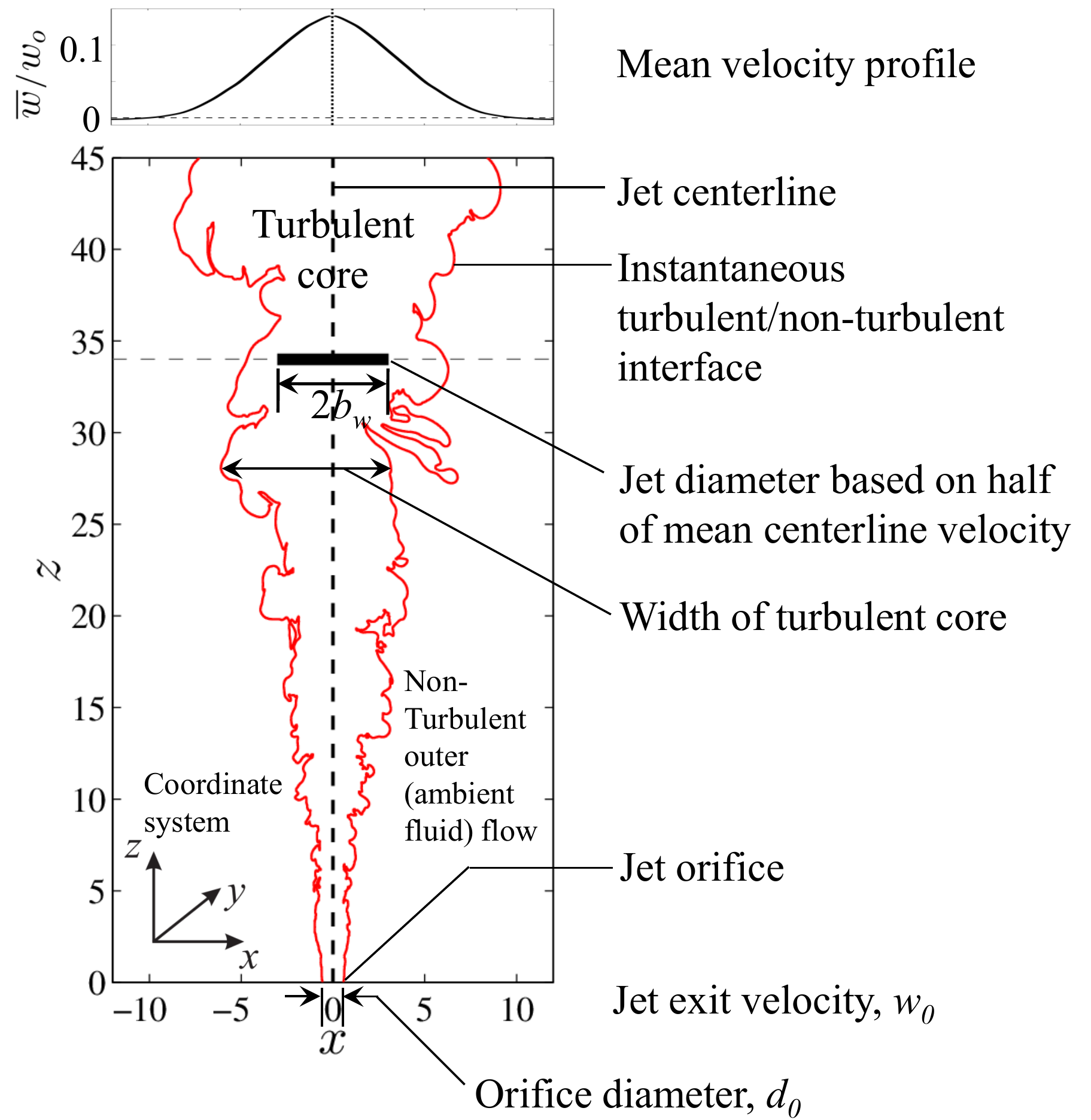}}
\end{center}
\caption{Schematic of an axial section of the jet flow to illustrate
the main terminologies and notations, see main text for additional
details.}
\label{fig:schmt}
\end{figure}

Figure~\ref{fig:schmt} shows a schematic of an axial section of the
jet, where $d_0$ represents the orifice diameter from which the jet
issues with a top hat velocity profile characterized by a uniform jet
exit mean velocity $\overline w_0$.  The flow is governed by the
incompressible Navier-Stokes equations (in non-dimensional form with
$\overline w_0$ and $d_0$ as velocity and length scales respectively),
\begin{equation}
\label{eq:cont}
\nabla \cdot \textit{\textbf{u}} = 0,
\end{equation}
\begin{equation}
\label{eq:NS}
(\partial{\textit{\textbf{u}}}/\partial{t}) + (\textit{\textbf{u}} \cdot \nabla)\textit{\textbf{u}} = - \nabla{p} + (1/Re)\nabla^2\textit{\textbf{u}},
\end{equation}

\noindent
where $\textit{\textbf{u}}$, $p$ represent the flow velocity vector
and pressure respectively, and $Re = \overline w_0 d_0/\nu$ is the
Reynolds number (= 2400 in the present simulations), and $\nu$ is the
kinematic viscosity of the fluid. Unless otherwise specified, all
distances, velocities and time presented in this article are
non-dimensionalized with $d_0$ and $\overline{w}_0$ as scales, and
vorticity with the time-averaged local scales at height $z$,
\textit{viz.}, mean centerline velocity $\overline{w}_{c}(z)$ and mean
half-velocity half-width $\overline{b}_{w}(z)$ at $w=\overline{w}_c/2$.

The Navier-Stokes equations are solved using an extension of Harlow \&
Welsh's scheme \cite[]{Harlow_PoF_1965} to non-uniform grids as
proposed by \cite{Verstappen_JCP_2003}, using a second-order finite
volume framework. The governing equations are integrated in time using
the second-order Adams-Bashforth scheme with a time step $\Delta t =
0.005$ flow unit. A non-uniform Cartesian mesh is used to resolve all
the relevant scales in the turbulent part of the flow-field. The
velocity and pressure variables are stored in a staggered arrangement
to prevent pressure-velocity decoupling. The pressure variable is
stored in the cell center and the velocity components are stored on
the cell faces \cite[]{Harlow_PoF_1965}. More details on the scheme
and the code are discussed in \cite{Prasanth_MS_Thesis}. The
computational domain extends up to $z = 55$ in the streamwise ($z$)
direction and $-50$ to $+50$ in the cross-stream directions ($x,
y$). The total grid size is $480 \times 480 \times 600$ in the $x$,
$y$ and $z$ directions respectively (yielding 138 million grid
points). In the streamwise direction, the grid size varies from 0.05
at the orifice to 0.18 at the outflow. In the cross-stream plane the
grid is uniform (of size 0.05) from $x$, $y$ = $-9.25$ to $+9.25$,
thus ensuring adequate resolution in the turbulent core; beyond this
the grid is gradually stretched towards the lateral side walls. The
simulation took up to $t = 900$ to attain a stationary state. Data for
computing flow statistics was acquired over the time interval
$900<t<2650$. The simulation was run on 108 nodes of the 360 TF
supercomputer at CSIR-4PI, Bangalore, India.

The bottom plane at $z = 0$ outside the orifice is treated as a
no-slip wall in the plane of the orifice exit. A fixed, uniformly
distributed noise of 5\% of $w_o$ with zero mean is superposed on the
top-hat velocity profile at the orifice in order to trip the flow to a
turbulent state. The noise is superposed only on the streamwise
velocity component, and at a random set of points chosen from a
uniform distribution over the orifice area. At the outflow boundary,
we use the zero normal derivative condition for all the variables,
with a layer of viscous padding from $z = 52$ to $z = 55$ to ensure
smooth exit of the turbulent flow from the computational domain. The
lateral boundaries of the computational domain are also treated as no
slip walls, but (at $x, y = \pm 50$) they are sufficiently far away
from the jet axis to have any significant effect on the momentum
balance. The main software used for the post-processing of the data is
Matlab-2014a.  Other details about the solver are discussed in
\cite{Prasanth_MS_Thesis}.

Notice that, unless otherwise specified exclusively, all the data will
be presented in non-dimensional form. The scales used for
non-dimensionalization are: the orifice diameter $d_{o}$ for the
distance, the jet exit velocity at the orifice $w_{0}$ for velocity,
and $d_{o}/w_{0}$ for time. Vorticity is non-dimensionalized by time
averaged local scales, \textit{viz.}, $\overline{w}_{c} /
\overline{b}_{w}$, where $\overline{w}_{c}$ is the local maximum
centerline velocity and $\overline{b}_{w}$ is the mean half-velocity half-width
of the jet.  Note that, however, for the sake of brevity, we have used
the variables without an asterisk (as a superscript) or not dividing
by the non-dimensional scale, in both the text as well as figures;
e.g. unless exclusively mentioned \textquoteleft{$q$}' represents a
non-dimensional form of a quantity \textquoteleft{$q$}'.

\subsection{Validation}
\label{sec:Validation}

\begin{figure}[!h]
\centering
\begin{overpic}
[width = 13.50 cm, height = 7.0 cm, unit=1mm]
{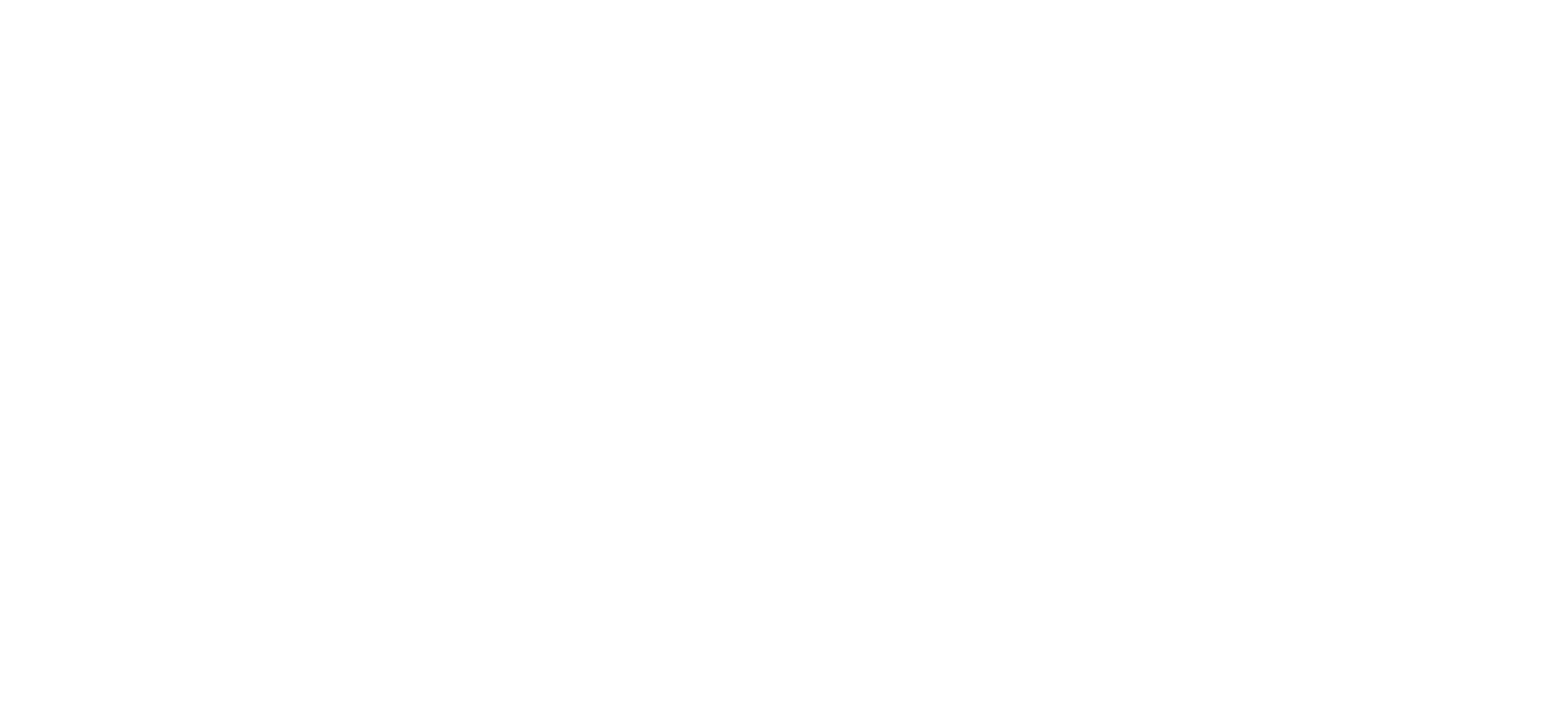}
\put(20,0){\includegraphics[width = 4.50 cm]{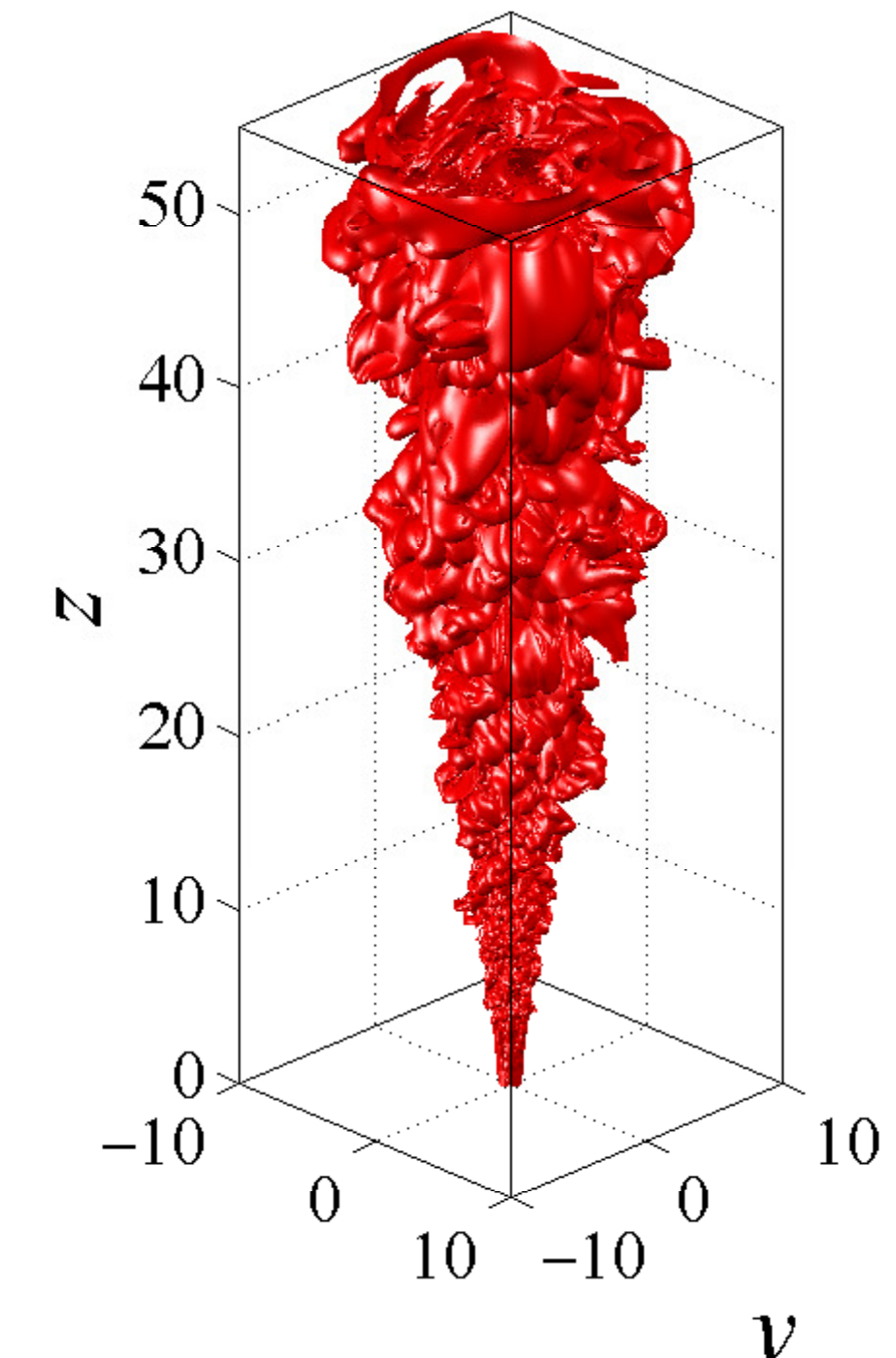}}
\put(80,0){\includegraphics[width = 4.50 cm]{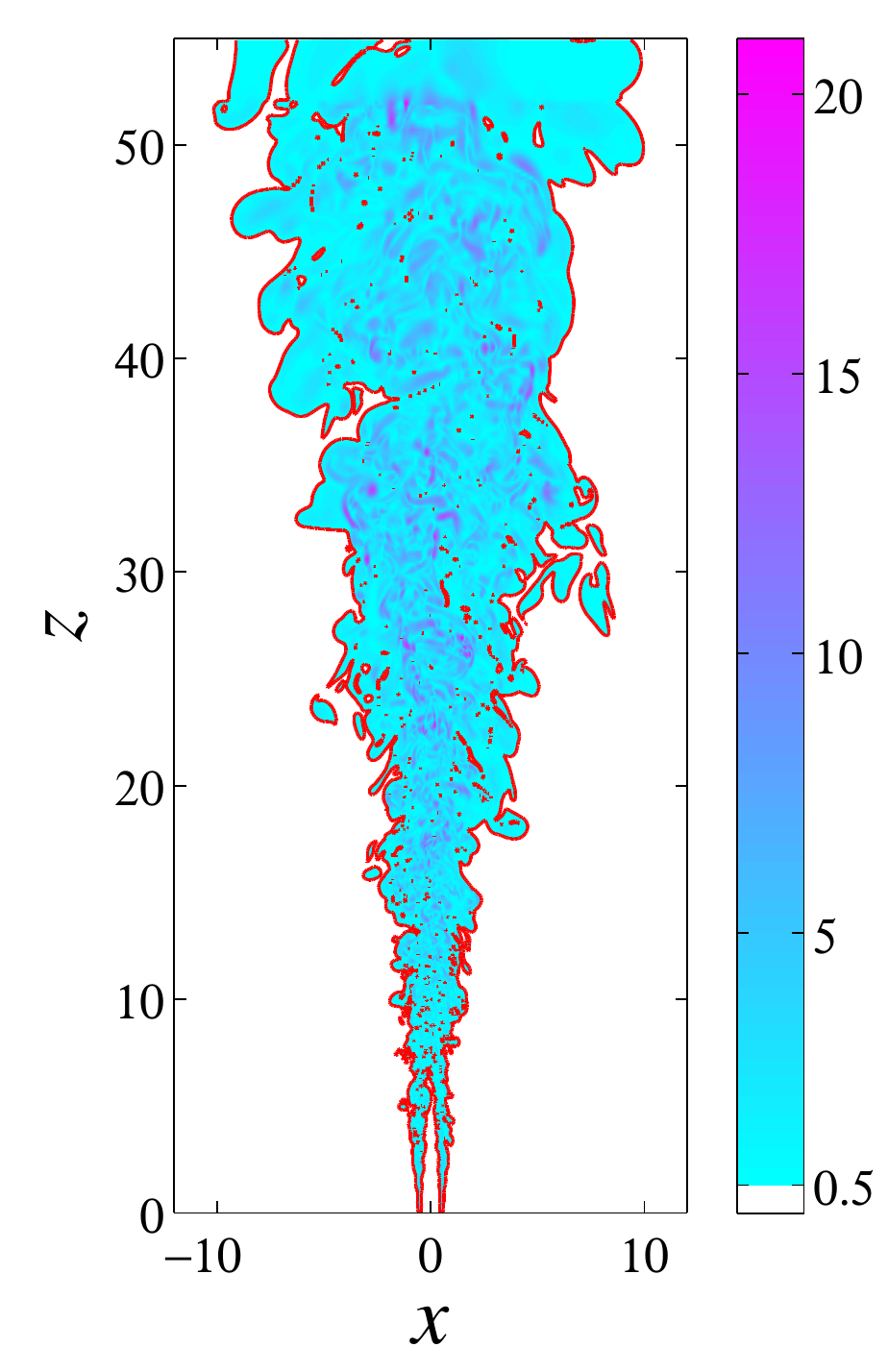}}
\put(31,0){\Large{\textit{x}}}
\put(18,66){(a)}
\put(75,66){(b)}
\end{overpic}
\caption{Instantaneous snapshot of vorticity field of a round
incompressible turbulent jet at time $t = 1995$ flow units since the
inception of the flow, obtained from the direct numerical
simulation. (a) Iso-surface of vorticity magnitude at a threshold of
0.5. (b) An axial section of the flow.}
\label{fig:jet-3D-thrld-0p5}
\end{figure}

Figure \ref{fig:jet-3D-thrld-0p5} shows the flow structure of the jet
at one instant. It is evident that the jet surface, and thus the jet
boundary, are highly convoluted.

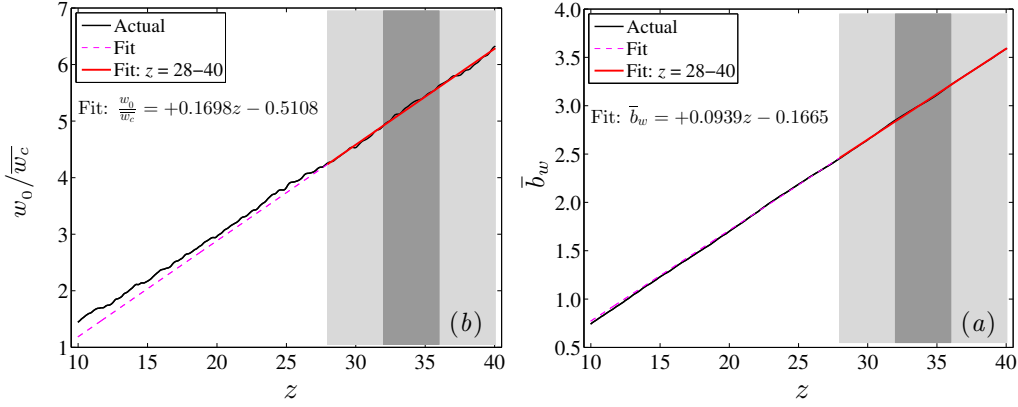
\begin{figure}
\centering
\begin{overpic}
[width = 13.50 cm, height = 5.5 cm, unit=1mm]
{Fig_box-eps-converted-to.pdf}
\put(7,3){\includegraphics[width = 6.0 cm]{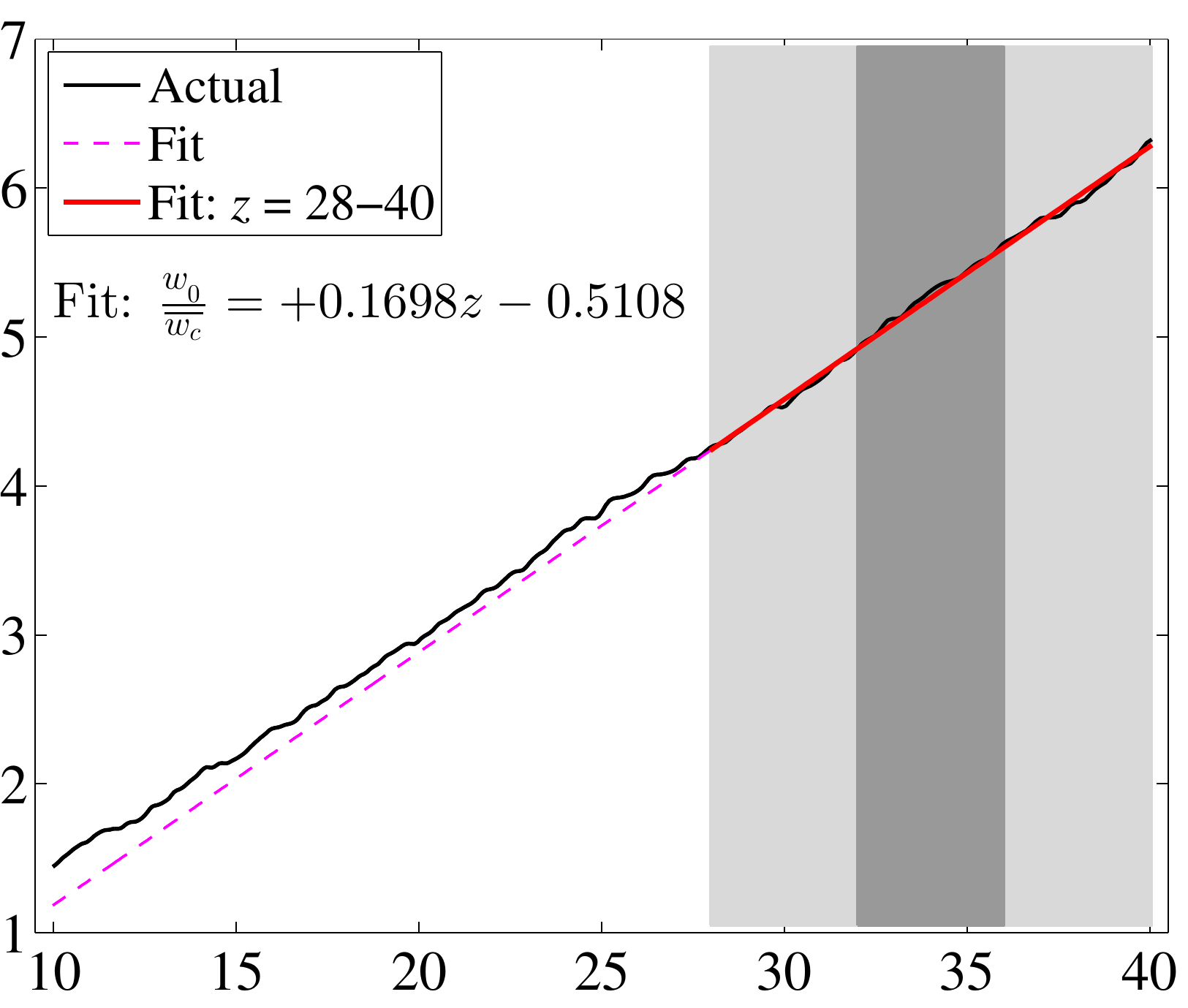}}
\put(73,3){\includegraphics[width = 6.2 cm]{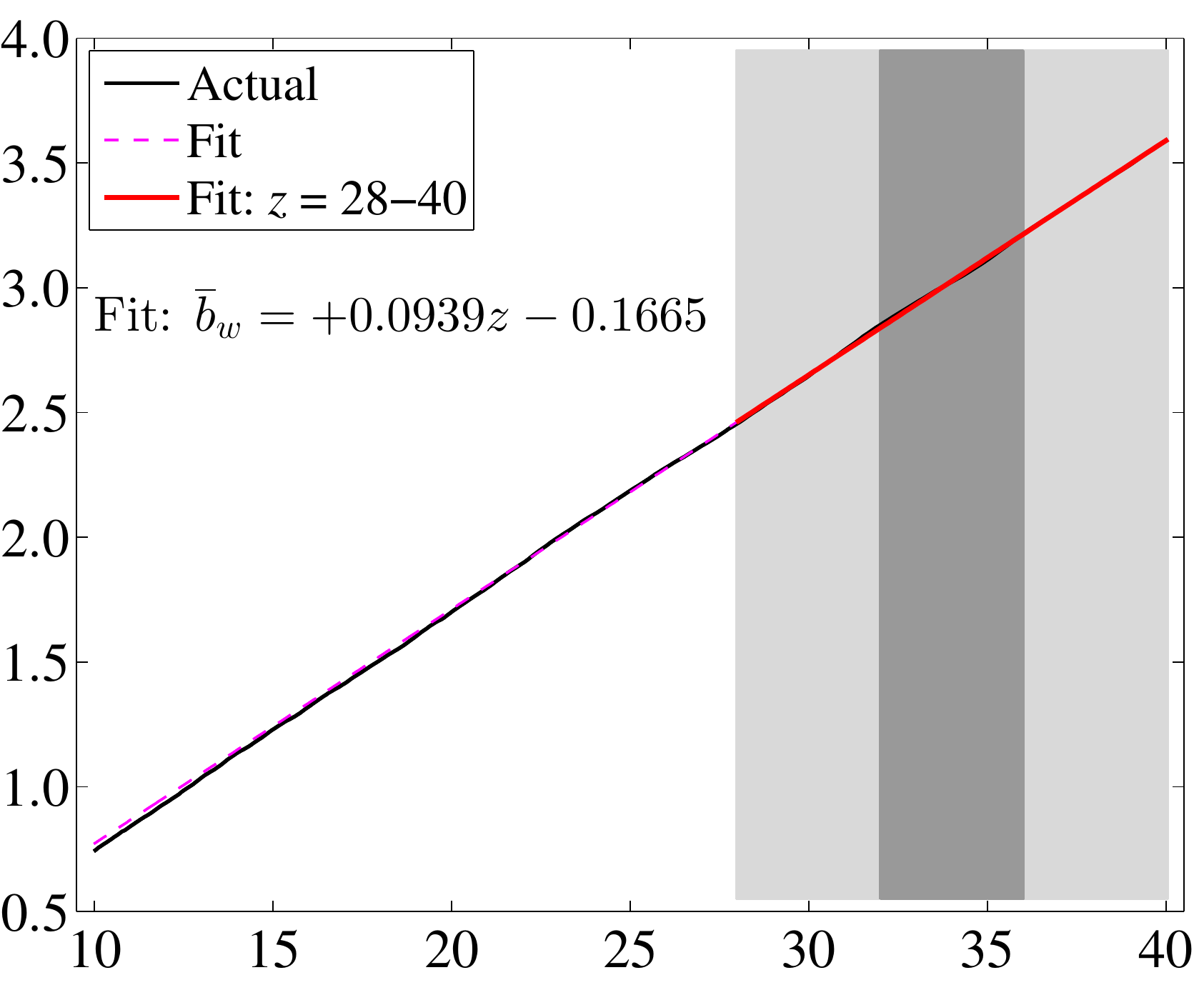}}
\put(37,0){\fontsize{11}{13.2}\selectfont \rotatebox{0}{$z$}}
\put(0,24){\fontsize{9}{10.8}\selectfont \rotatebox{90}{$w_{_{0}}\big/ \overline{w_c}$}}
\put(105,0){\fontsize{11}{13.2}\selectfont \rotatebox{0}{$z$}}
\put(68.5,28){\fontsize{10}{12}\selectfont \rotatebox{90}{$\overline{b}_w$}}
\put(59,9){(\textit{b})}
\put(127,9){(\textit{a})}
\end{overpic}
\caption{Variation with the streamwise distance of the time-azimuthal
averaged flow data: (\textit{a}) inverse centerline velocity;
(\textit{b}) jet half-velocity width. The light and dark grey
rectangle indicate respectively the self-similar and self-preservation
region (to be discussed in section \ref{sec:Self-preservation}).}
\label{fig:uc_width}
\end{figure}

Figures\,\ref{fig:uc_width}(a) and \ref{fig:uc_width}(b) show the
streamwise variation of the inverse of the mean centerline velocity
and the mean-half-velocity half-width respectively. We see that the
inverse of the centerline velocity increases linearly with the
streamwise distance. The slope from the data shown in
Fig.~\ref{fig:uc_width}(a) is 0.17, which is comparable to the
experimental values of 0.16 \cite[]{Panchapakesan_JFM_1993} and 0.17
\cite[]{hussein1994velocity}. Figure~\ref{fig:uc_width}(b) shows that
the half-velocity half-width of the jet increases linearly with
streamwise distance and the rate of increase is 0.094, which is
comparable to the experimentally measured value of 0.097 by
\citet{Westerweel_JFM_2009}.

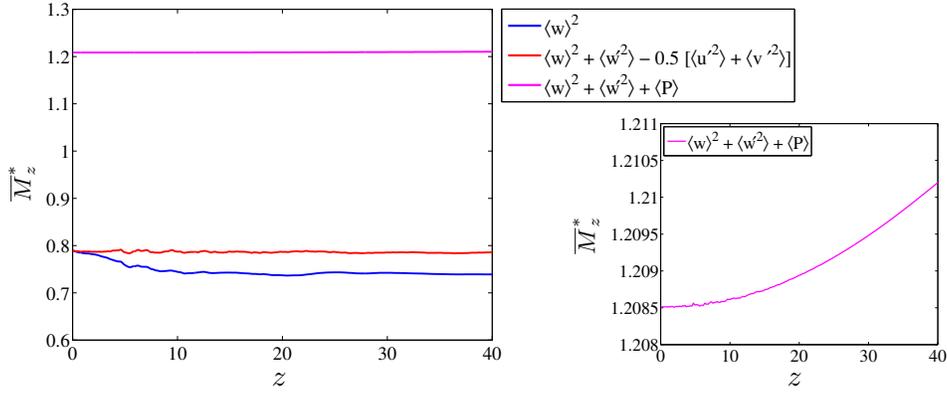
\begin{figure}
\centering
\begin{overpic}
[width = 13.50 cm, height = 5.1 cm, unit=1mm]
{Fig_box-eps-converted-to.pdf}
\put(10,3){\includegraphics[width = 6.1 cm]{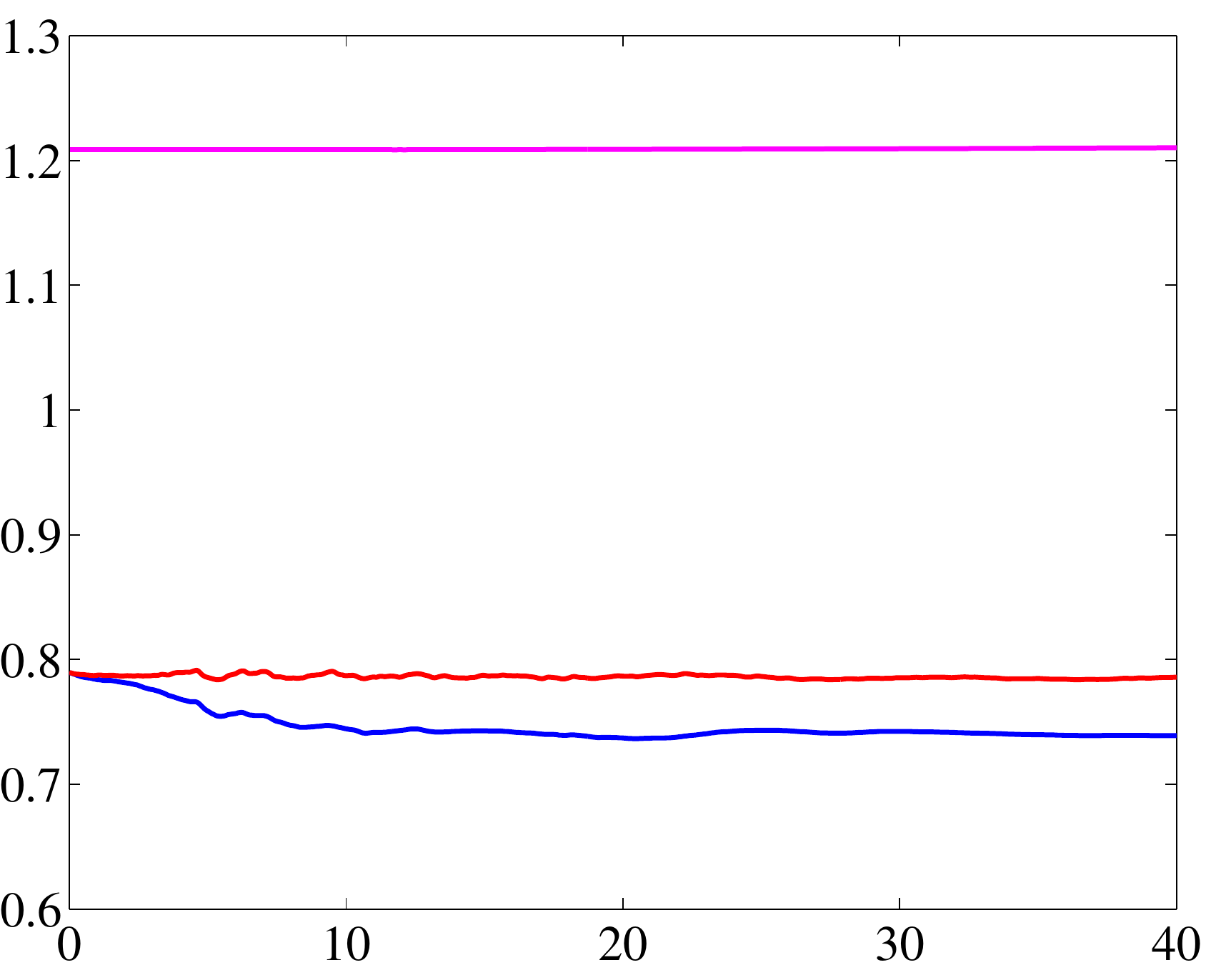}}
\put(70,37.7){\includegraphics[width = 4.0 cm]{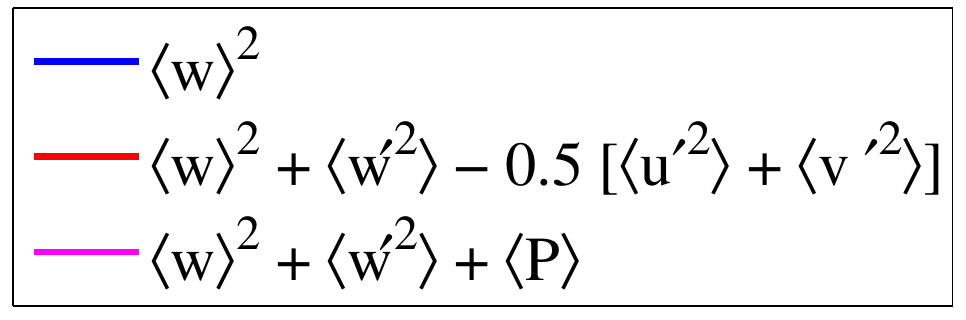}}
\put(85,3){\includegraphics[width = 4.5 cm]{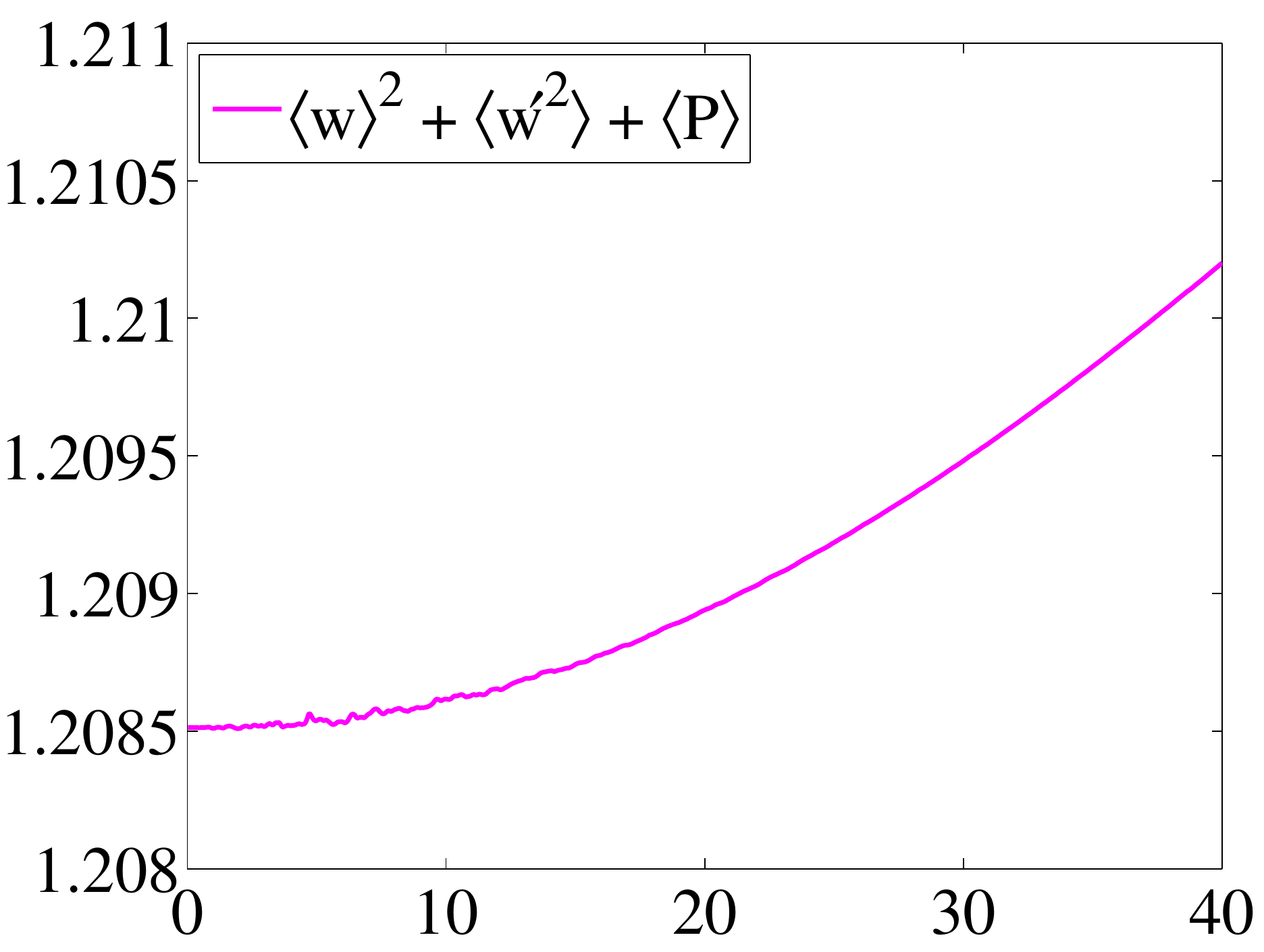}}
\put(40,0.5){\fontsize{11}{13.2}\selectfont \rotatebox{0}{$z$}}
\put(5,25){\fontsize{9}{10.8}\selectfont \rotatebox{90}{$\overline{M}_z^*$}}
\put(108.5,0.5){\fontsize{11}{13.2}\selectfont \rotatebox{0}{$z$}}
\put(80,18){\fontsize{9}{10.8}\selectfont \rotatebox{90}{$\overline{M}_z^*$}}
\end{overpic}
\caption{Variation with the streamwise distance of the mean streamwise
momentum flux in the jet normalized by the inlet conditions, defined
in different ways. On the right, the plot shows the zoomed-in view of
the momentum flux including the pressure term calculated directly from
the DNS data. The variation in normalized momentum flux over a
downstream distance of 40 orifice diameters is about 0.0017.}
\label{fig:mom}
\end{figure}

Figure \ref{fig:mom} shows the variation of streamwise momentum flux
with the streamwise distance.  The net momentum flux \( \int
(\overline{w}^2 + \overline{{w^\prime}^2}
-0.5[\overline{{u^\prime}^2}+\overline{{v^\prime}^2}] )dS\), where the
overbar denotes time average, prime denotes fluctuation
\cite[]{hussein1994velocity}, and the domain of integration is $-50\le
x, y\le50$, is conserved to better than 99\% over the whole domain in
the $z$ direction.

\subsection{Self-preserving flow}
\label{sec:Self-preservation}

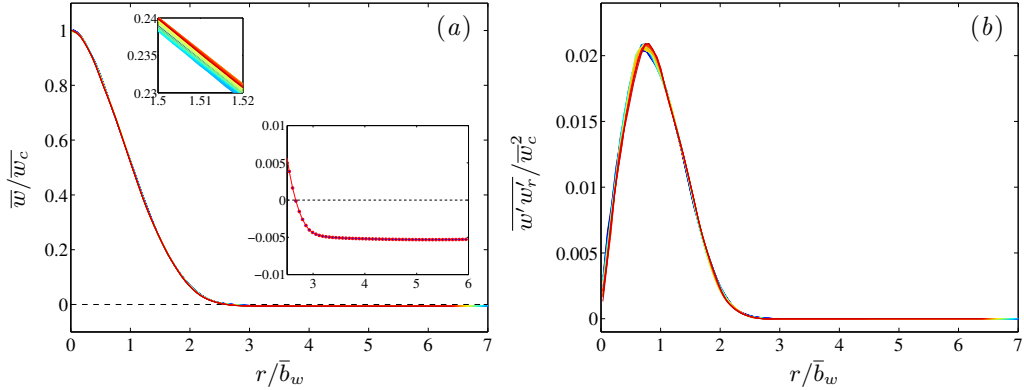
\begin{figure}[!h]
\centering
\begin{overpic}
[width = 13.50 cm, height = 5.1 cm, unit=1mm]
{Fig_box-eps-converted-to.pdf}
\put(5,3){\includegraphics[width = 6.0 cm]{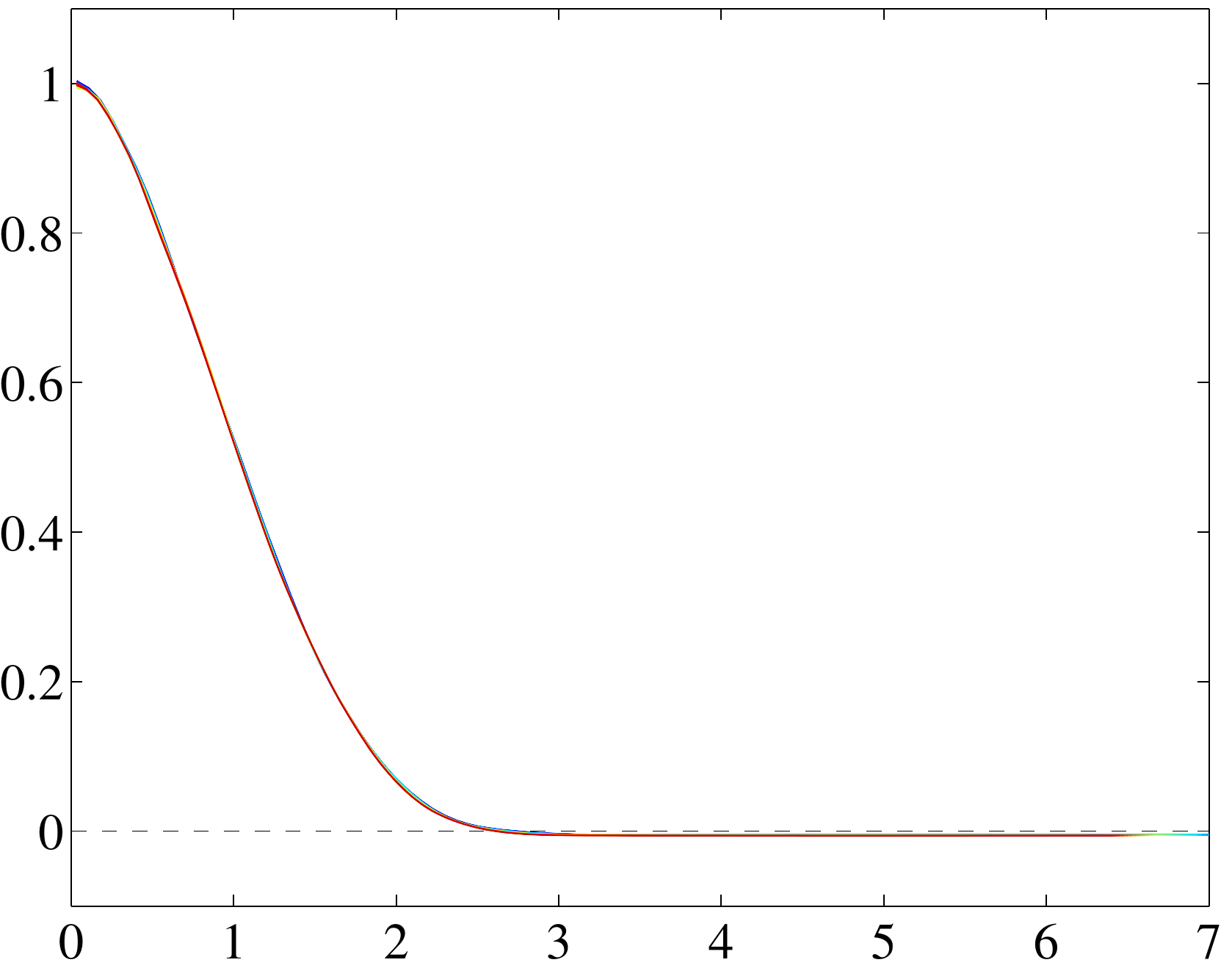}}
\put(16,36.5){\includegraphics[width = 1.7 cm]{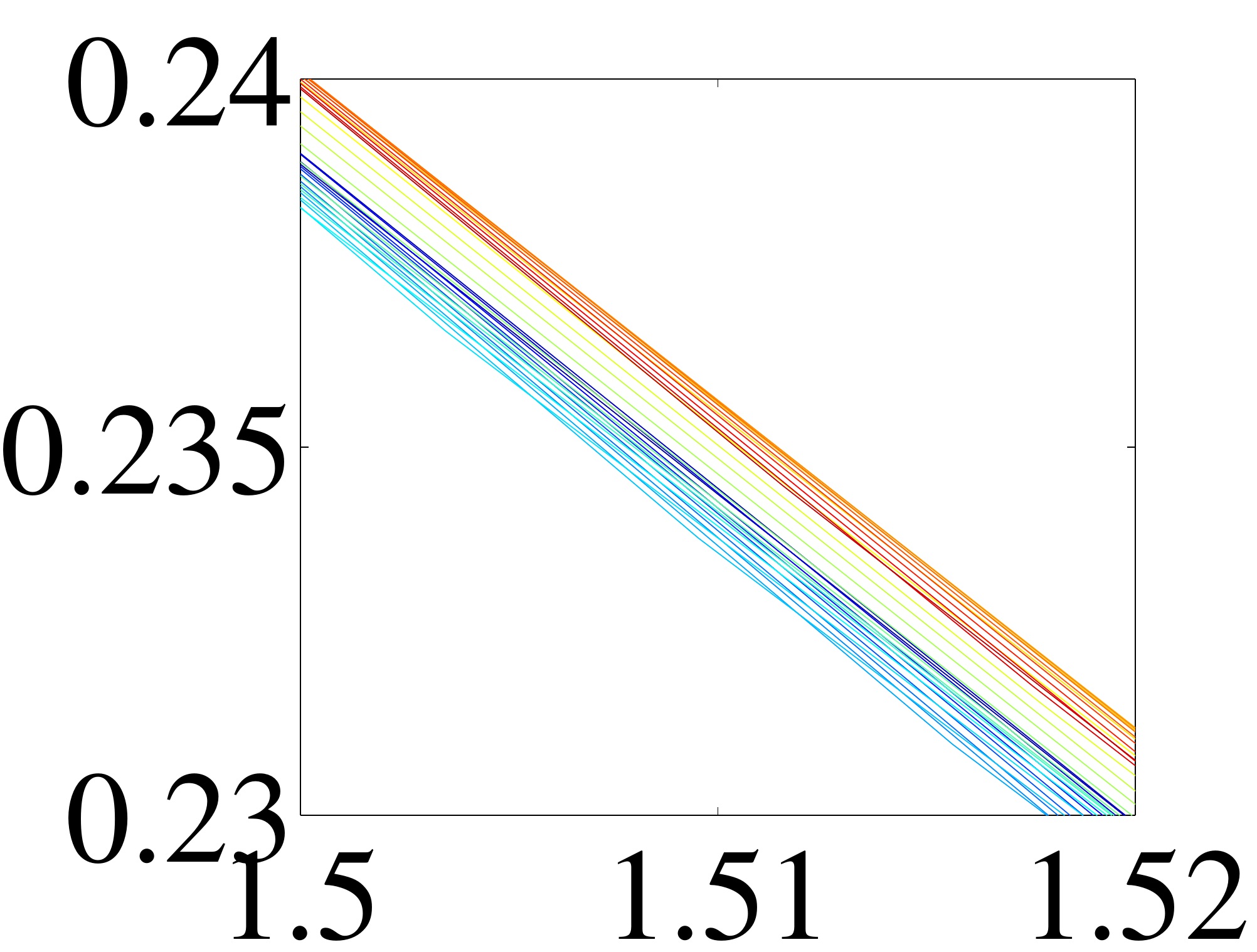}}
\put(32,12){\includegraphics[width = 3.0 cm]{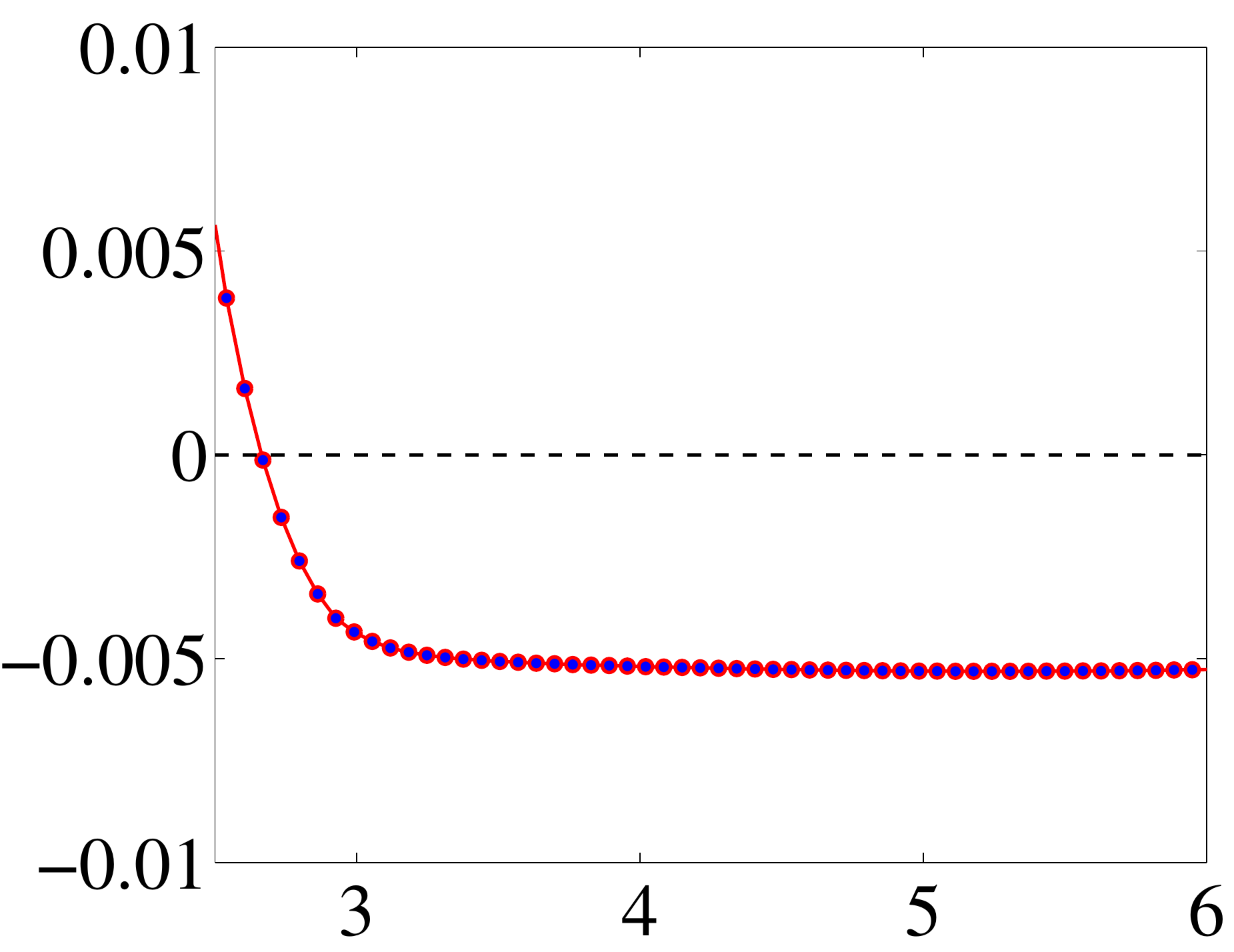}}
\put(73,3){\includegraphics[width = 6.25 cm]{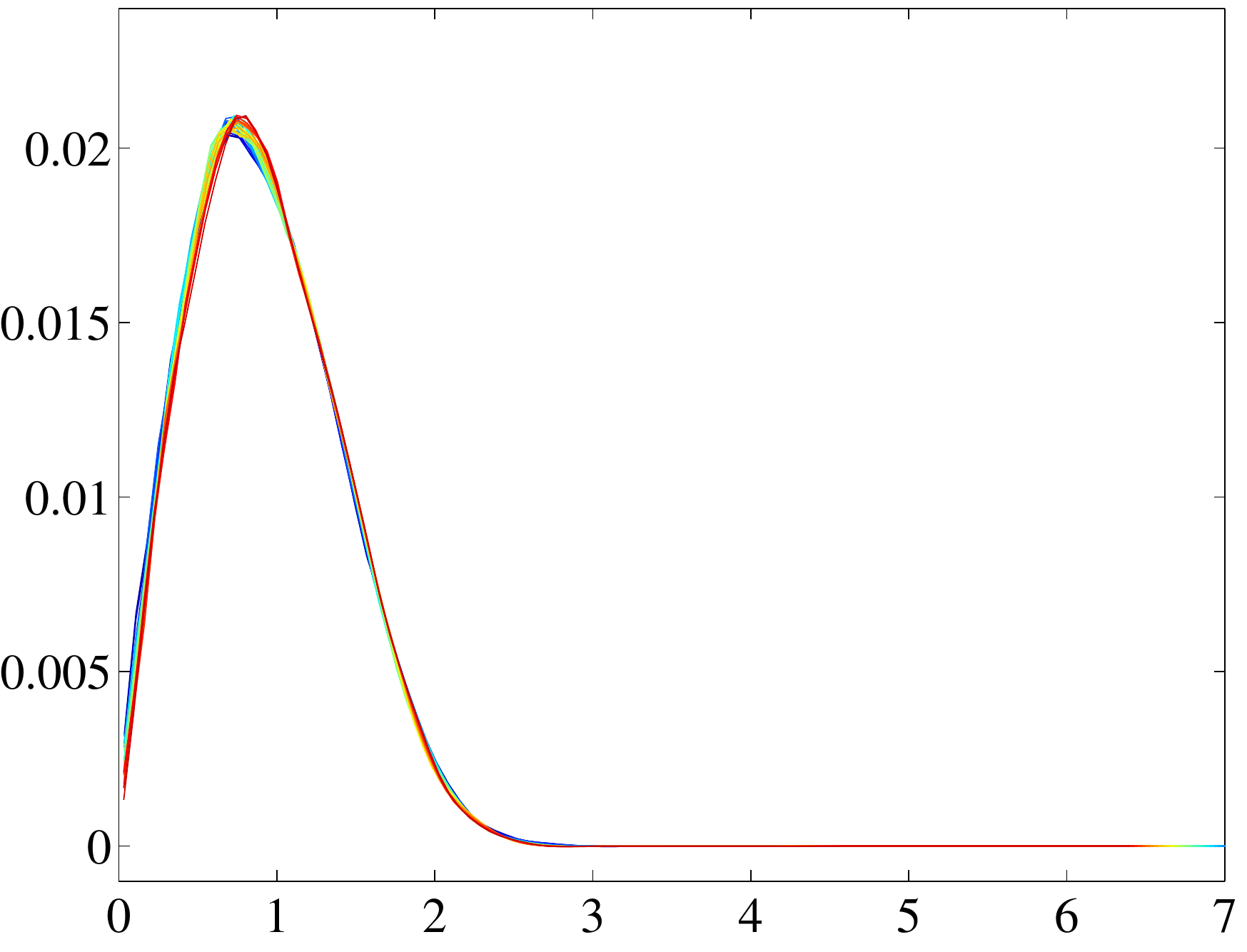}}
\put(33,0){\fontsize{9}{10.8}\selectfont \rotatebox{0}{$r / \overline{b}_w$}}
\put(0,23){\fontsize{9}{10.8}\selectfont \rotatebox{90}{$\overline{w} / \overline{w_c}$}}
\put(104,0){\fontsize{9}{10.8}\selectfont \rotatebox{0}{$r / \overline{b}_w$}}
\put(67,20){\fontsize{9}{10.8}\selectfont \rotatebox{90}{$\overline{w^{\prime} w_r^{\prime}} \big/ \overline{w}_c^2$}}

\put(57,46){(\textit{a})}
\put(128,46){(\textit{b})}

\end{overpic}
\caption{Variation in the radial direction of the time and azimuthally
averaged flow data. The data are plotted for all the 33 stations
spanning the self-preservation region, namely, from $z = 32$ to $36$.
(\textit{a}) Axial velocity profile. Top inset shows a zoomed-in view
of a particular portion of the velocity profiles to emphasize that
there are 33 lines which fall on top of each other. Bottom inset shows
the zoomed-in view between 2.5 to 6 on the abscissa, showcasing
primarily the negative axial velocities outside the core of the jet;
notice that the data here is averaged over all 33
stations. (\textit{b}) Reynolds shear stress profiles.}
\label{fig:meanvel_restress}
\end{figure}

We observed that the jet attains a state of self-similarity at a
distance of $z/d_{_{0}} \approx 28$. The flow remains self-similar up
to $z/d_{_{0}} \approx 40$. In this region, the streamwise velocity
profiles show a remarkable collapse when normalized with local
centerline velocity and $\overline{b}_{w}$ as shown in
Fig.~\ref{fig:meanvel_restress}(a). It is found that the jet is in a
self-similar state over the range $28 \leqslant z \leqslant 40$ to
within $0.6\%$ in the mean velocity
($\overline{w}/\overline{w}_c$). However, for the flow to be
``self-preserving'', both streamwise velocity profiles as well as the
Reynolds shear stress should show a close collapse, when normalized
with the same local scales, i.e., the centerline velocity and
$\overline{b}_{w}$. We found that the jet is in an equilibrium or
self-preserving state, \textit{i.e.}, mean velocity and Reynolds
stress are self-similar over the range $32 \leqslant z \leqslant 36$,
the maximum in $(\overline{{w^\prime}^2}/\overline{w}^2_c)^{0.5}$ (not
shown here) and Reynolds shear stress $- \overline{w^\prime
w_{r}^\prime}/\overline{w}^2_c$ (Fig.~\ref{fig:meanvel_restress}(b))
vary less than $1\%$ and $1.5\%$ respectively.  Therefore, we will
confine our analysis to the self-preserving region. We observed that
there are negative velocities outside the core of the jet due to
recirculation; however they are negligibly small, the maximum being
$-0.0053$ of the centerline velocity as shown in the bottom inset of
Figure \ref{fig:meanvel_restress}(a).

\section{Defining the jet boundary: Two boundaries?}
\label{sec:Two boundaries}

\begin{figure}[!h]
\centering
\begin{overpic}
[width = 13.50 cm, height = 6.25 cm, unit=1mm]
{Fig_box-eps-converted-to.pdf}
\put(10,2.5){\includegraphics[width = 5.0 cm]{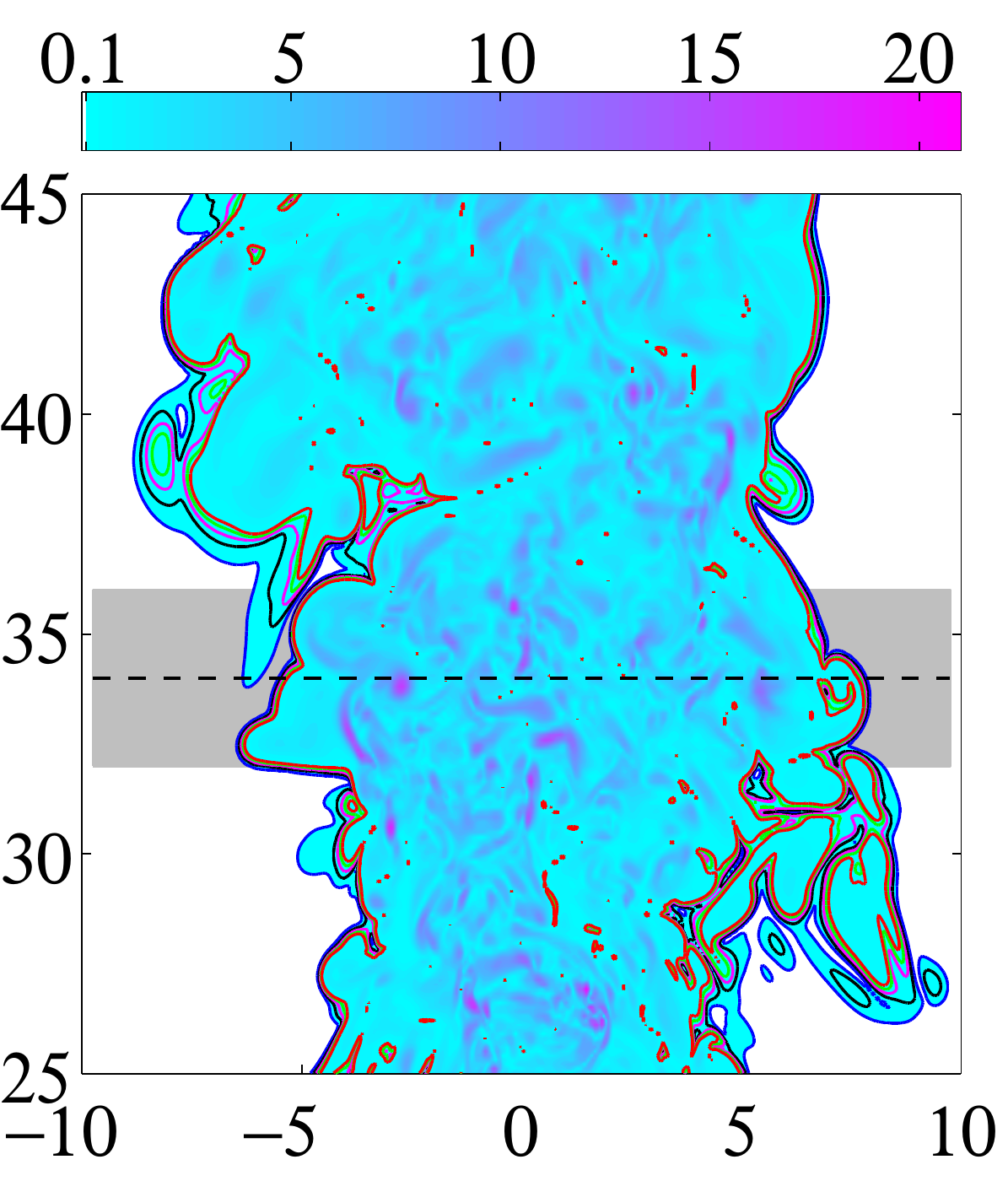}}
\put(70,2.5){\includegraphics[width = 5.20 cm]{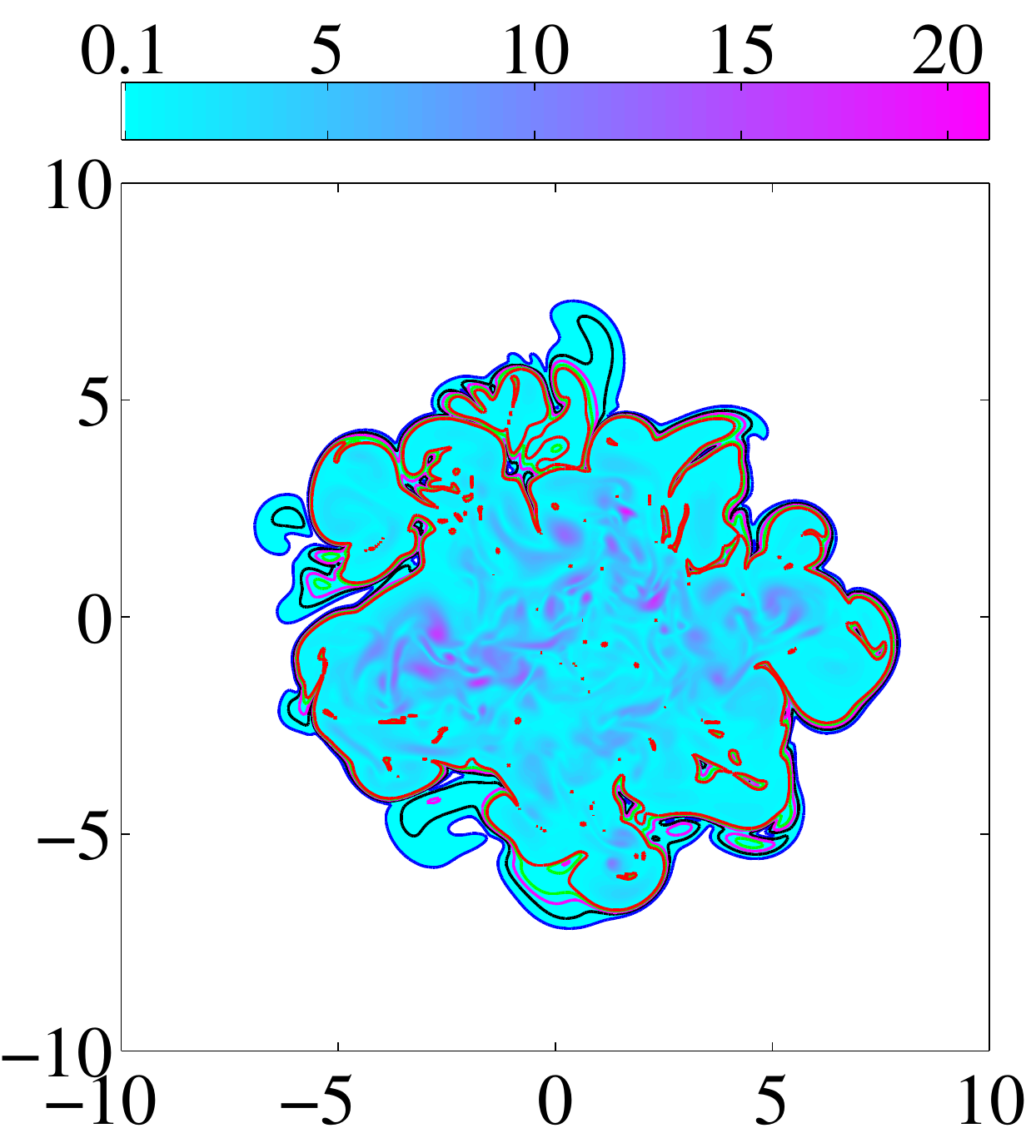}}
\put(115,30){\includegraphics[width = 2.0 cm]{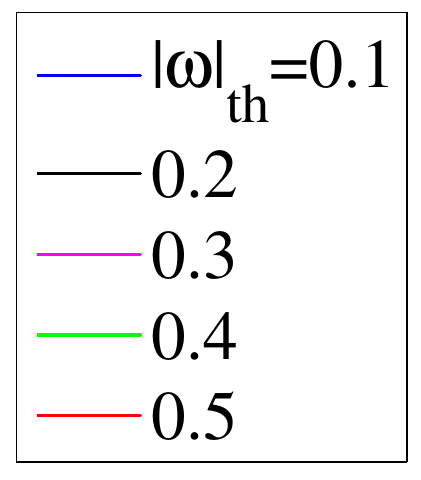}}
\put(35,0.5){\fontsize{10}{12}\selectfont \rotatebox{0}{$x$}}
\put(5,28){\fontsize{11}{13.2}\selectfont \rotatebox{90}{$z$}}
\put(98,0.5){\fontsize{10}{12}\selectfont \rotatebox{0}{$x$}}
\put(68,28){\fontsize{10}{12}\selectfont \rotatebox{90}{$y$}}
\put(16,10){(\textit{a})}
\put(78,10){(\textit{b})}
\end{overpic}
\caption{Instantaneous vorticity snapshot at $t = 1995$ flow units
since the start of the flow, for vorticity magnitude threshold of
0.1. (\textit{a}) shows an axial section at $\Phi = 0^{\circ}$ and
(\textit{b}) shows a diametral section at $z = 34$. Also shown are the
jet boundaries at various vorticity magnitude thresholds from 0.1 to
0.5 in steps of 0.1. In (\textit{a}), the gray rectangle highlights
the self-preservation region, and the dotted black line at $z =34$
shows the location which lies in the middle of the self-preservation
region, and where the vorticity contours in (b) are drawn.}
\label{fig:jet-edges-many thresholds}
\end{figure}
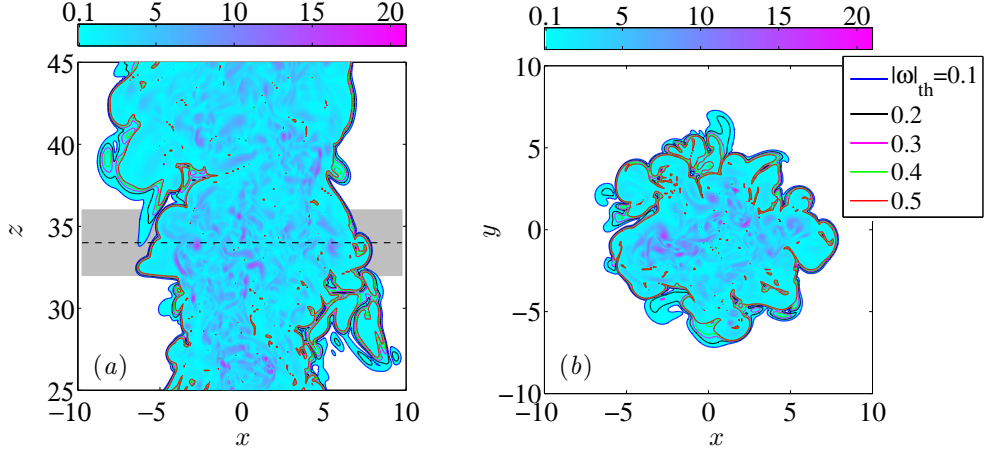

The boundary of the turbulent jet is defined by putting threshold on
the vorticity modulus $|\bm{\omega}|$. Figure \ref{fig:jet-edges-many
thresholds} shows the contours of vorticity magnitude within the core
of the jet in axial (Figure \ref{fig:jet-edges-many thresholds}(a))
and diametral (Figure \ref{fig:jet-edges-many thresholds}(b))
planes. In both the axial and diametral planes, the jet boundaries are
plotted at five different thresholds on $|\bm{\omega}|$, ranging from 0.1 to 0.5.

\begin{figure}
\centering
\begin{overpic}
[width = 14.0 cm, height = 16.5 cm, unit=1mm]
{Fig_box-eps-converted-to.pdf}
\put(0,110){\includegraphics[width = 6.75 cm]{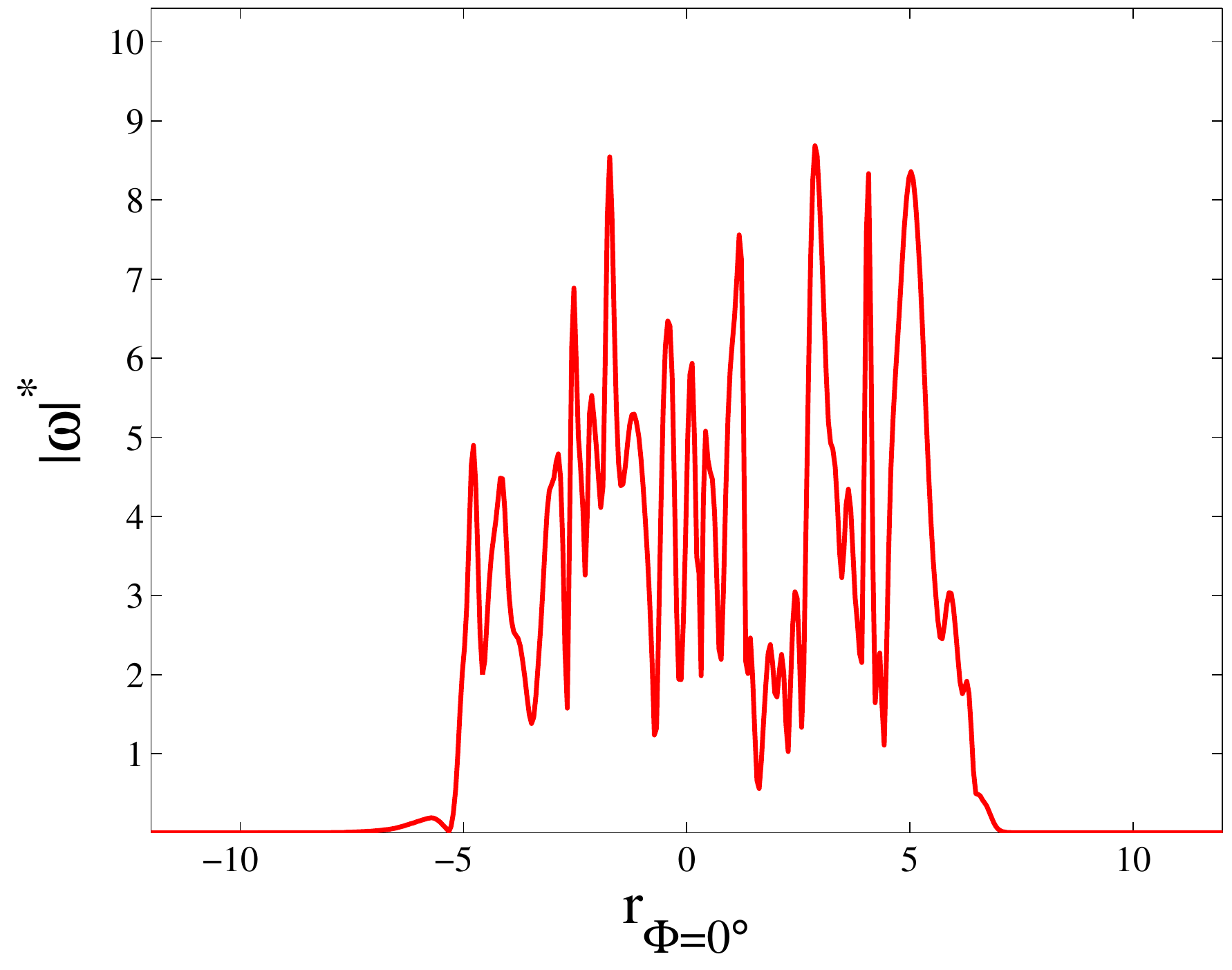}}
\put(70,110){\includegraphics[width = 6.75 cm]{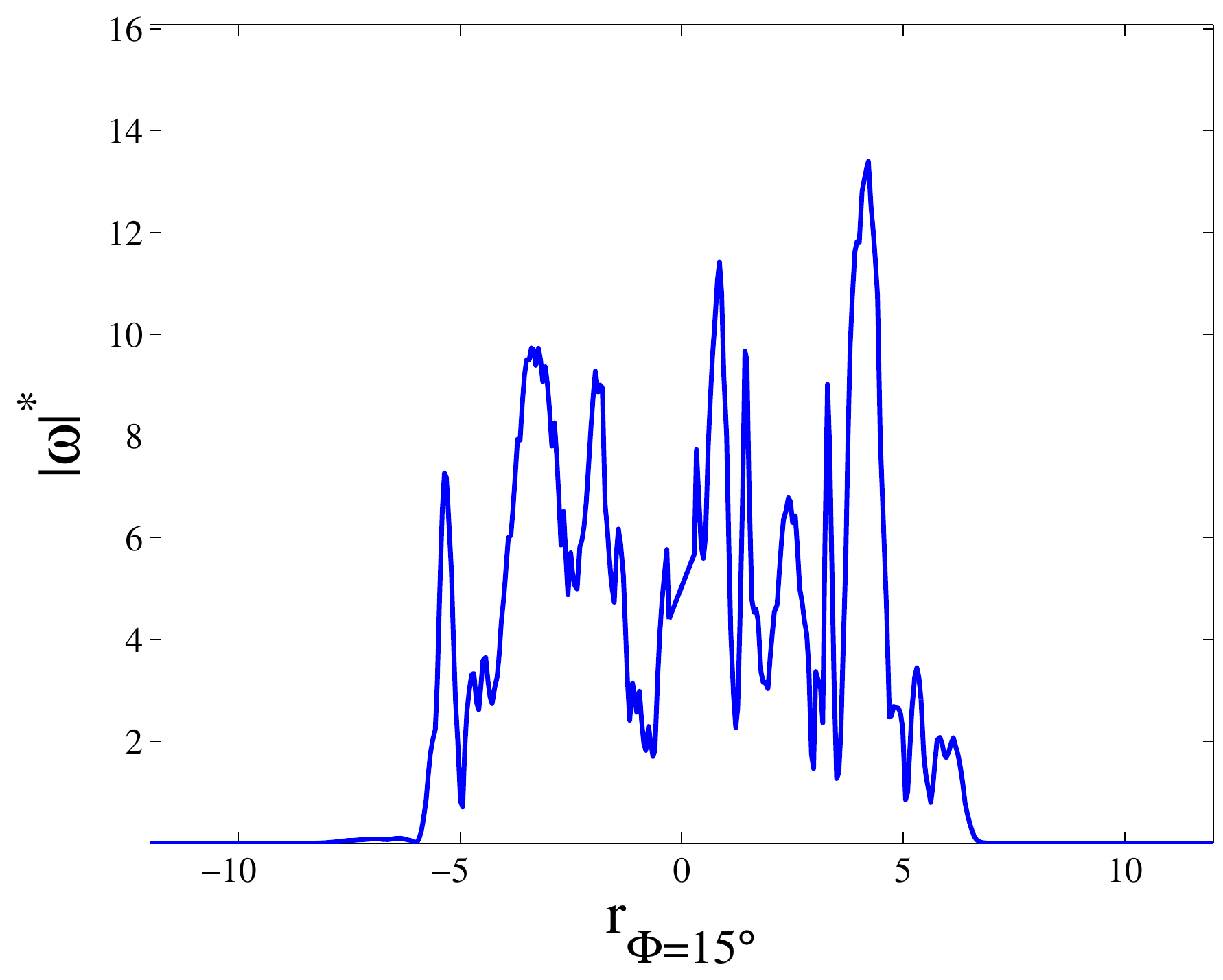}}
\put(0,55){\includegraphics[width = 6.75 cm]{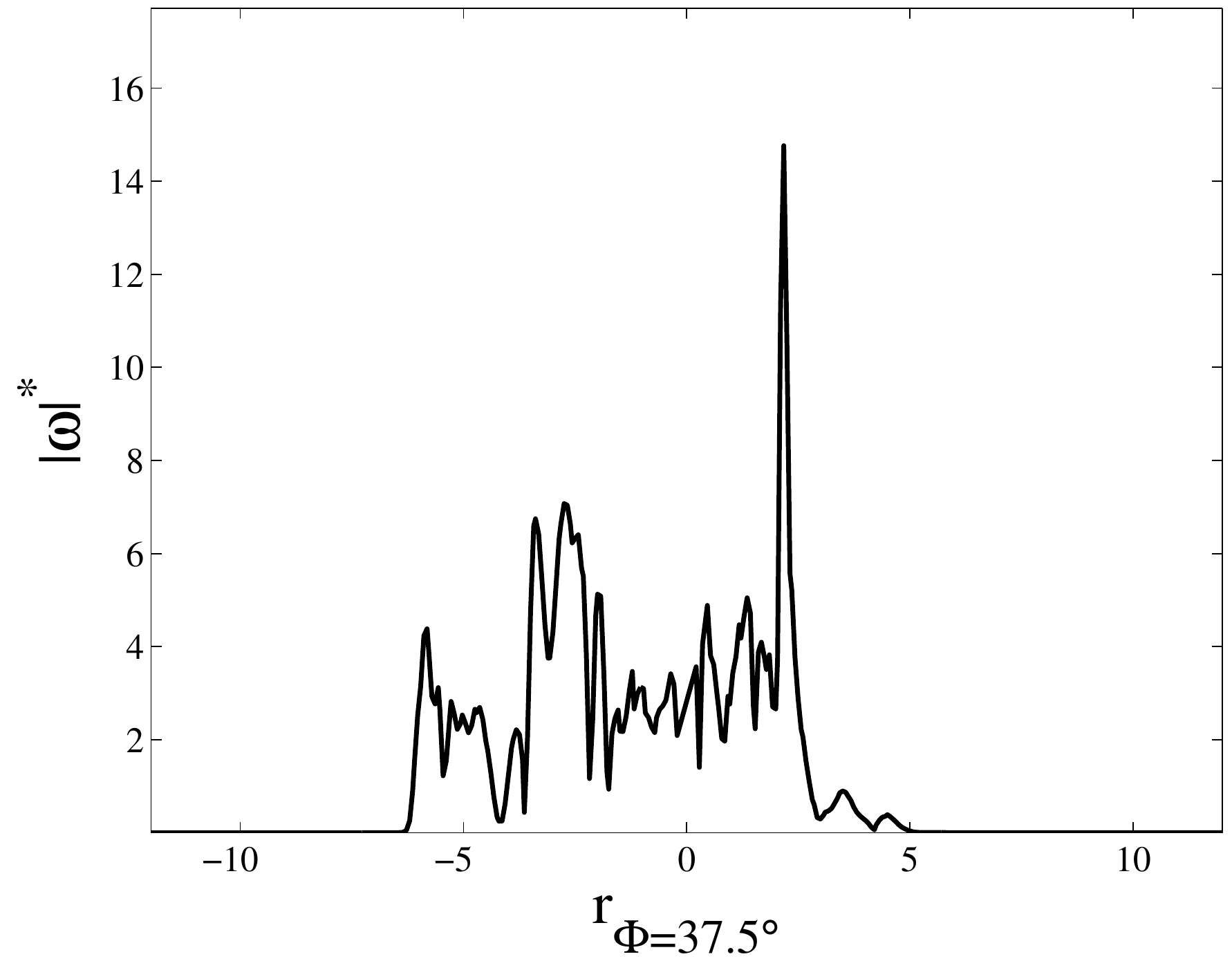}}
\put(70,55){\includegraphics[width = 6.75 cm]{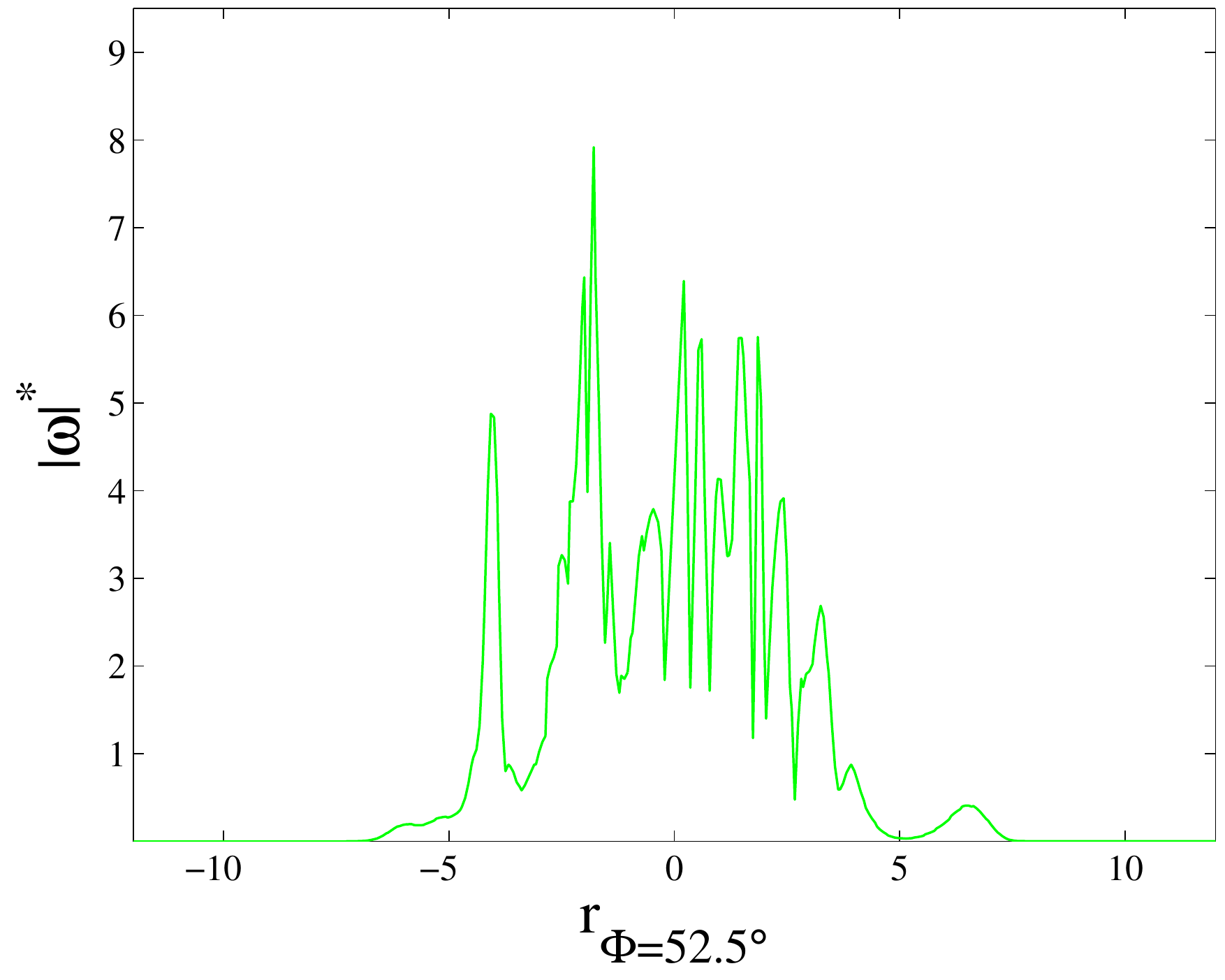}}
\put(0,0){\includegraphics[width = 6.75 cm]{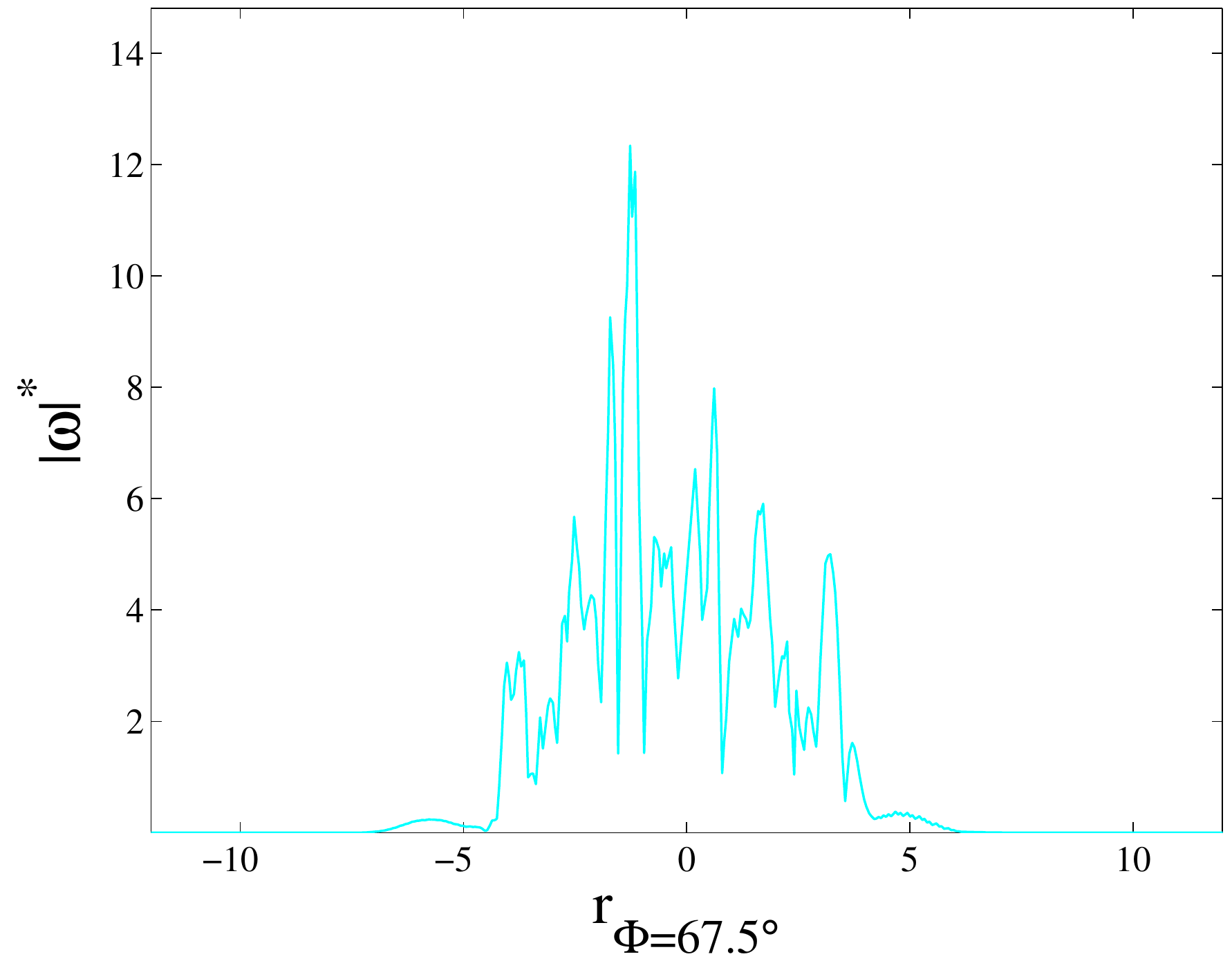}}
\put(70,0){\includegraphics[width = 6.25 cm]{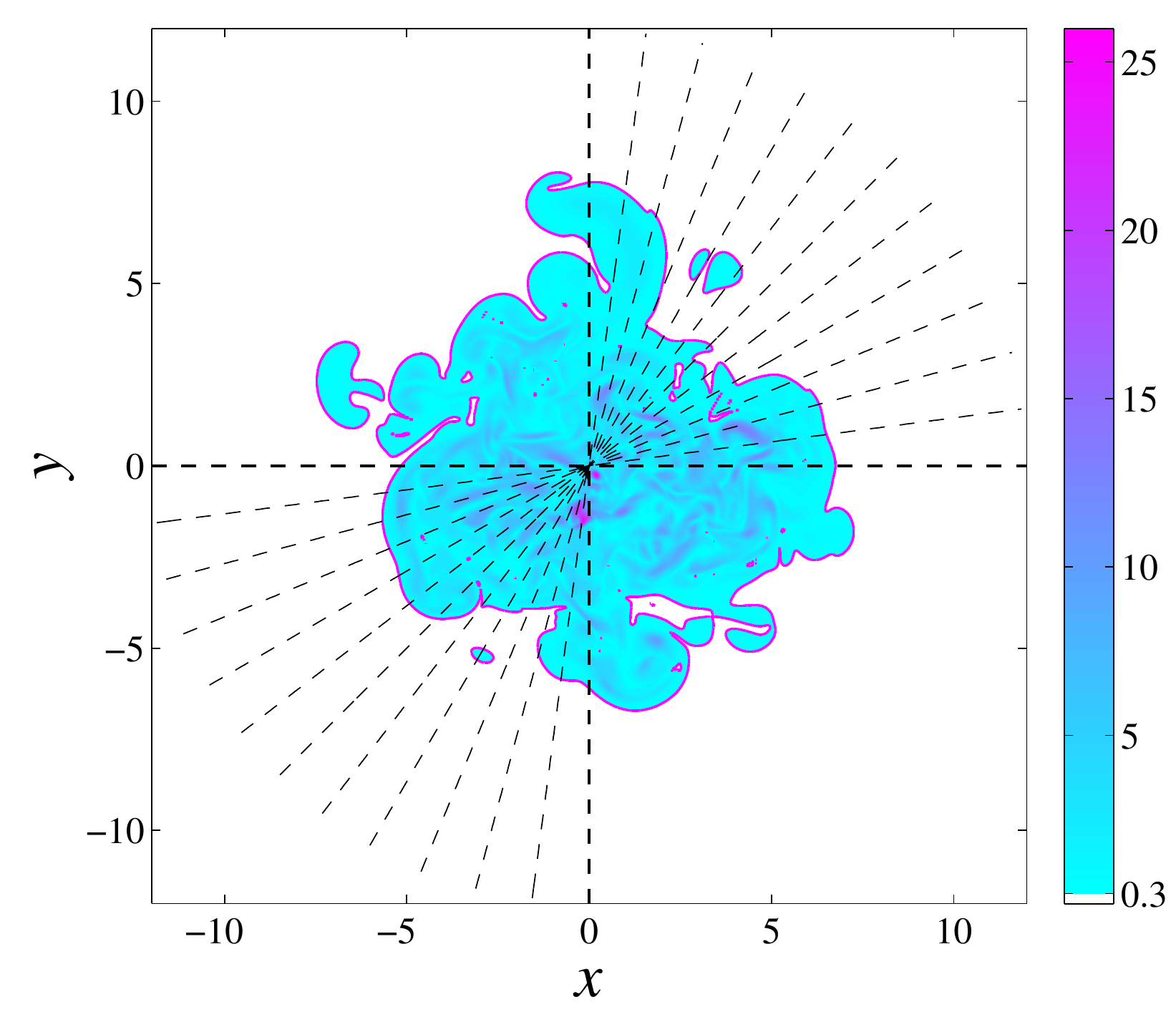}}
\put(0,160){(a)}
\put(70,160){(b)}
\put(0,105){(c)}
\put(70,105){(d)}
\put(0,49){(e)}
\put(70,49){(f)}
\end{overpic}
\caption{(a - e): Instantaneous vorticity magnitude profiles in axial
sections at $z = 35$ and $t = 1995$: (a) $\Phi = 0^{\circ}$; (b) $\Phi
= 15^{\circ}$; (c) $\Phi = 37.5^{\circ}$; (d) $\Phi = 52.5^{\circ}$;
(e) $\Phi = 67.5^{\circ}$. Vorticity is non-dimensionalized using the
local scales, i.e. $\overline{w}_c / \overline{b}_w$. (f) shows the
vorticity contours in a diametral plane at $z = 35$; the jet edge is
plotted at a threshold of $|{\omega}|_{\mathrm{th}=0.3}^*$. The broken
radial lines are drawn in an interval of $7.5^{\circ}$ between $\Phi =
0^{\circ} - 90^{\circ}$. The vorticity profiles in (a) to (e) are
plotted along the dotted lines at respective radial locations.}
\label{fig:inst vort mag prof}
\end{figure}

In order to locate the boundary of a turbulent jet, we look at the
variation of the vorticity modulus $(|\bm{\omega}|)$ across the jet
diameter; for example, Figure \ref{fig:inst vort mag prof} shows that
the instantaneous variation of vorticity magnitude plotted along the
radial direction at various radial locations. It is evident from the
figure that the irrotational and rotational fluid regions are
separated by a very thin and sharp interface. Vorticity shows wild
fluctuations within the core of the jet.

The definition of the edge is determined as a characteristic value of
$|\bm{\omega}|$ connected with the interface layer of relatively rapid
change that is encountered as a test point moves inward toward the jet
core from the ambient (nearly) irrotational flow.

\begin{figure*}
\centering
\begin{overpic}
[width = 16.50 cm, height = 6.4 cm, unit=1mm]
{Fig_box-eps-converted-to.pdf}
\put(85,0){\includegraphics[width = 4.10 cm]{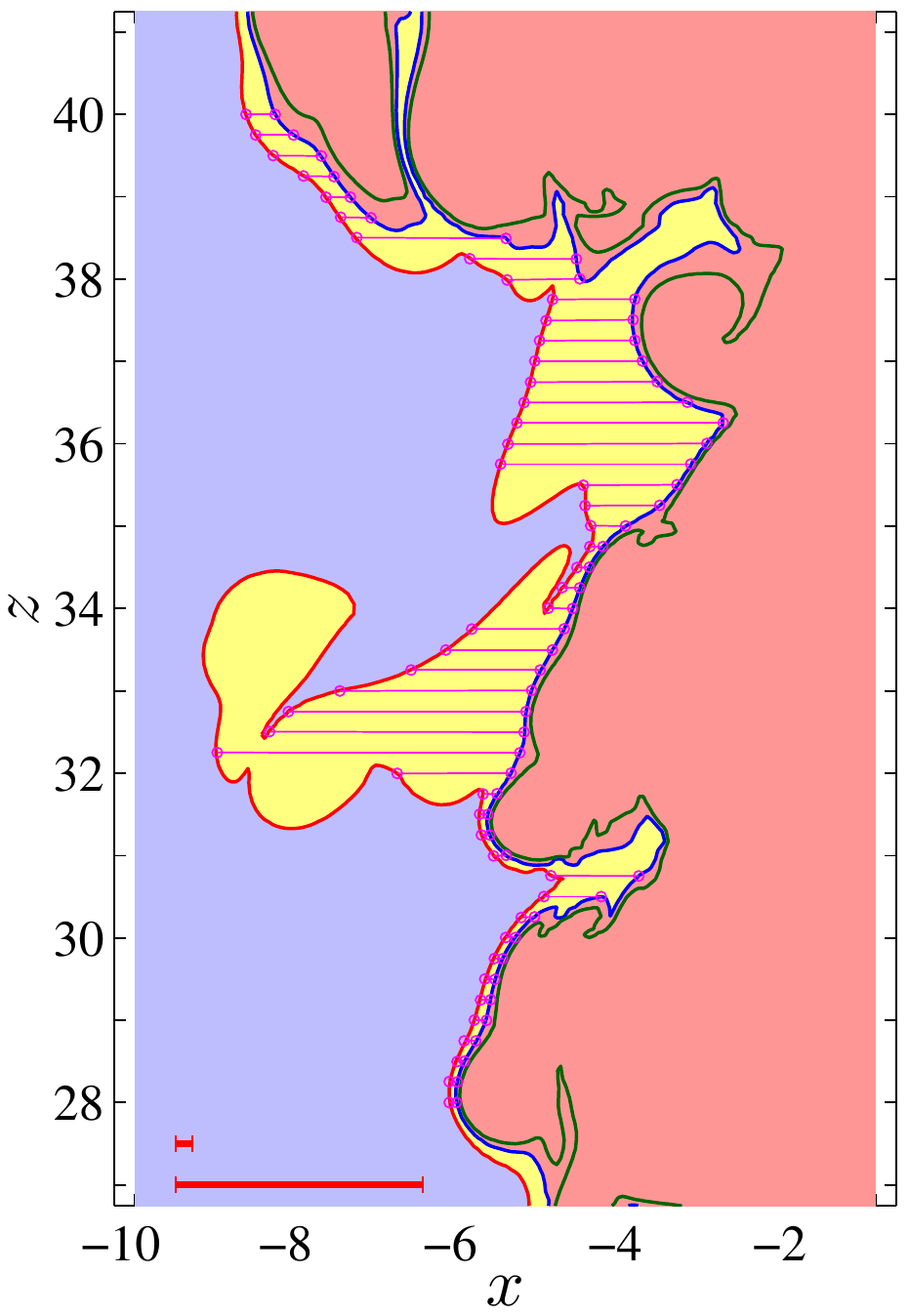}}
\put(126,0){\includegraphics[width = 3.75 cm]{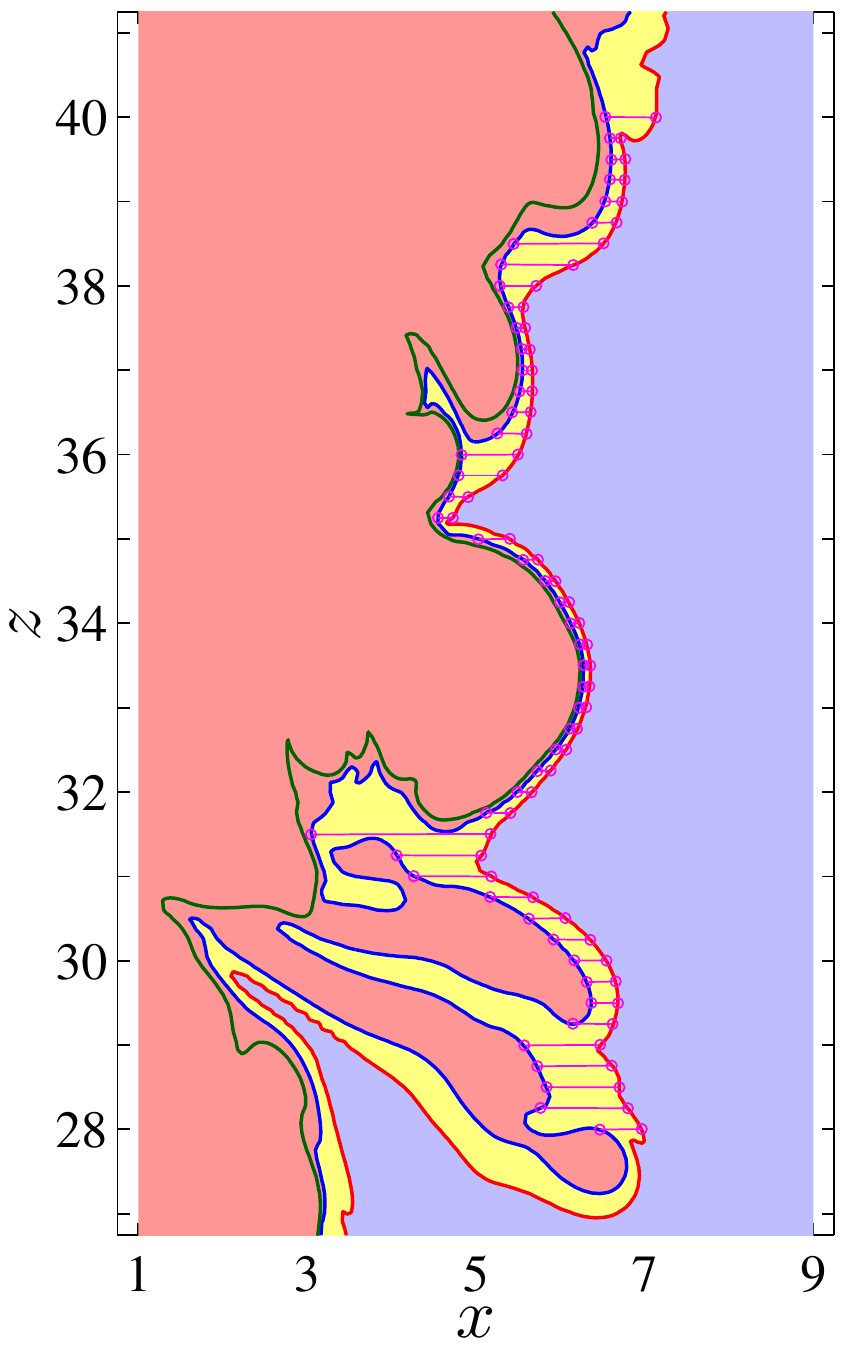}}
\put(0,0){\includegraphics[width = 8.40 cm]{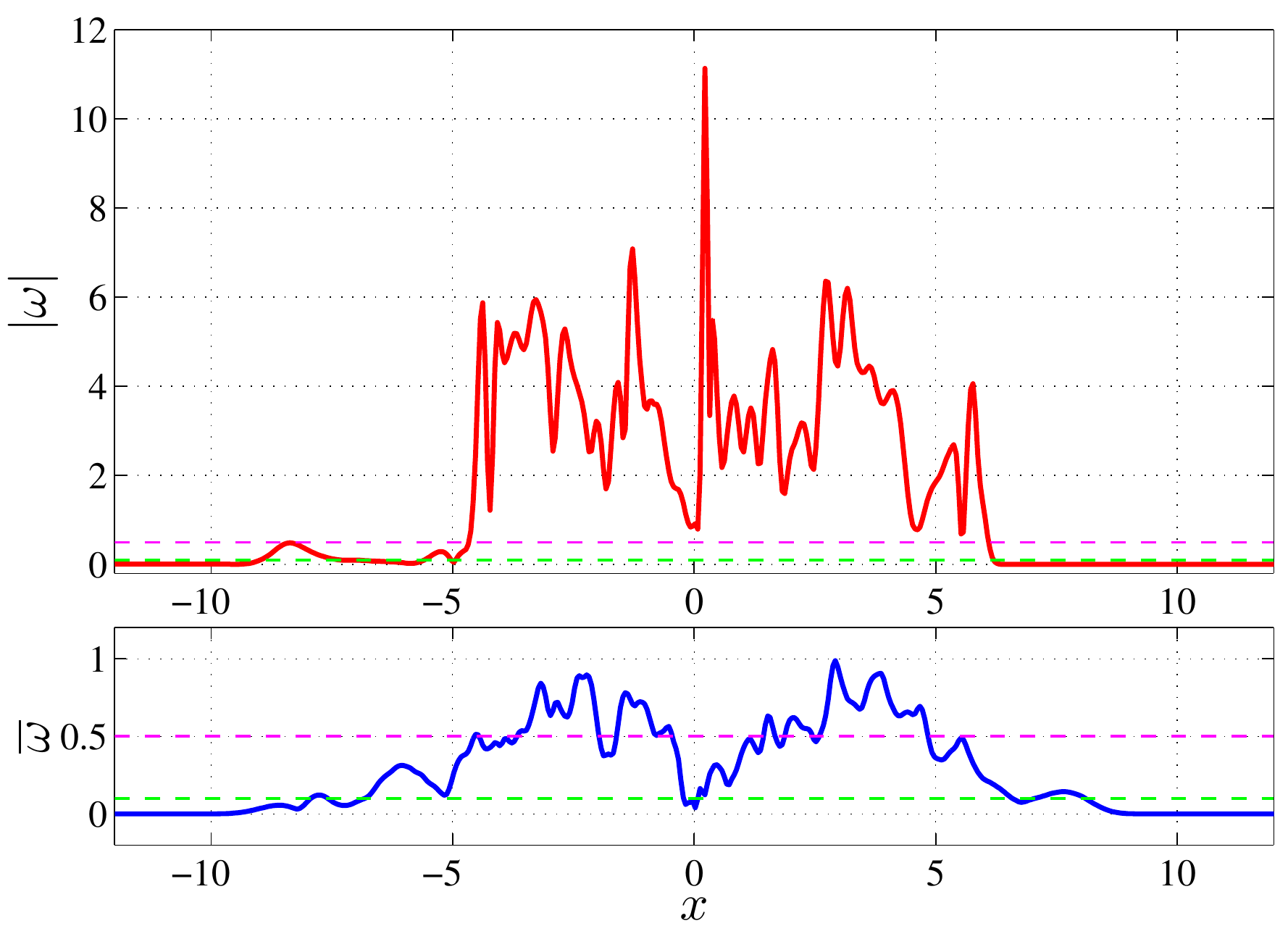}}
\put(1,61){(a)}
\put(88,61){(b)}
\end{overpic}
\caption{(a) Typical variation of the normalized instantaneous
vorticity modulus $|\bm{\omega}|$ (in red) at $t = 925$, compared with
the time-averaged vorticity modulus, $|\overline{\bm{\omega}}|$ (in
blue). The broken horizontal lines are marked at 0.1 (green) and 0.5
(pink). The bumps within $-10<x<-5$ correspond to the two
intersections of the line $z = 34.05$ with the outer viscous buffer
zone (yellow region in the left panel of b).  (b) Axial section in the
$xz$ plane at $t = 925$. The $|\bm{\omega}| = 0.1$ and $0.5$ contours
in the axial cross section of the jet showing the wide
variability in the separation distance between the two contours that
bound the yellow region (viscous buffer zone); lighter reddish region
is the turbulent core, lighter lavender region is the ambient. The
red, blue and dark green boundaries are marked respectively at
thresholds $|\bm{\omega}| = 0.1, 0.5, 1$. The thick red horizontal
lines at the bottom in the left panel show $5\eta$ (short line) and
$5\lambda$ (long line).}
\label{fig:vort-prof} 
\end{figure*}

Figure \ref{fig:vort-prof}(a) displays the variation of the
total vorticity modulus $|\bm{\omega}|$ across the jet diameter at $z
= 34.05$. For comparison the magnitude of the mean flow vorticity is
also shown.  It is first of all seen that the fluctuating vorticity
$|\bm{\omega|}$ in the core can be an order of magnitude larger than
the time mean flow vorticity.

\begin{figure}
\centering
\begin{overpic}
[width = 13.50 cm, height = 7.0 cm, unit=1mm]
{Fig_box-eps-converted-to.pdf}
\put(30,0){\includegraphics[width = 7.0 cm]{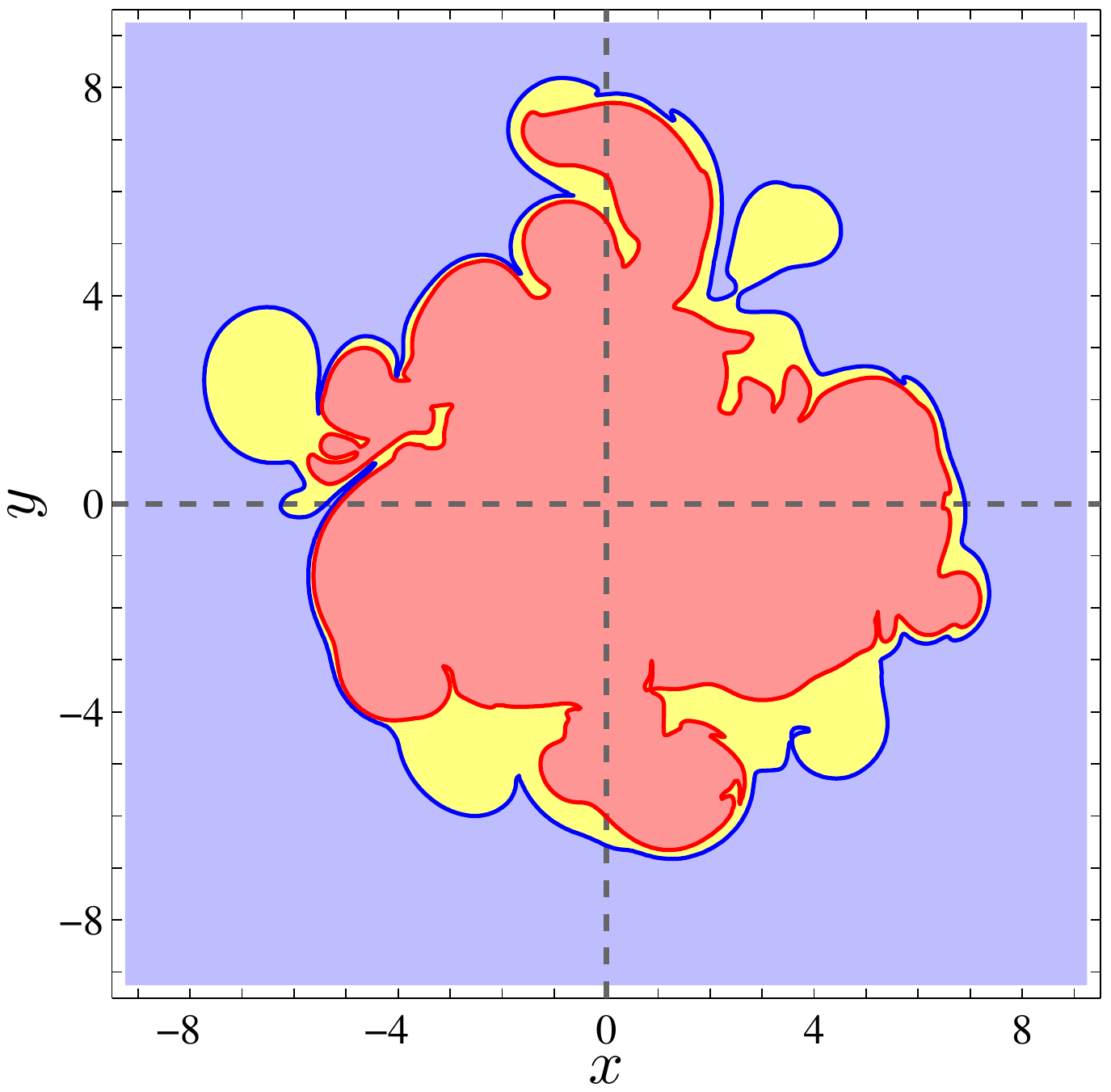}}
\end{overpic}
\caption{Instantaneous sectional view of the jet in a diametral plane
at $z = 35$ showing the boundary of the jet at $t = 1995$ for two
thresholds, namely, 0.1 (blue) and 0.5 (red). These boundaries
demarcate the flow into three regions: Irrotational (violet), Viscous
(yellow) and Turbulent (light red).}
\label{fig:inst diam jet boudaries}
\end{figure}

If we take a test point that moves towards the turbulent core from the
ambient in the right in \ref{fig:inst vort mag prof}(a), we see that
$|\bm{\omega}|$ remains nearly constant till the test point crosses $x
\approx 7$ where it encounters a sudden and relatively rapid change in
$|\bm{\omega}|$ from nearly 0.1 to 4, thereafter showing wild
fluctuations till it reaches $x \approx -4.5$ where $|\bm{\omega}|$
drops steeply all the way to about 0.5, then becomes less steep, and
finally takes the form of two noticeable but smooth bumps, the one
farther away being the bigger, beyond which $|\bm{\omega}|$ remains
nearly constant at a small value. We will shortly demonstrate (from
comparison with Figure \ref{fig:vort-prof} (b)) that these two bumps
correlate with the outer buffer zone with its small but apparently
non-turbulent vorticity.  For several vorticity profiles studied, it
is observed that the $|\bm{\omega}|$ invariably shows a rapid and
sudden increase beyond $|\bm{\omega}| = 0.5$; however, up to
$|\bm{\omega}| = 0.5$ the vorticity profile shows both rapid increase
(e.g. at $x \approx 6.5$) as well as slow increase (e.g. $x \approx
-5$). In order to explore the physical picture of the flow near the
jet boundary, we plot the vorticity $|\bm{\omega}|$ field in axial
sections near the left and right boundaries of the jet as shown in
Figure \ref{fig:vort-prof}(b). We also draw three boundaries defined
by a threshold on $|\bm{\omega}| = 0.1, 0.5, 1.0$ respectively. We
observe that the two boundaries for $|\bm{\omega}| = 0.5, 1$ mostly
lie close to each other except in a small region $2<x<7$ and
$30.5<z<31$. However, the separation between boundaries defined by
$|\bm{\omega}| = 0.5$ and $|\bm{\omega}| = 0.1$ show a different
trend: the separation between these two boundaries can be very small
at some places, but can vary appreciably at certain other locations
along the boundary of the jet; for example, in
Fig.~\ref{fig:vort-prof}(b) the two boundaries are close to each other
in some areas (e.g., $-8<x<-6$ and $38<z<45$) and well separated in
others (e.g., $-8<x<-3$ and $31<z<38$). Figure \ref{fig:inst diam jet
boudaries} shows the vorticity field and the jet boundaries defined
based on two thresholds (namely 0.1 and 0.5) in a diametral
plane. Just as in the case of the axial plane, it is clearly visible
in Figure \ref{fig:inst diam jet boudaries} that the two boundaries
are quite close to each other at some places and widely separated at
others.

In the region bounded by $|\bm{\omega}| = 0.5$ and $|\bm{\omega}| =
0.1$, vorticity shows interesting behaviour - while at some places it
increases rapidly, it shows very slow spatial frequency of variations
some other places.  This lays the basis to define two boundaries for
the jet instead of a single boundary.

Based on extensive studies of these vorticity profiles across the
edge, it is found convenient to introduce two distinct edges or
boundaries of the jet.  The first is located at $|\bm{\omega}| = 0.5$,
beyond which $|\bm{\omega}|$ varies rapidly to 1.0 or above in the
turbulent core. This defines the turbulent / non-turbulent (T/NT)
\textquoteleft{inner}' boundary of the turbulent jet.  For reasons
that will be clear shortly, we find it necessary to define an
\textquoteleft{outer}' boundary at $|\bm{\omega}| = 0.1$, which
similarly locates the rotational / irrotational (R/IR) boundary of the
jet. The contour surfaces $|\bm{\omega}| = 0.1$ and $0.5$ will thus be
referred to in the sequel as the \textquoteleft{outer}' (R/IR) and
\textquoteleft{inner}' (T/NT) boundaries of the jet respectively.  As
expected, the edges of the jet are highly convoluted (especially the
inner boundary, which is a fractal curve \citep{Sreenivasan_JFM_1986,
sreenivasan1989mixing}). When the separation between the inner and
outer boundaries is large, we call the region between them an
\textquoteleft{outer buffer zone}'.

It would be worthwhile to study the variation of separation distance
between inner and outer boundaries. Figure \ref{fig:vort-prof}(b)
shows how the distance between the two boundaries varies along the
inner boundary of the jet.  One simple measure of this distance is the
separation along the $x$ axis from the inner to the outer boundary on
either side of the jet. It is seen that this inter-boundary separation
varies from $0.090$ to $3.67$ in the left boundary and from $0.077$ to
$2.13$ in the right boundary. For comparison, estimates of the two
relevant scales in the problem, namely, the Kolmogorov length $(\eta)$
and the Taylor microscale $(\lambda)$ at $z\approx33$, $x,y=0$ are
shown in Fig.~\ref{fig:vort-prof}(b). It is seen that inter-boundary
separation varies from the order of the Kolmogorov scale to the Taylor
microscale; however, the changes in $|\bm{\omega}|$ from 0.5 toward
1.0 and above in the turbulent core is almost always sharp, and has a
length scale comparable to the Kolmogorov scale.  In the light of the
many discussions on this issue in the literature
\citep{Silva_PoF_2010, da2011role, da2008invariants,
Westerweel_JFM_2009, hunt2006mechanics}, we conclude that both scales
seem to have a role to play, from the sharp T/NT interface layer of
thickness $\mathcal{O}(\eta)$ to the larger lateral extent of the
viscous buffer zone with a dimension of order $\lambda$.
Understanding the role of these different scales, and the question of
whether this difference in scales persists at higher Reynolds numbers
requires further investigation.

Note that there have been other proposals as well for defining two
boundaries, but, using slightly different criteria; for example,
\citet{Reeuwijk_JFM_2014} used thresholds on enstrophy to define inner
and outer boundaries for a temporal plane jet. Their inner boundary,
that separates the turbulent core from the buffer region, represents
the enstrophy threshold for which the interface propagation velocity
$v_n$ is zero, and the outer boundary, that separates the viscous
superlayer (VSL) from the buffer region, represents the threshold
below which the enstrophy production is negligible. (Also, they define
a buffer zone which possibly is inward of our inner boundary, and will
here be called the \textquoteleft{inner buffer zone}'.)  Note that the
thresholds used by \citet{Reeuwijk_JFM_2014} in the case of a temporal
plane jet at $Re = 5000$ for defining the boundaries are not
applicable in our case as the thresholds are a function of Reynolds
number and also of the dimensionality and nature of the flow.

\subsection{Nature of vorticity field between T/NT and R/IR interfaces}
\label{sec:vorticity from T/NT to R/IR}

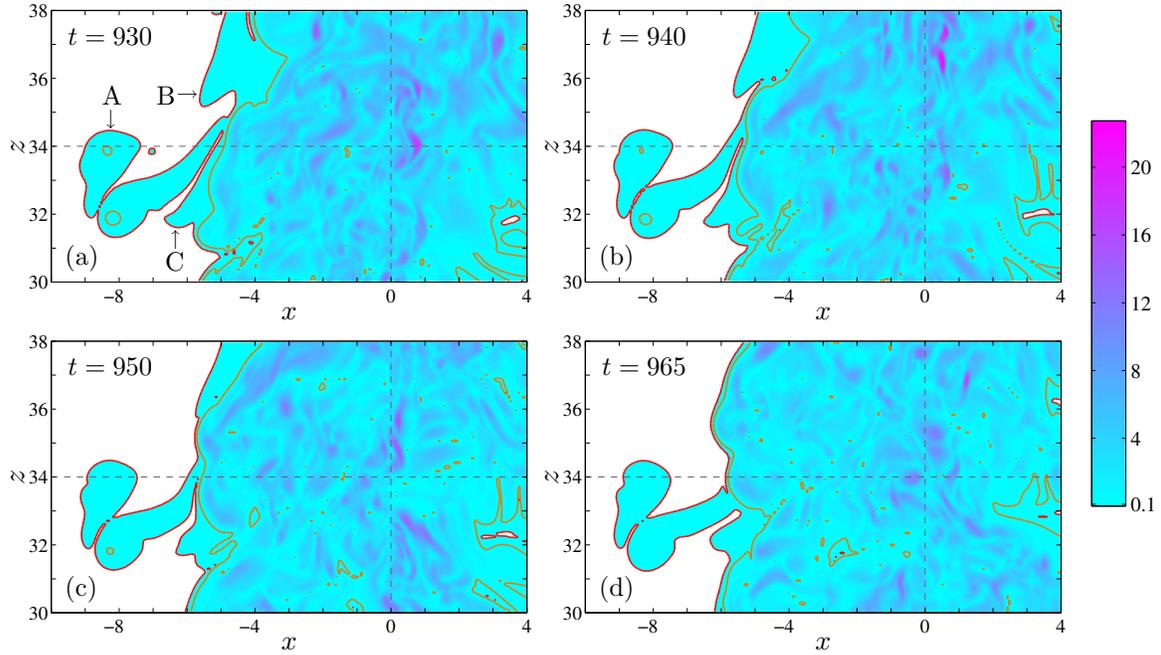
\begin{figure}
\centering
\begin{overpic}
[width = 16.0 cm, height = 8.7 cm, unit=1mm]
{Fig_box-eps-converted-to.pdf}
\put(5,44){\includegraphics[width = 7.0 cm]{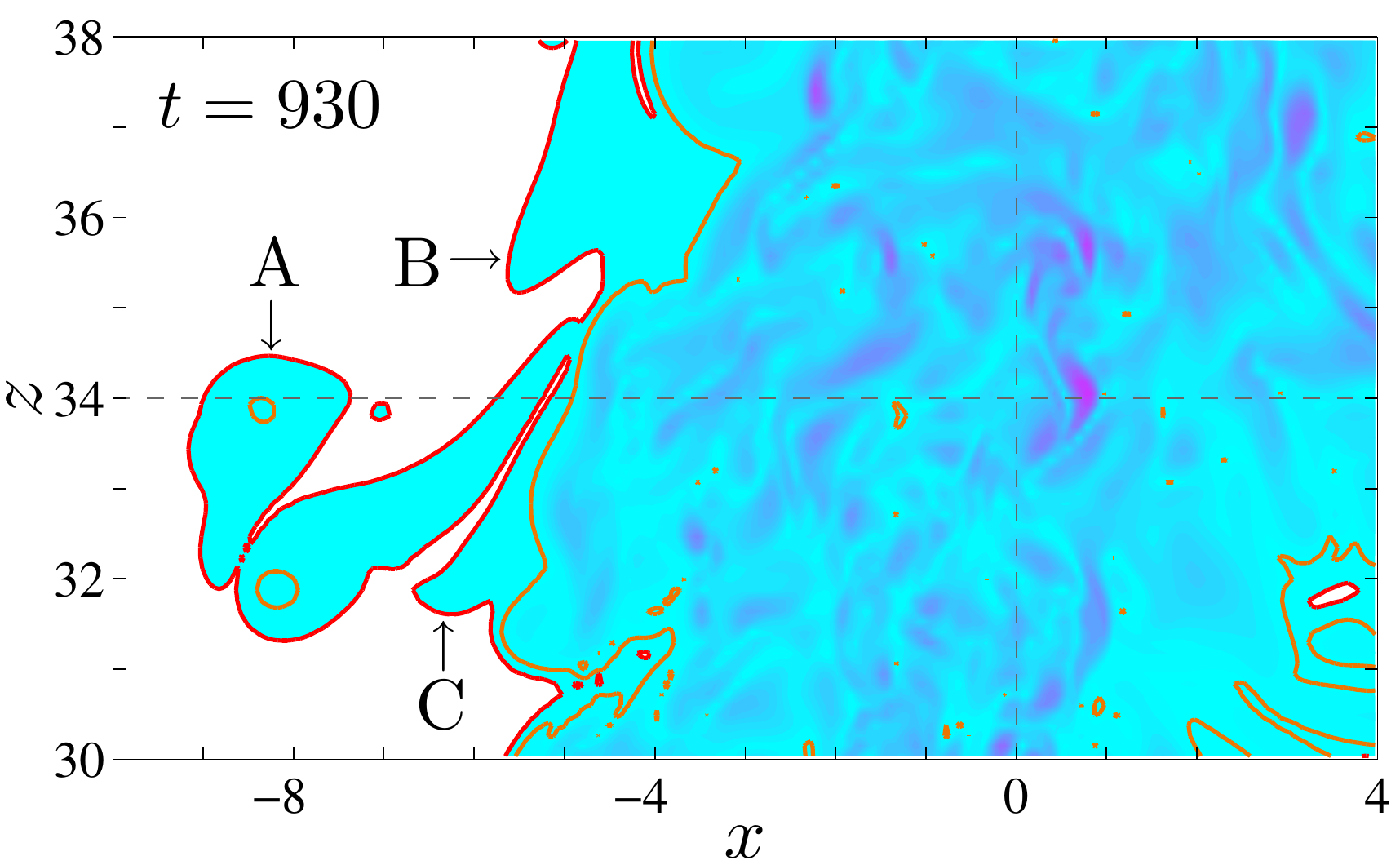}}
\put(76,44){\includegraphics[width = 7.0 cm]{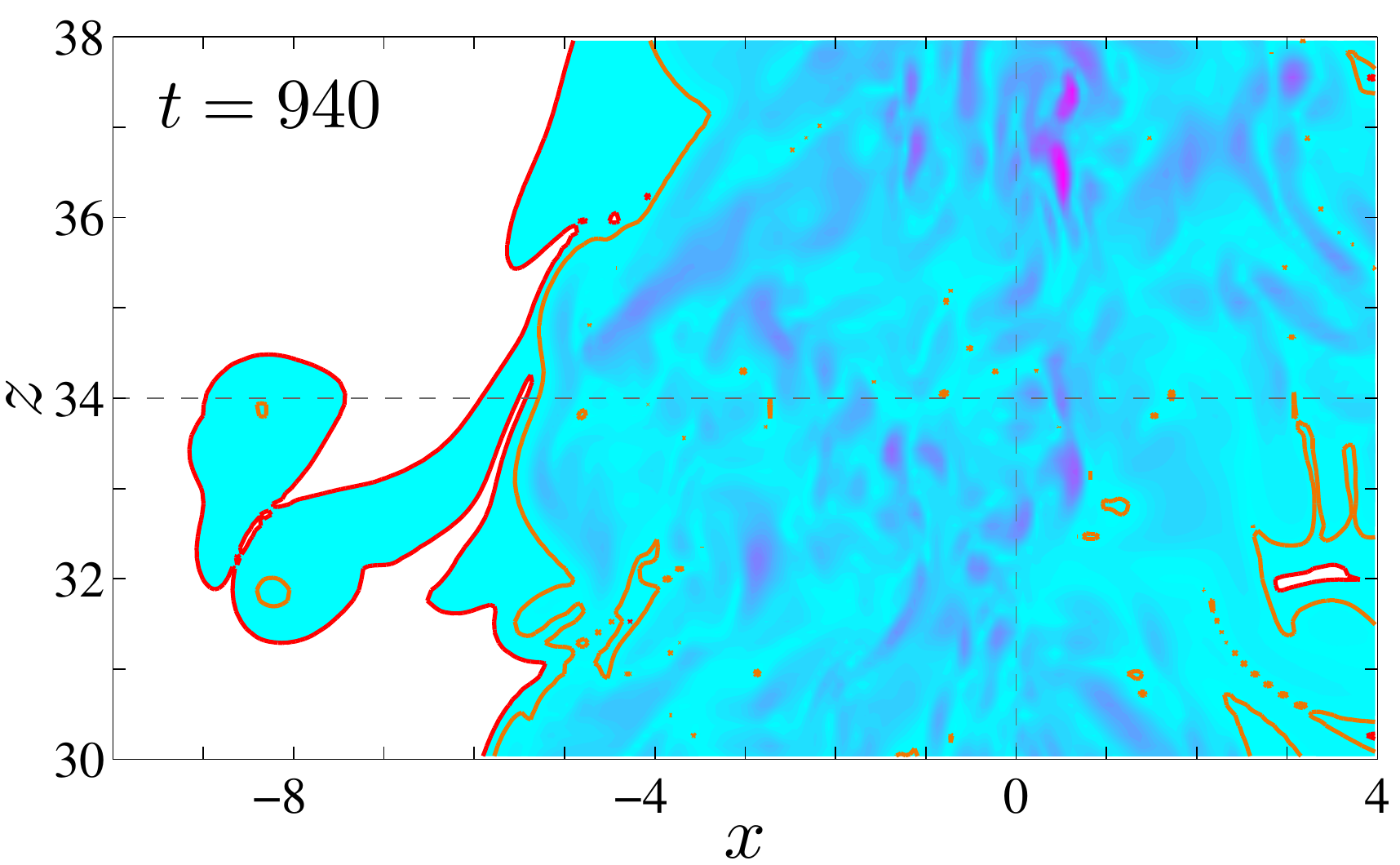}}
\put(5,0){\includegraphics[width = 7.0 cm]{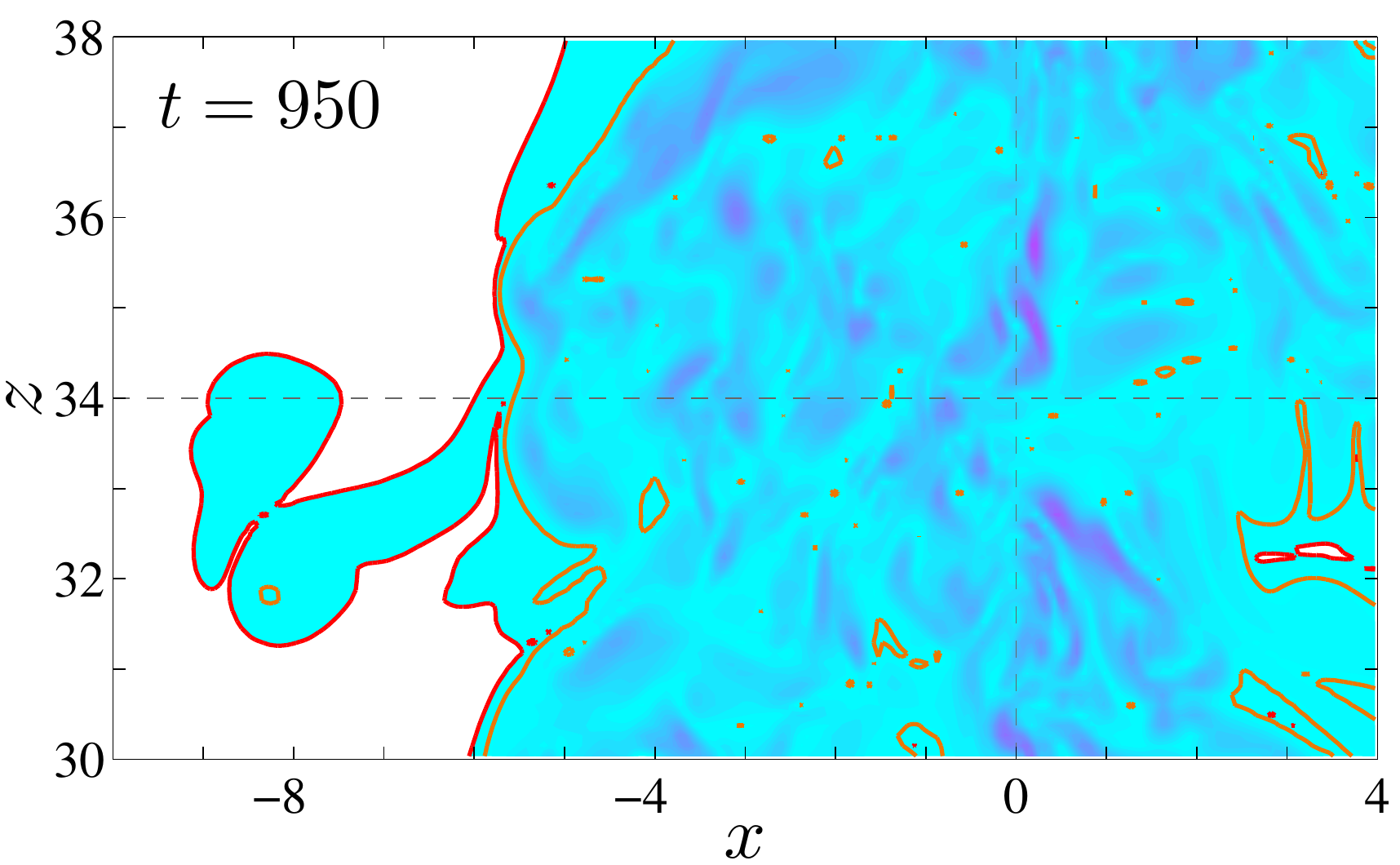}}
\put(76,0){\includegraphics[width = 7.0 cm]{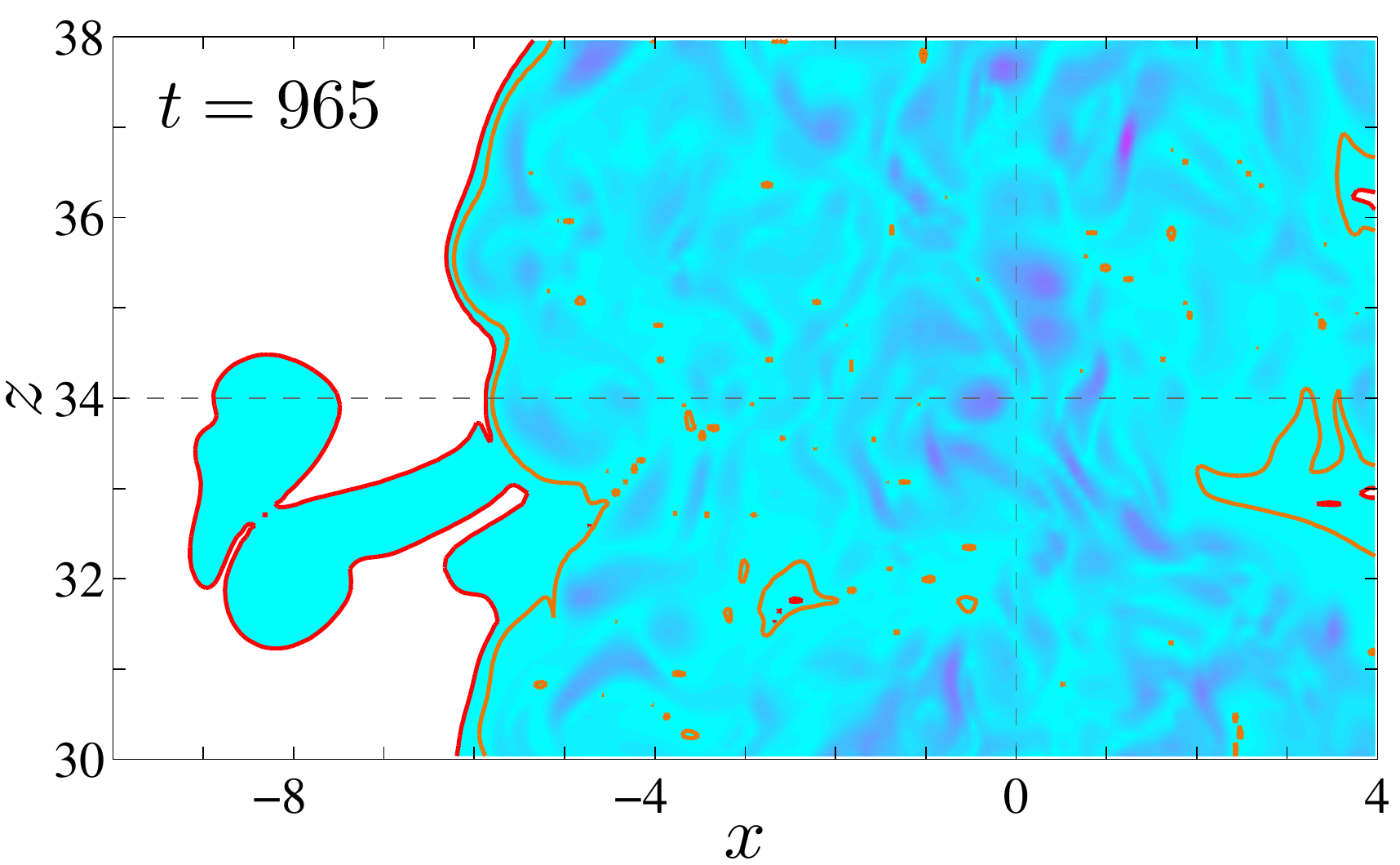}}

\put(148,17){\includegraphics[width = 1.0 cm]{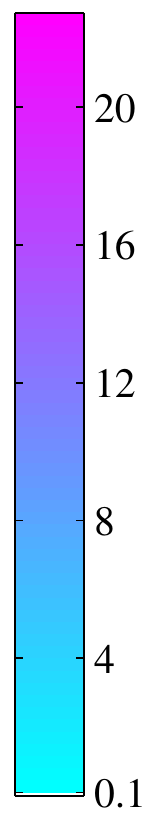}}
\put(12.5,51.5){\fontsize{10}{12}\selectfont {(a)}}
\put(83.5,51.5){\fontsize{10}{12}\selectfont {(b)}}
\put(12.5,7.5){\fontsize{10}{12}\selectfont {(c)}}
\put(83.5,7.5){\fontsize{10}{12}\selectfont {(d)}}
\end{overpic}
\caption{Temporal evolution of a part of the jet. Three areas in the
figure labeled as A, B and C illustrate important features of the
evolution (see text for discussion).}
\label{fig:vort-mag-axial}
\end{figure}

Figure \ref{fig:vort-mag-axial} shows the time evolution (over $930
\leqslant t \leqslant 965$) of a part of the jet $(32<z<36)$ that is
accurately self-preserving. Region A hardly changes in shape or in
vorticity range over this whole time interval.  On the other hand,
region B keeps continuously shrinking, and has disappeared at $t =
965$.  This disappearance is largely due to the outward motion of the
inner boundary till, in (d), the two boundaries are very close to each
other in the interface above $z = 34$.  Region C undergoes minor
changes in shape, in part again because of the outward movement of the
inner boundary. Between (a) and (d) it is seen that the turbulent core
has expanded, with a general reduction in the larger vorticity-sparse
areas seen near the boundaries in the earlier images.

\begin{figure}
\centering
\begin{overpic}
[width = 14.0 cm, height = 5.2 cm, unit=1mm]
{Fig_box-eps-converted-to.pdf}
\put(35,0){\includegraphics[width = 6.5 cm]{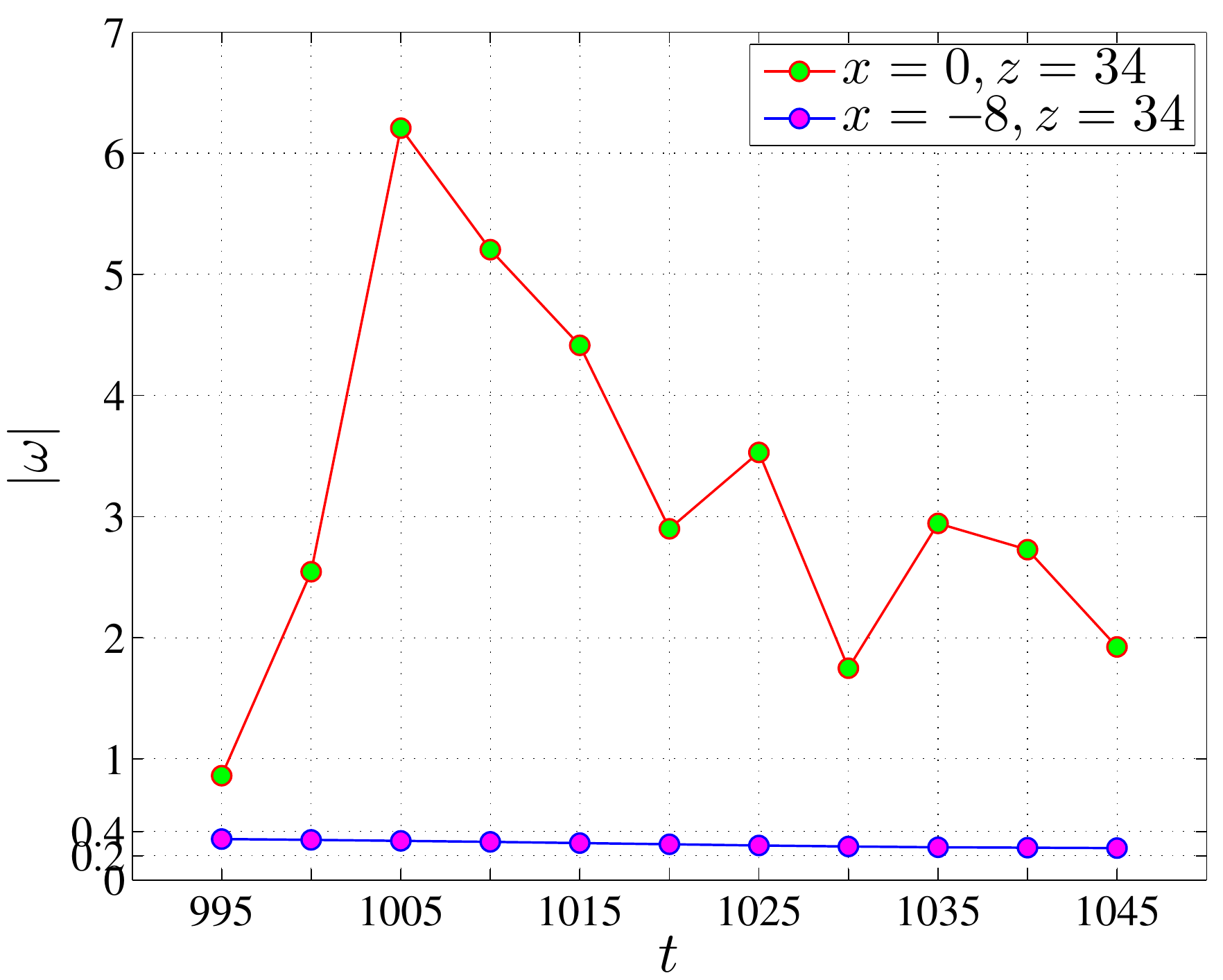}}
\end{overpic}
\caption{Temporal variation of the total vorticity modulus at two
points in the plane $z = 34$, one on the centerline $(0, 0, 34)$ and
the other at $(-8, 0, 34)$ in the outer buffer region between the
inner and outer boundaries.}
\label{fig:vort-mag-prof}
\end{figure}

Figure \ref{fig:vort-mag-prof} compares the variation of
$|\bm{\omega}|$ with time at two points. The vorticity at the center
line shows substantial variation in time, whereas that in area A in
the outer buffer zone is more than an order of magnitude lower and
exhibits a slow and gentle decay with hardly any fluctuation that can
be associated with turbulence. This justifies why such a buffer zone
can be characterized as \textquoteleft{viscous}', and the structure at
zone A in Fig.~\ref{fig:vort-mag-axial} may be called a
\textquoteleft{viscous tongue}', most probably a relic or fossil from
an earlier excursion of in- or out-of-plane vorticity from the core or
buffer zone into an ambient nearly at rest.

\subsection{Interface thickness}
\label{Interface thickness}

We are interested in finding out the thickness of the jet interface,
i.e., the interface which separates the rotational mass, which
constitutes the turbulent jet fluid, and the irrotational mass which
is outside the jet core.

\begin{figure}[!h]
\centering
\begin{overpic}
[width = 13.50 cm, height = 5.3 cm, unit=1mm]
{Fig_box-eps-converted-to.pdf}
\put(0,0){\includegraphics[width = 6.75 cm]{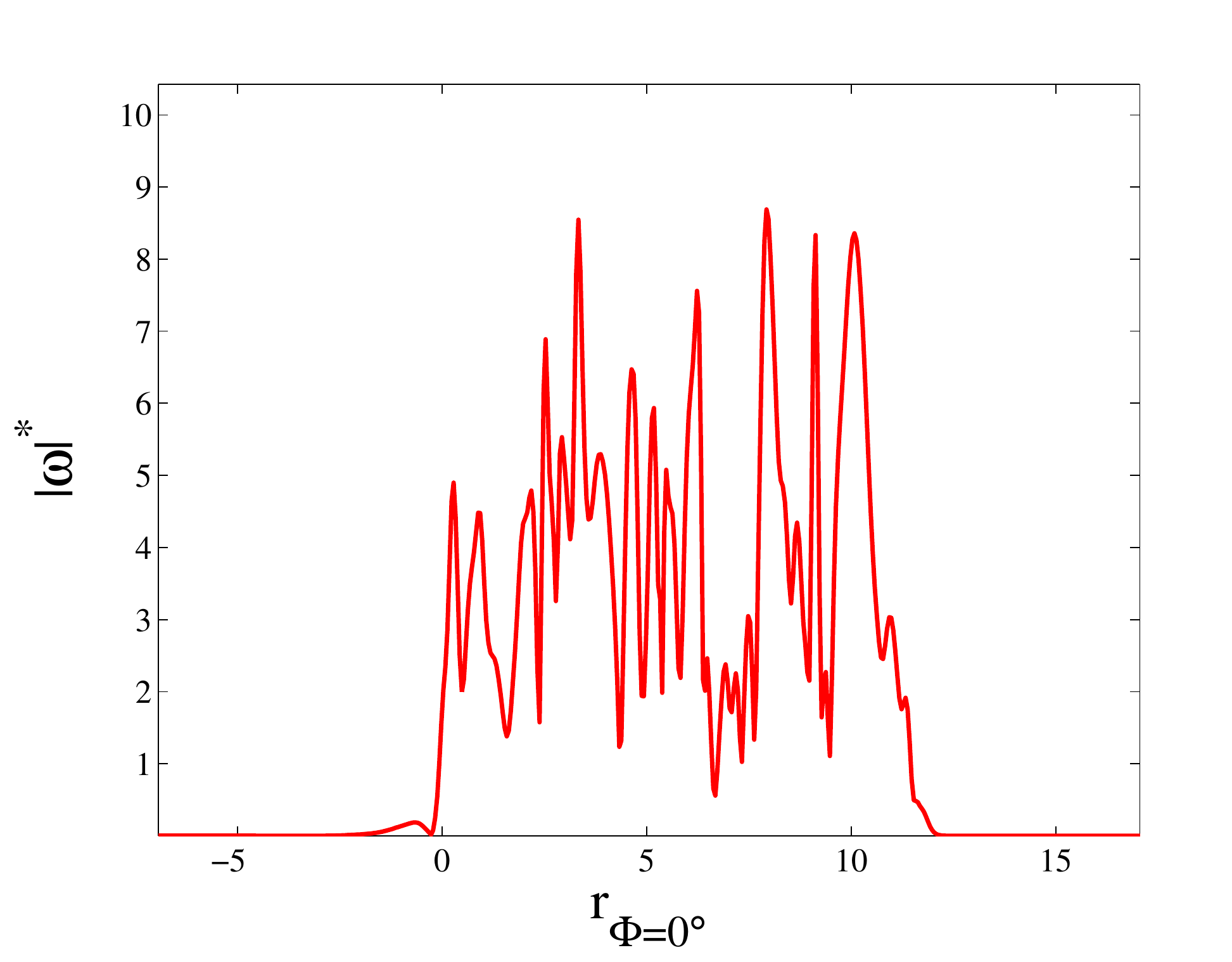}}
\put(70,0){\includegraphics[width = 6.75 cm]{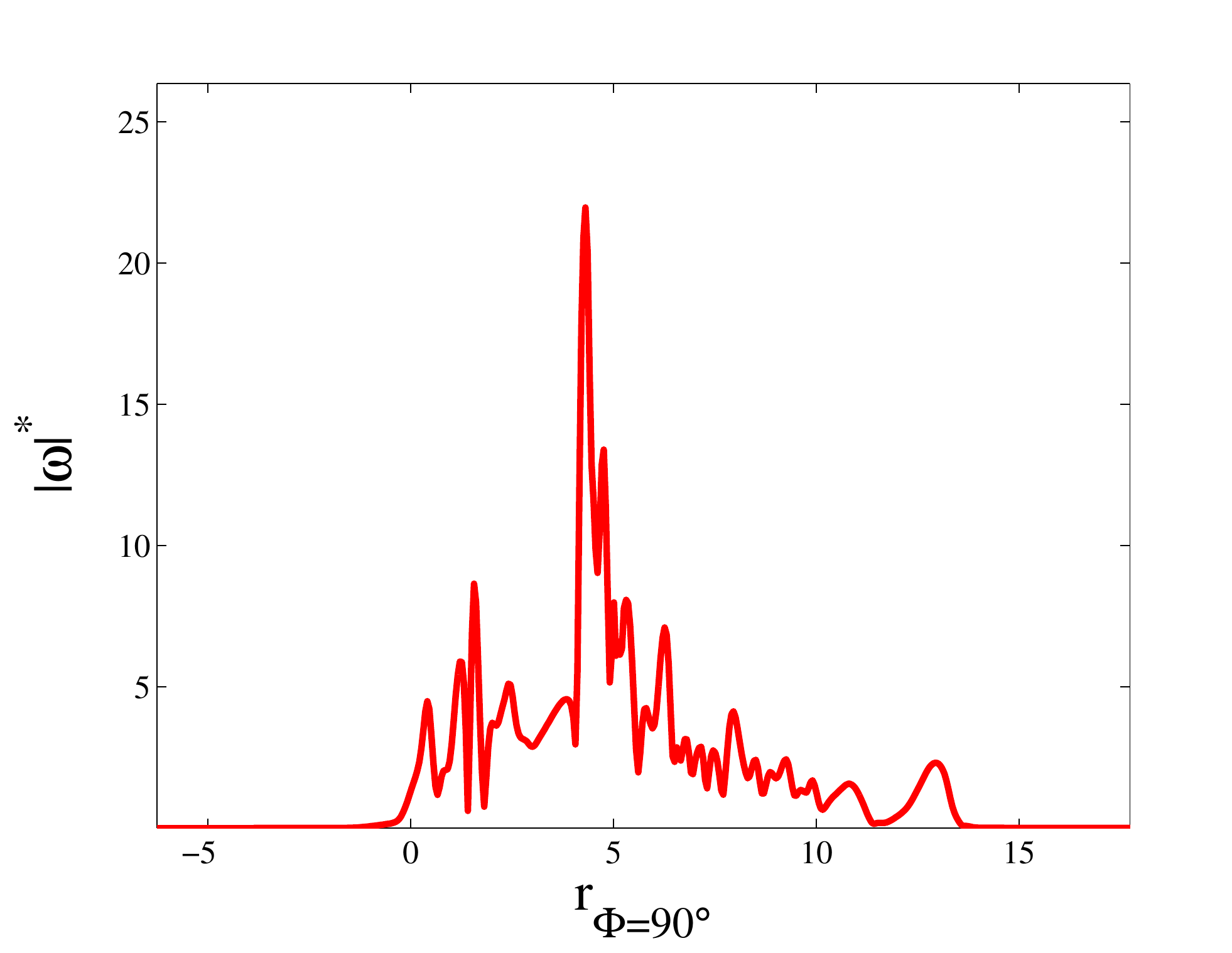}}
\put(0,49){(a)}
\put(70,49){(b)}
\end{overpic}
\caption{Instantaneous vorticity magnitude profiles at $z = 35$ in two
axial sections: (a) $\Phi = 0^{\circ}$; (b) $\Phi = 90^{\circ}$; $t =
1995$. Vorticity is non-dimensionalized using the local scales,
i.e. $\overline{w}_c / \overline{b}_w$. Note that for doing the
edge-based averaging, the origin of the coordinates system in the
radial direction is shifted to the edge of the jet on the left side.}
\label{fig:inst vort mag prof-Int_Ori_LHS}
\end{figure}

\begin{figure}[!h]
\centering
\begin{overpic}
[width = 16 cm, height = 17 cm, unit=1mm]
{Fig_box-eps-converted-to.pdf}
\put(20,0){\includegraphics[width = 12.5 cm]{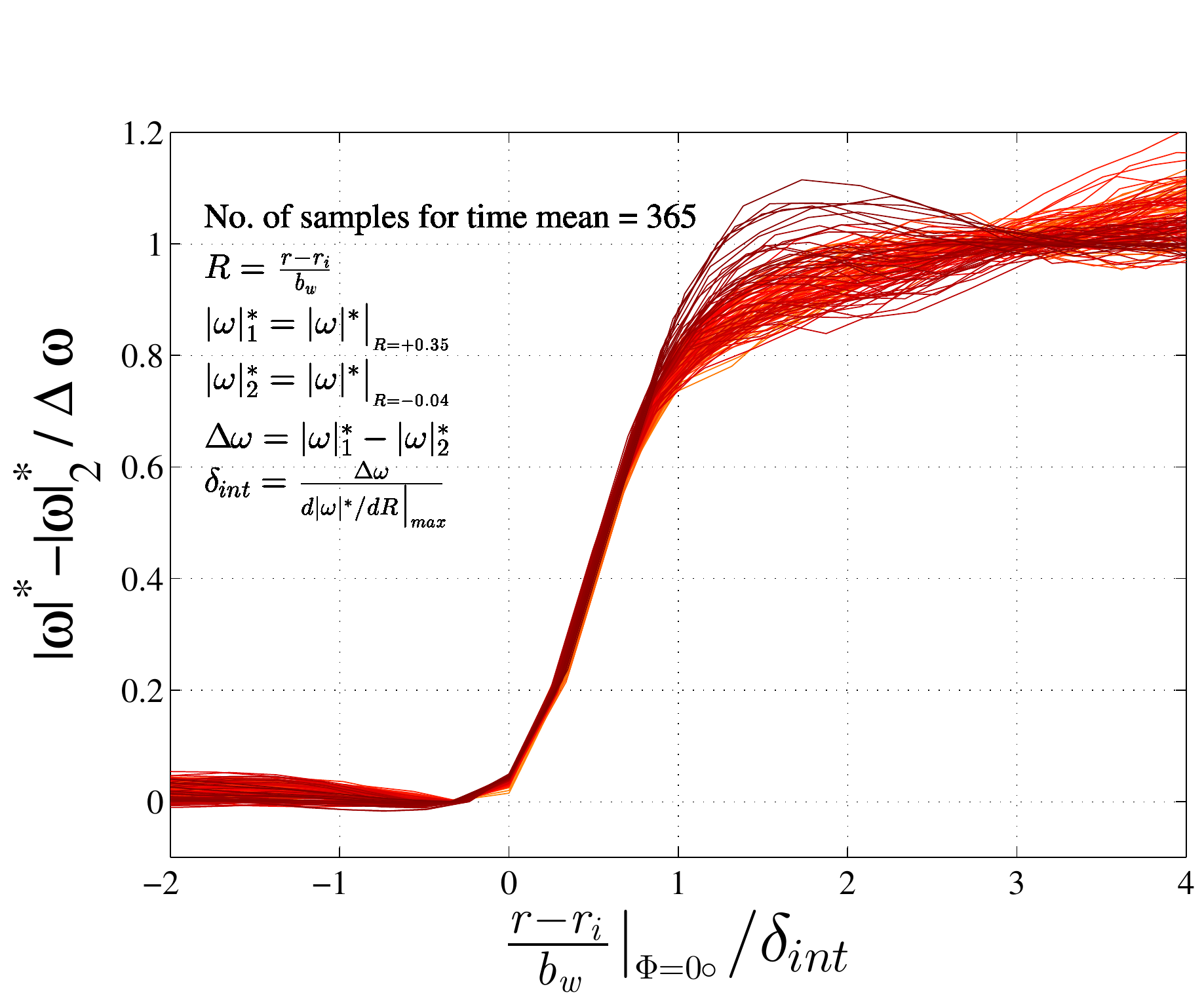}}
\put(0,102){\includegraphics[width = 8.0 cm]{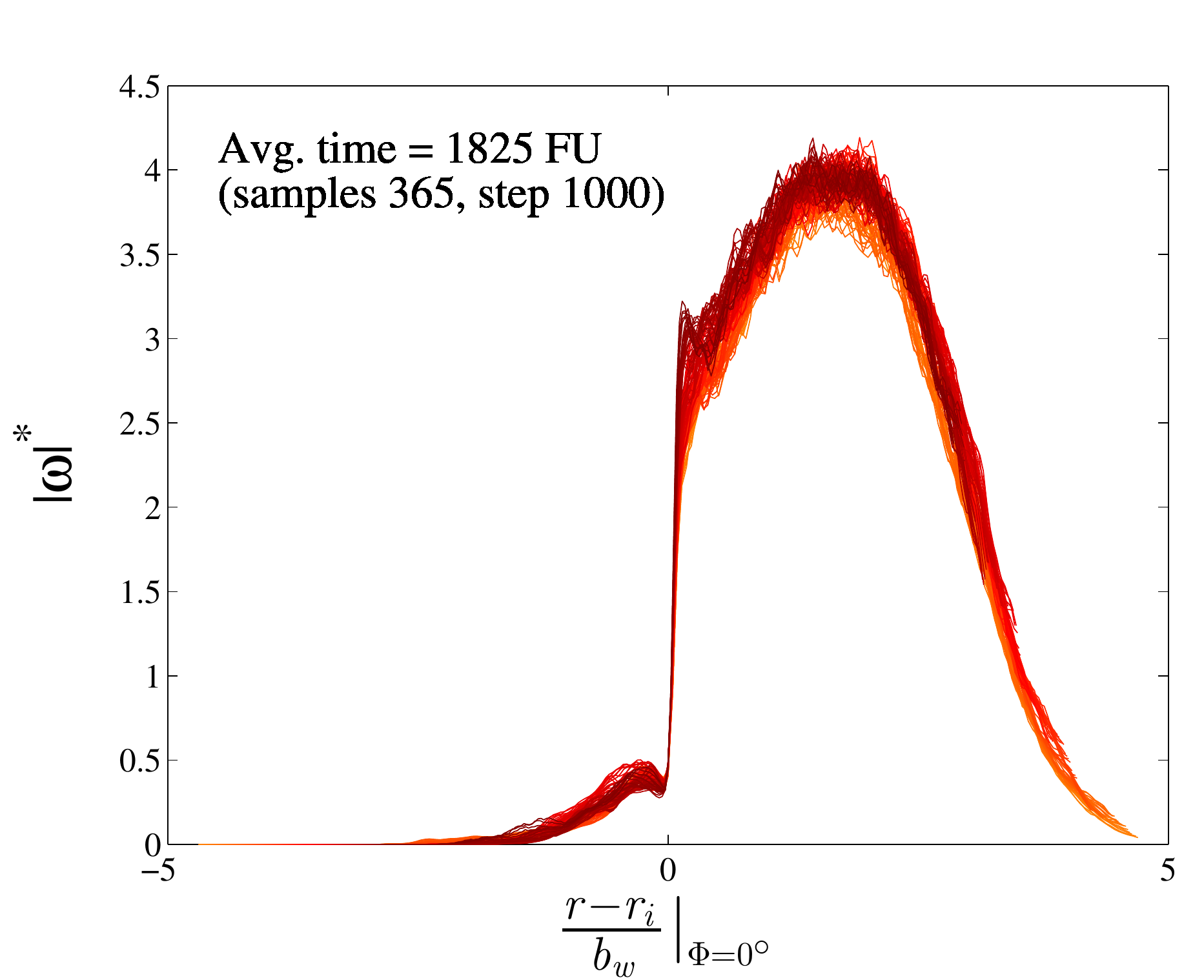}}
\put(80,102){\includegraphics[width = 8.0 cm]{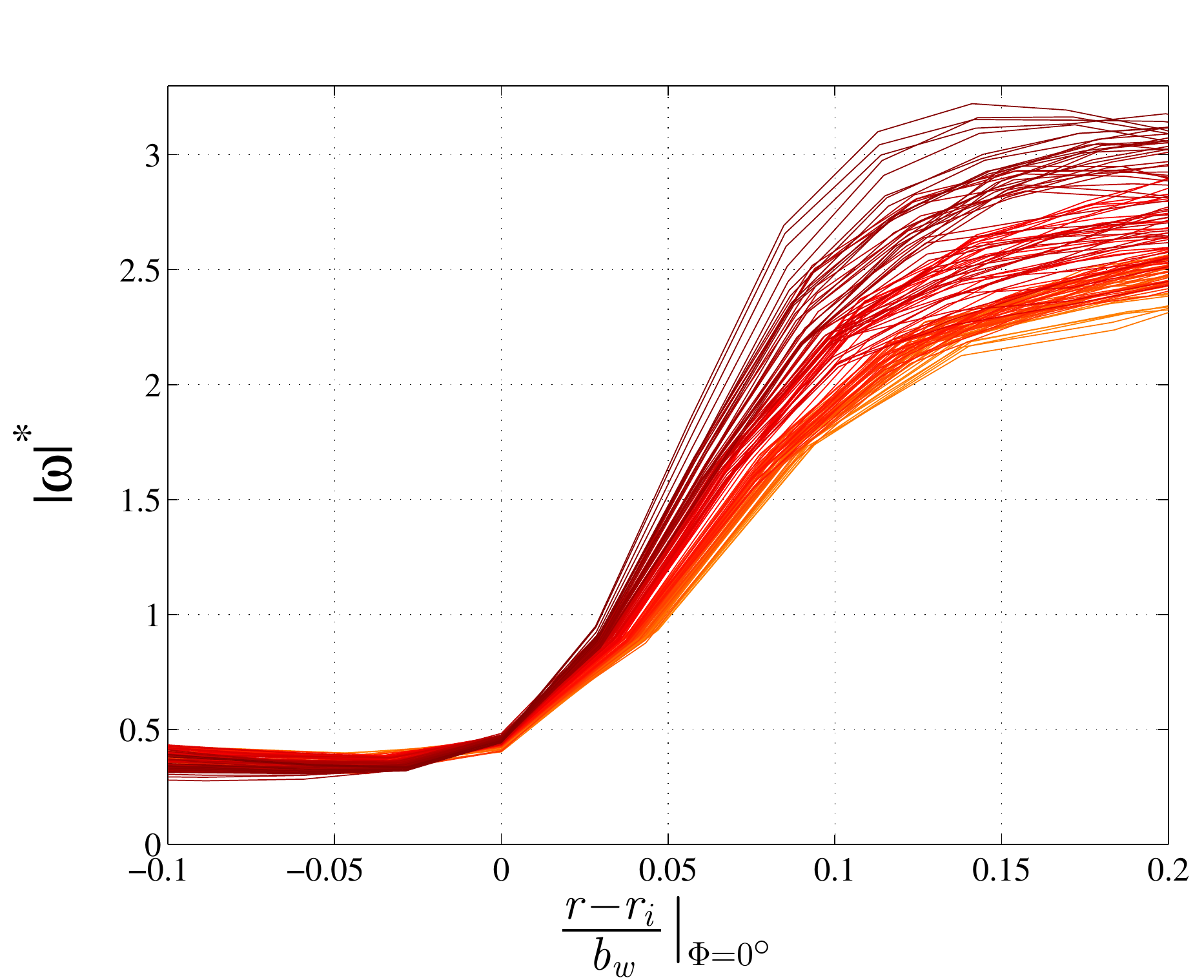}}
\put(0,167){(a)}
\put(83,167){(b)}
\put(18,90){(c)}
\end{overpic}
\caption{Edge-based time mean data (see text for details); $z = 25 -
40, \Phi = 0^{\circ}$. The data is averaged over 1825 flow units. The
edge is defined at $|\omega|_{th} = 0.5$. Vortcity is normalized with
the local scales, namely, $\overline{w}_{c}$ and $\overline{b}_{w}$
Radial profiles of vorticity magnitude profiles: (a) across the entire
section of the jet; (b) zoomed in near the interface; (c) zoomed in
view with different normalization as mentioned in the legend.}
\label{fig:edge based time mean - vort prof}
\end{figure}
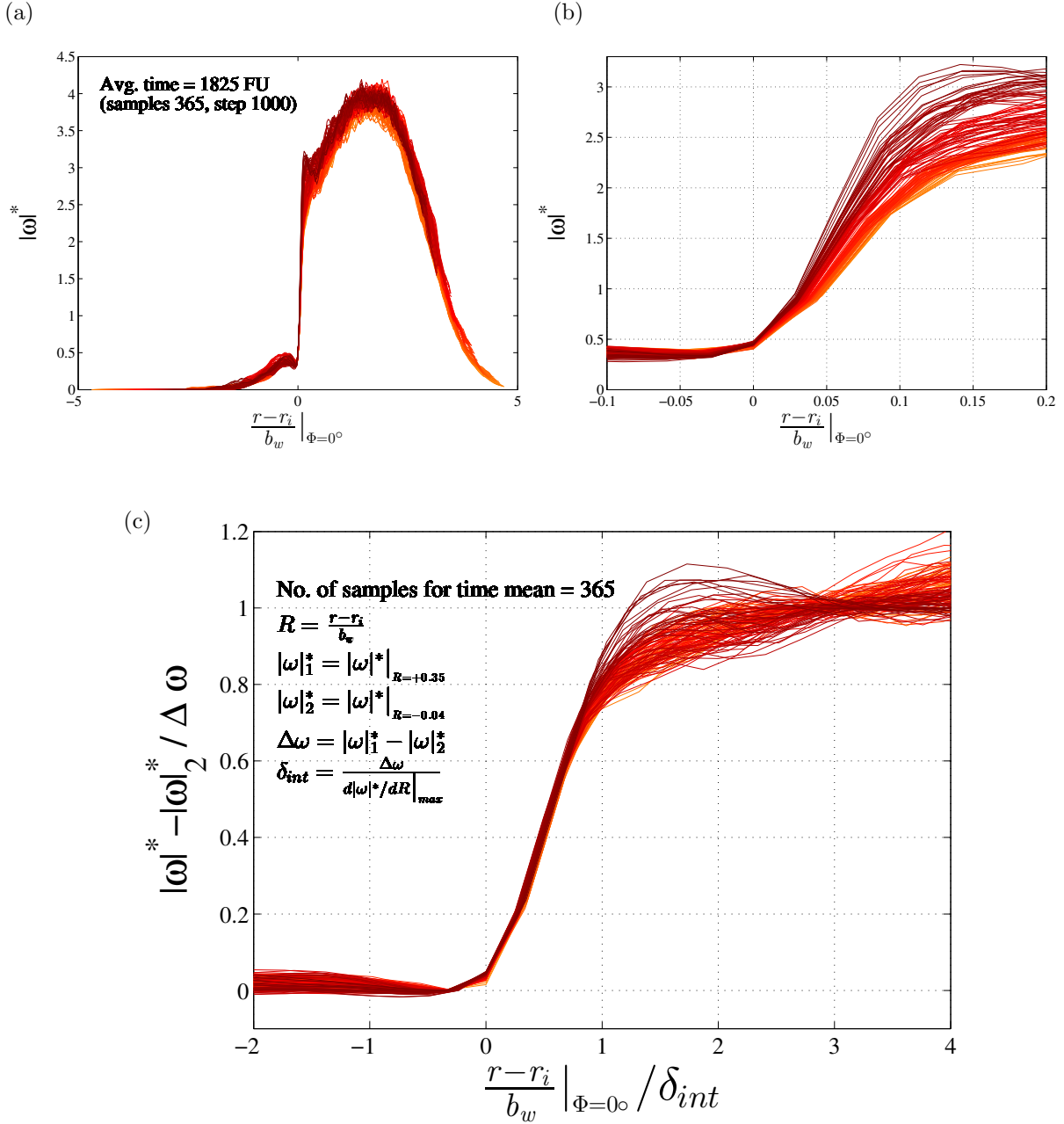

For this purpose, we do edge-based time averaging of the vorticity,
which essentially means shifting the origin of the coordinate system
in radial direction to the jet interface where the vorticity profile
shows a very sharp and steep decline, and then perform the averaging;
for example, see Figure \ref{fig:inst vort mag prof-Int_Ori_LHS}
wherein the origin of the coordinate system in radial direction is
shifted to the left interface of the jet. Figure \ref{fig:edge based
time mean - vort prof} shows the vorticity profiles for the edge-based
time averaged data wherein the edge is fixed at a vorticity magnitude
threshold of 0.5. similar kind of vorticity profiles are presented in
\cite{Westerweel_JFM_2009}. Figure \ref{fig:edge based time mean -
vort prof}(a) shows that the vorticity profiles show a close collapse
at the interface when scaled with the half-velocity width. Figure
\ref{fig:edge based time mean - vort prof}(b) shows a zoomed-in view
of figure \ref{fig:edge based time mean - vort prof}(a), focused near
the interface. If we normalize the vorticity profiles in a way similar
to that in the case of mixing layers, then we observe an even tighter
collapse in the interface region, as seen in Figure \ref{fig:edge
based time mean - vort prof}(c). Based on the characteristics of the
vorticity profiles near the interface, but towards the core of the
jet, we identified three regimes, namely overshooting, rising and flat
profiles, as shown in Figure \ref{fig:edge based time mean - vort prof
- 3 Regimes}.

\begin{figure}[!h]
\centering
\begin{overpic}
[width = 16.0 cm, height = 13.0 cm, unit=1mm]
{Fig_box-eps-converted-to.pdf}
\put(0,65){\includegraphics[width = 8.0 cm]{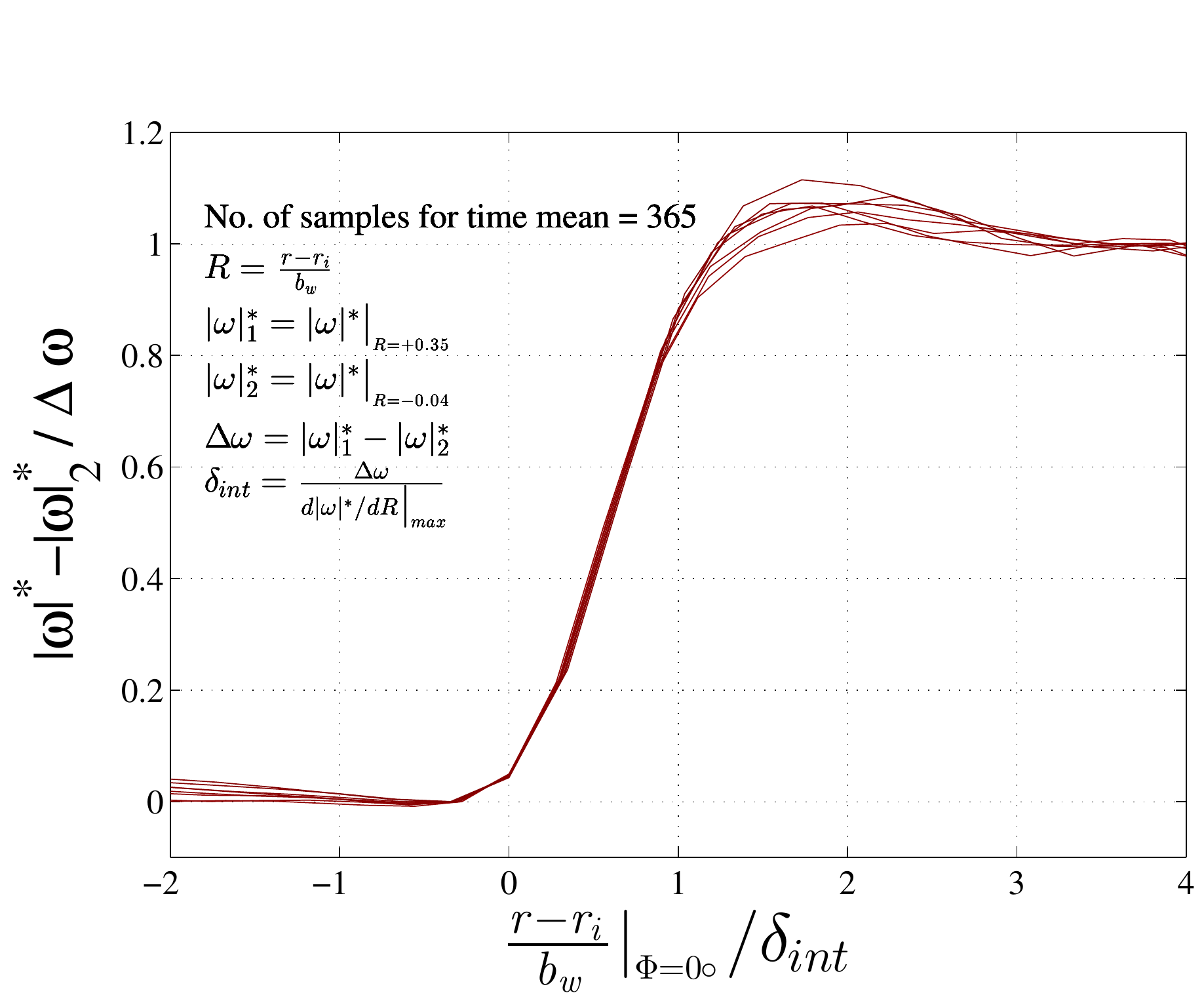}}
\put(80,65){\includegraphics[width = 8.0 cm]{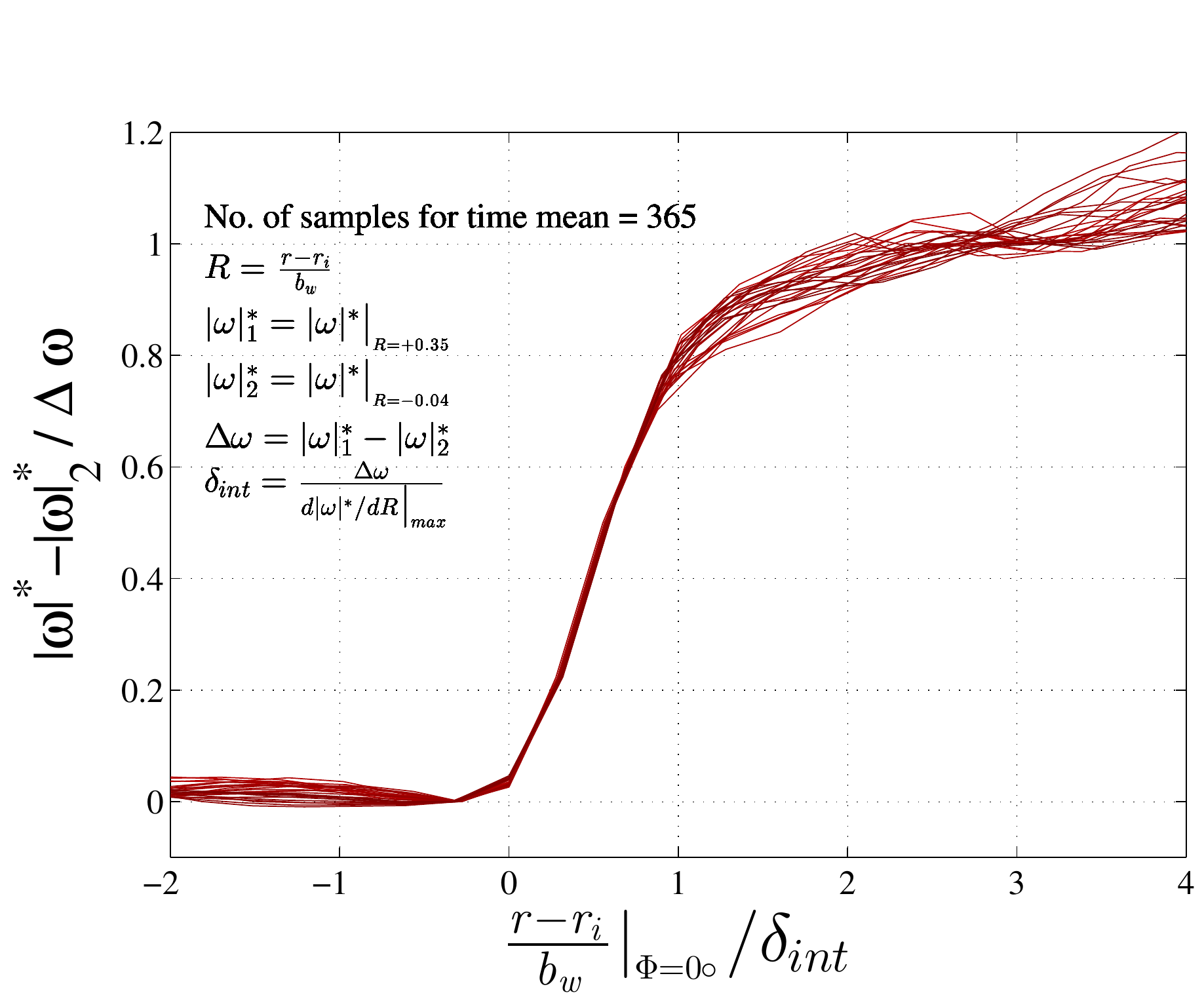}}
\put(40,0){\includegraphics[width = 8.0 cm]{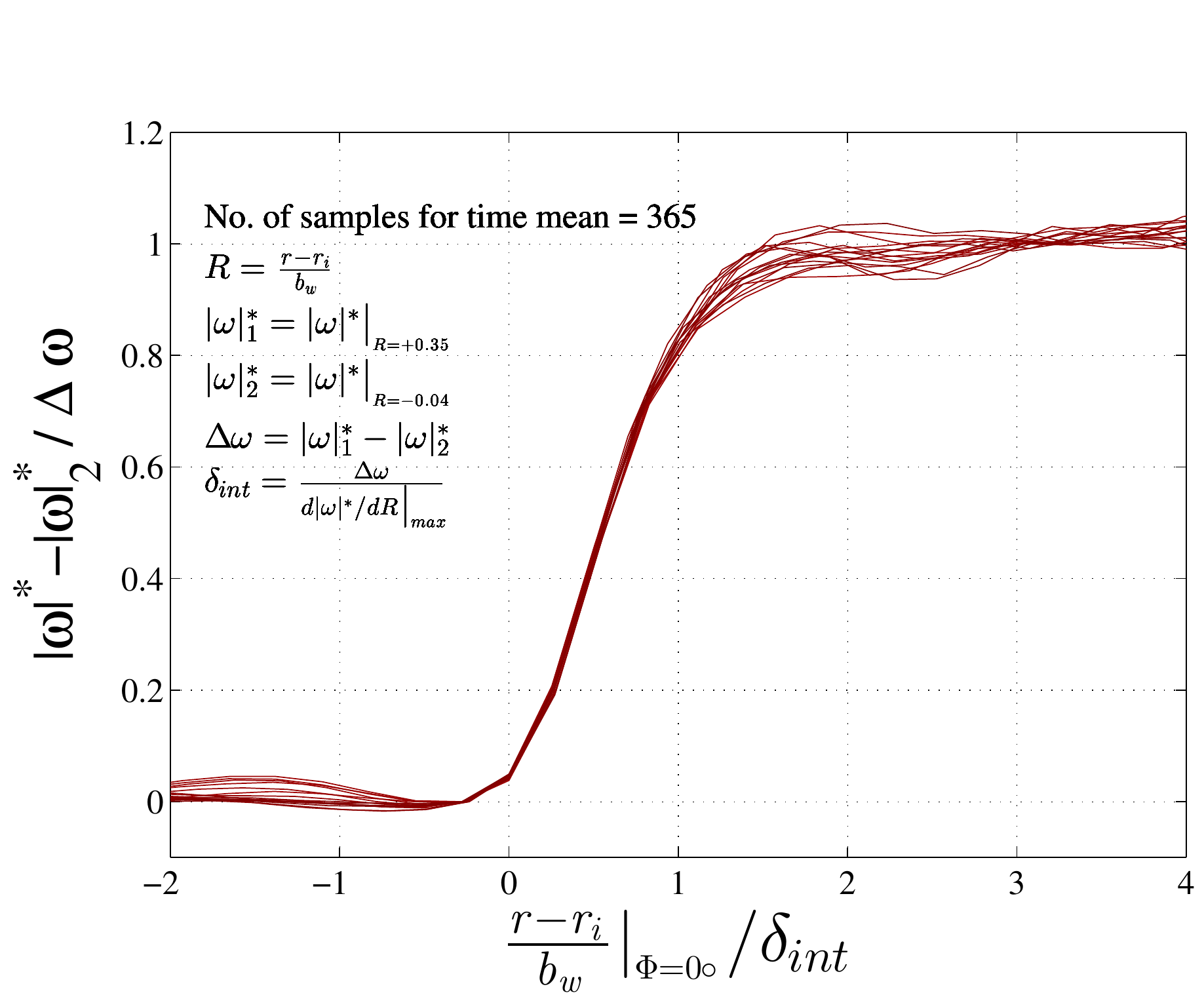}}
\put(0,125){(a)}
\put(81,125){(b)}
\put(40,58){(c)}
\end{overpic}
\caption{Three regimes of the vorticity magnitude profiles shown in
figure \ref{fig:edge based time mean - vort prof}(c): (a) overshoot:
$z = 39 - 40$; (b) rising: $z = 29 - 32$; (c) flat: $z = 37 - 39$.}
\label{fig:edge based time mean - vort prof - 3 Regimes}
\end{figure}

\begin{figure}[!h]
\centering
\begin{overpic}
[width = 13.50 cm, height = 5.5 cm, unit=1mm]
{Fig_box-eps-converted-to.pdf}
\put(30,0){\includegraphics[width = 7.0 cm]{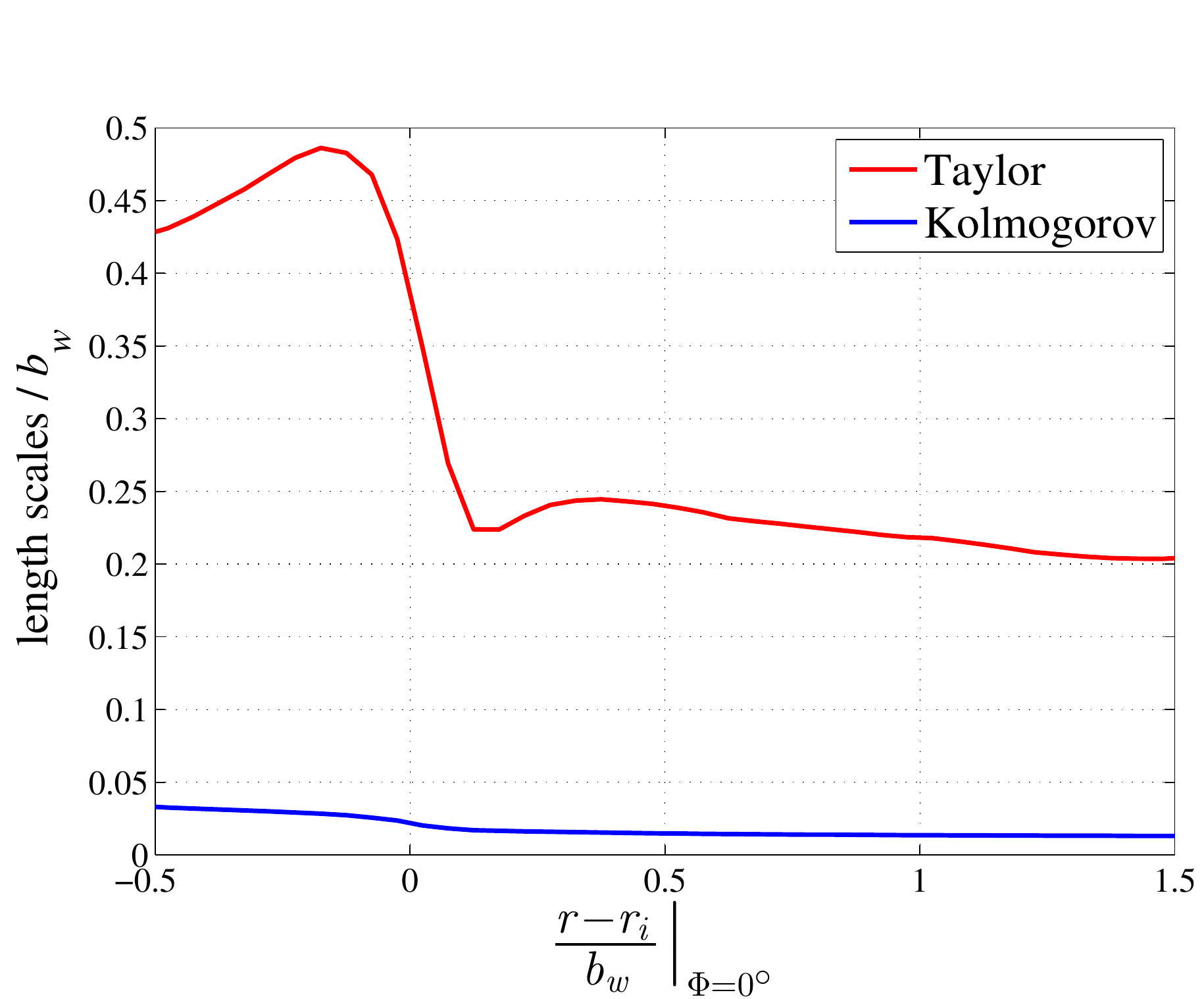}}
\end{overpic}
\caption{Variation of Kolmogorov and Taylor length scales at $z \approx 33$.}
\label{fig:Taylor Kolmogorov}
\end{figure}

The interface width in our case is about 0.1 of the half-velocity
width (see Figure \ref{fig:edge based time mean - vort prof}). It is
important to see how much is the interface thickness when compared
with the Taylor and Kolmogorov scales. The important question to ask
is: among the two scales, with what does the interface thickness
scale? The Kolmogorov and Taylor scales are shown in Figure
\ref{fig:Taylor Kolmogorov}. Our simulations showed that the interface
thickness is about double the Kolmogorov scale whereas it is about one
fifth of the Taylor microscale.

\section{Interface tracking and entrainment}

\begin{figure}
\centering
\begin{overpic}
[width = 13.50 cm, height = 11.50 cm, unit=1mm]
{Fig_box-eps-converted-to.pdf}
\put(10,0){\includegraphics[width = 12.0 cm]{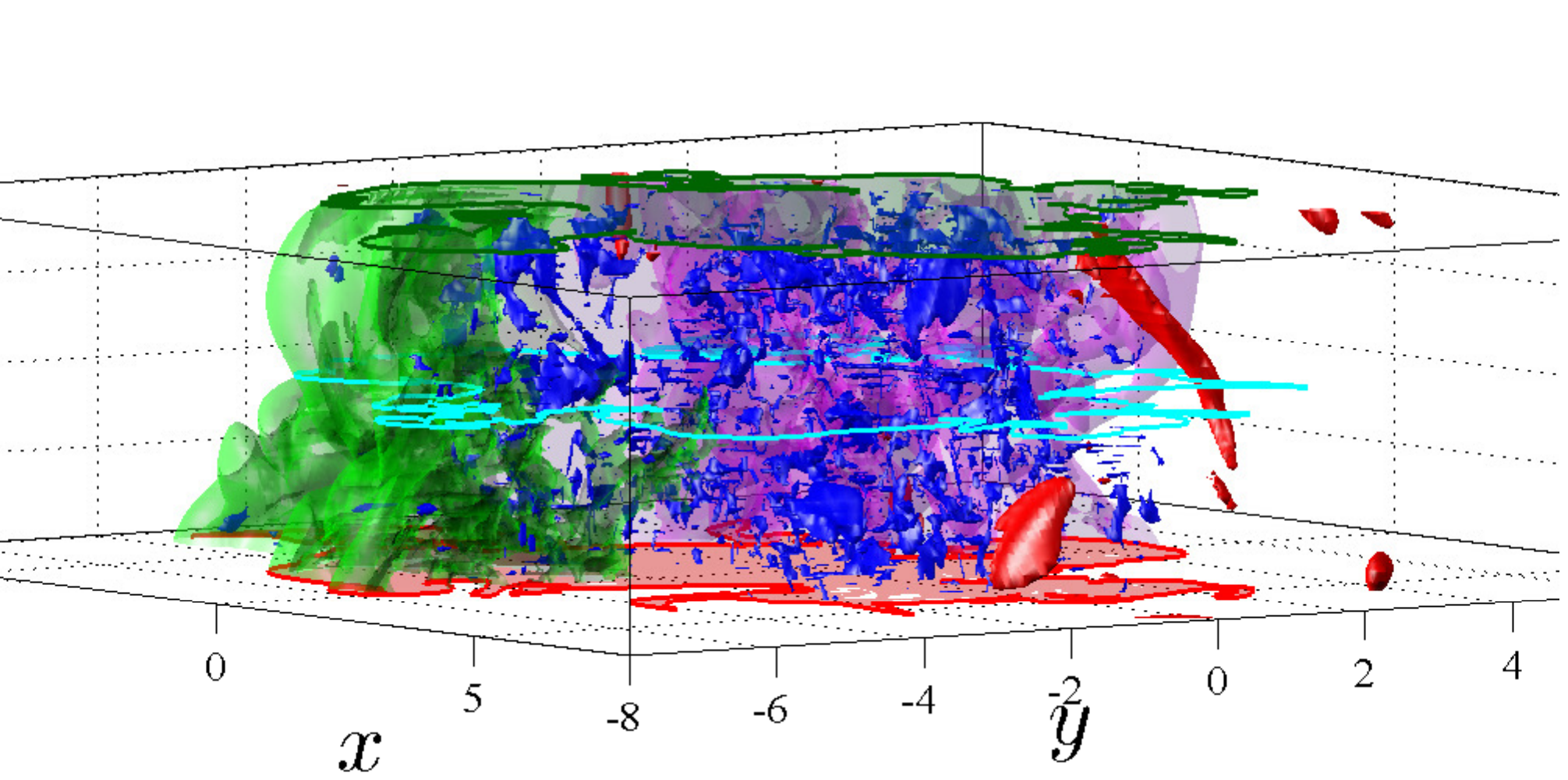}}
\put(10,60){\includegraphics[width = 12.0 cm]{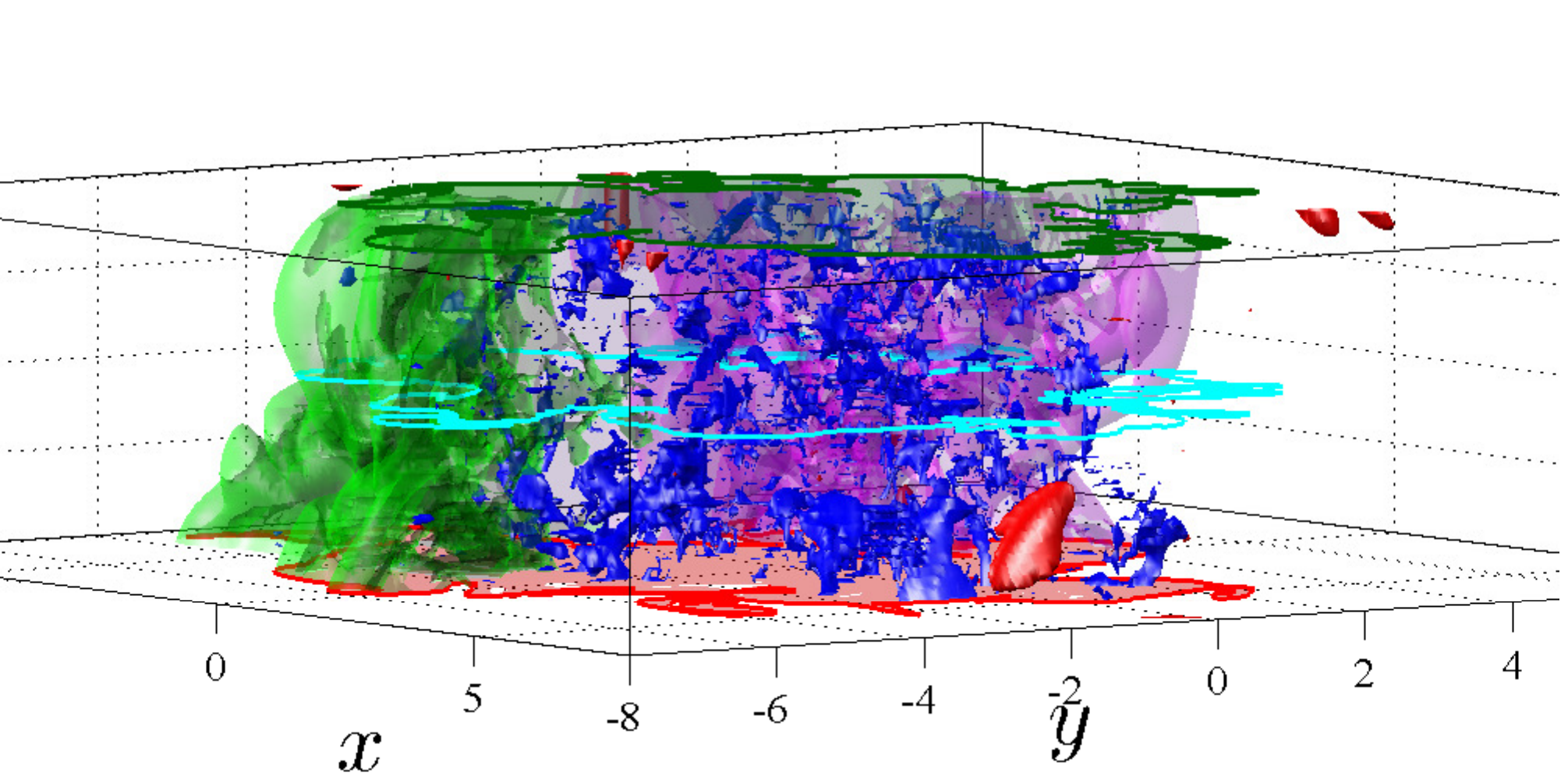}}
\put(5,28){\huge{\textit{z}}}
\put(5,88){\huge{\textit{z}}}
\put(0,110){(\textit{a})}
\put(0,50){(\textit{b})}
\end{overpic}
\caption{Instantaneous 3D sectional view of Jet depicting Islands
(red) and Lakes (blue). The jet envelope defined by a vorticity
fluctuations magnitude threshold at 1.0 is shown by pink and green
surface in the first and second quadrants. (\textit{a}) $t = 399000$;
(\textit{b}) $t = 399550$.}
\label{fig:Islands and Lakes}
\end{figure}
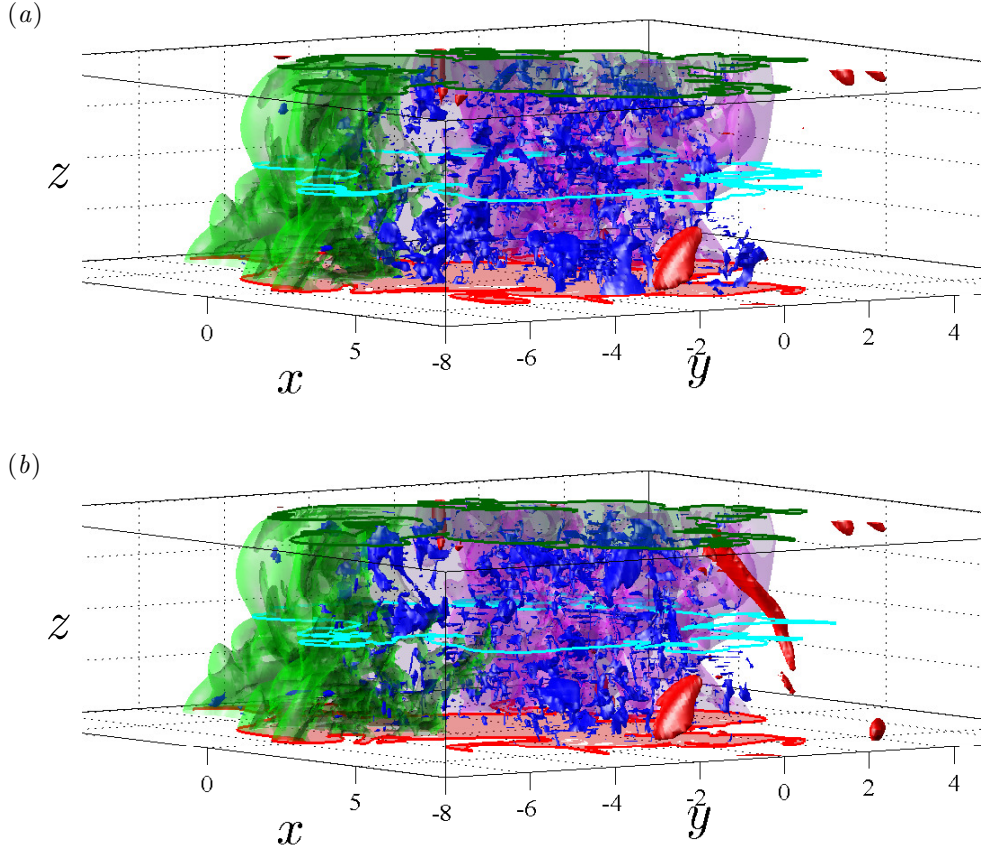

Figure \ref{fig:jet-3D-thrld-0p5}(a) shows a three-dimensional view of
the round turbulent jet at one time instant. Figure \ref{fig:Islands
and Lakes} shows the 3-D view of the jet in the self-preserving
region, especially showing the internal structure of the jet. The
threshold used for identifying the jet boundary is $|\bm{\omega}| =
0.5$.  Both the figures show that the jet boundary is quite jagged and
convoluted. Figure \ref{fig:Islands and Lakes} shows that there are
some isolated patches of vorticity within the core of the jet as well
as in the ambient. In order to identify these vorticity patches, we
define the following terms.

\begin{itemize}

\item \textbf{Jet boundary/edge}: This is an iso-surface (in three
dimensions) or a iso-contour (in two dimensions) which is defined
based on a threshold set on the vorticity magnitude that marks the
boundary on the turbulent jet.

\item \textbf{Well/Gulf/Incursion}: A narrow region of the
non-turbulent outside fluid making deep penetration into the turbulent
jet.

\item \textbf{Lake}: Isolated region of non-turbulent fluid present
within the core of the turbulent jet.

\item \textbf{Island}: Isolated region of turbulent fluid present within
the expanse of outside non-turbulent fluid.

\end{itemize}

Figure \ref{fig:Islands and Lakes} shows an isometric view of a
section of the jet at two different instants. It is clearly evident
from the figures that there are isolated \textquoteleft{lakes}' of
non-turbulent fluid marked with blue color in the figures. Also, the
presence of isolated \textquoteleft{islands}' of turbulent fluid can
be seen outside the boundary of the jet. Note that, these islands can
be a part of the turbulent jet which looks isolated in the section
chosen here (and hence qualified as islands), but might connected to
the jet somewhere else along the axial direction; for example, the two
red patches seen near the upper surface on the right side in Figures
\ref{fig:Islands and Lakes}(\textit{a,b}). However, these figures also
reveal that there are totally disconnected and isolated blobs of
turbulent fluid lying outside the jet boundary; for example, the one
red blob near the bottom surface in Figure \ref{fig:Islands and
Lakes}(\textit{a}), two red blobs near bottom surface and one long red
tongue, starting near the upper surface and running all the way
towards the bottom surface in Figure \ref{fig:Islands and
Lakes}(\textit{b}) are the example of islands. The lakes and islands
will appear as the disconnected areas in a two-dimensional
cross-section; see for example Figure \ref{fig:islands and lakes -
axial plane}. To get the true turbulent mass flux in a jet, the
calculation should exclude the contributions from the irrotational
mass in islands and lakes as they do not constitute turbulent flow.

Tracking the jet boundary in three dimensions is indeed
challenging. Therefore, we began with a two-dimensional (2D) case
having fairly simple geometry, and then extended the methodology to
the complex situations in 2D. Once we ensured that the methodology
works fine for the two-dimensional case, we have extended it in
three-dimensions (3D). However, note that it is not a straight forward
extension from 2D to 3D.

\begin{figure}
\centering
\begin{overpic}
[width = 14.0 cm, height = 6.25 cm, unit=1mm]
{Fig_box-eps-converted-to.pdf}
\put(0,3){\includegraphics[width = 3.25 cm]{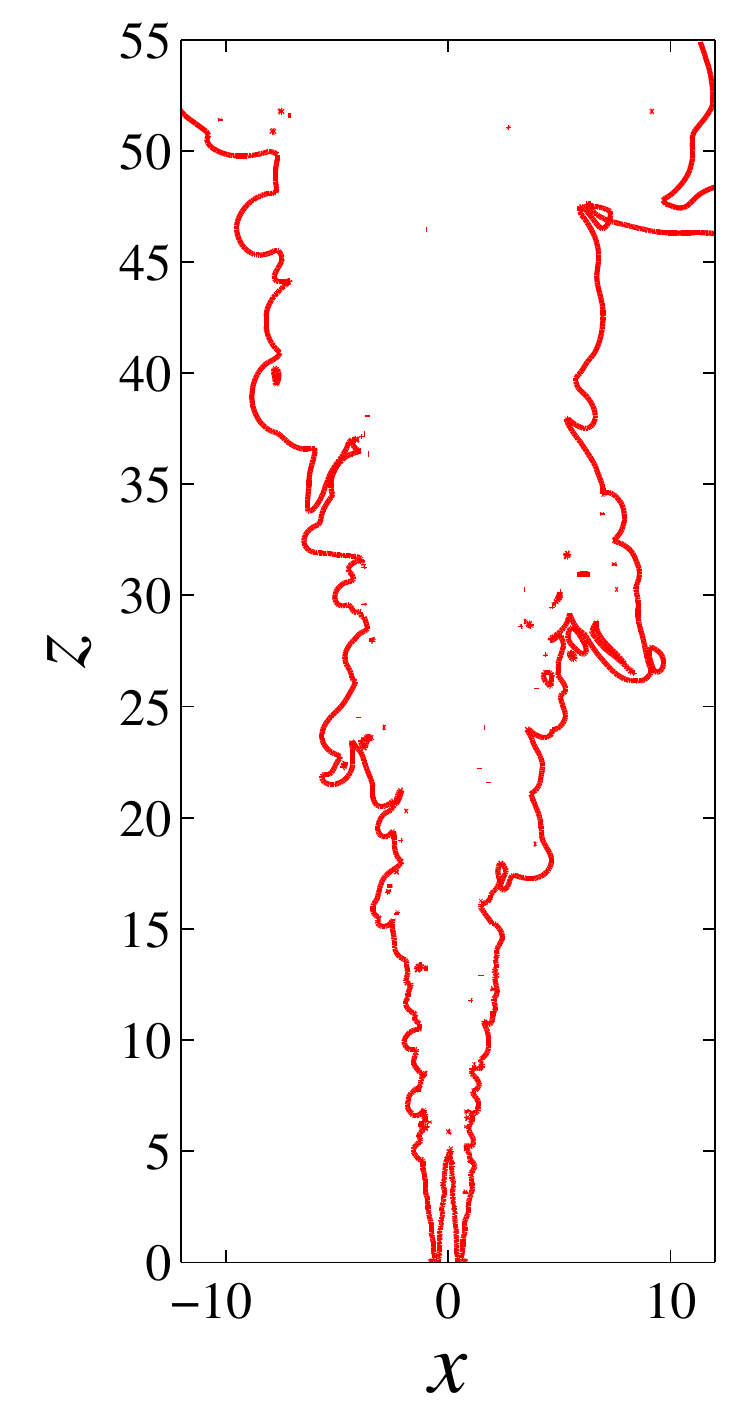}}
\put(35,3){\includegraphics[width = 3.25 cm]{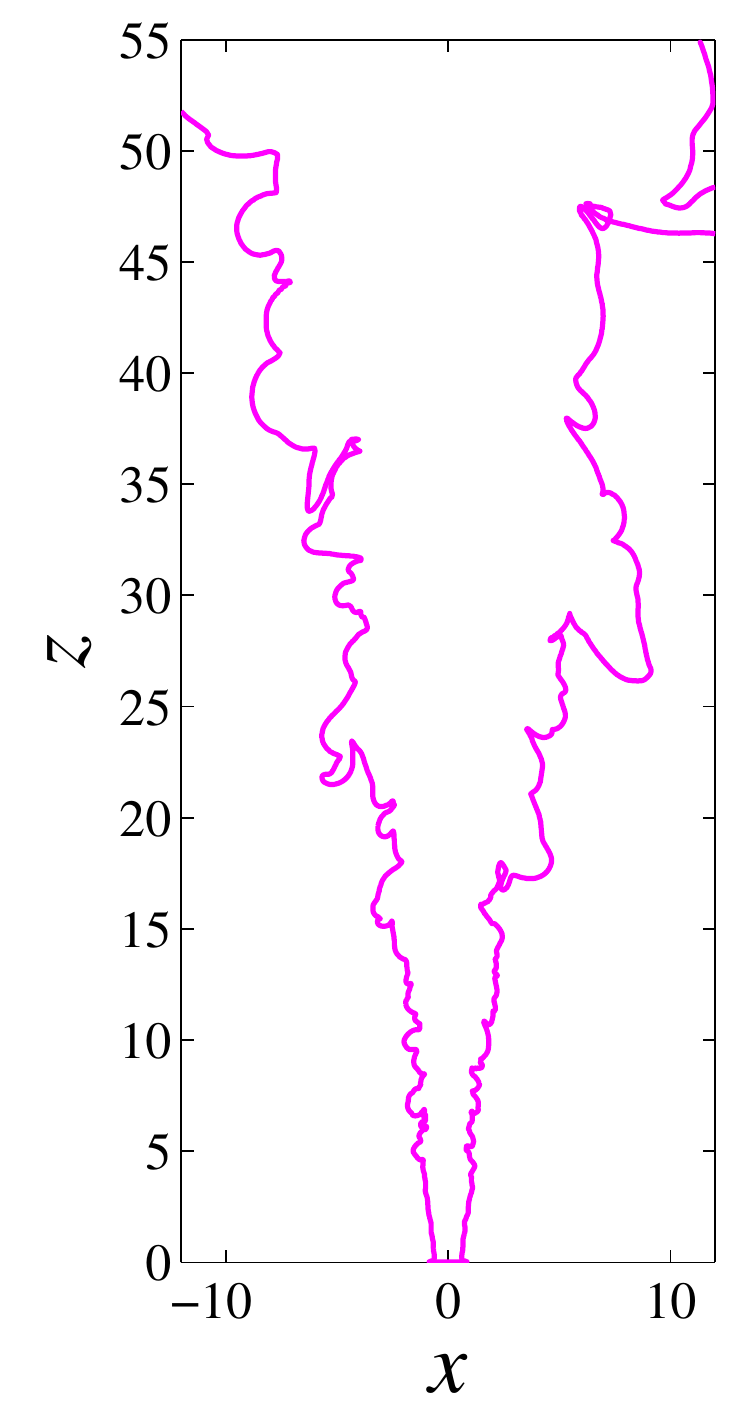}}
\put(70,3){\includegraphics[width = 3.25 cm]{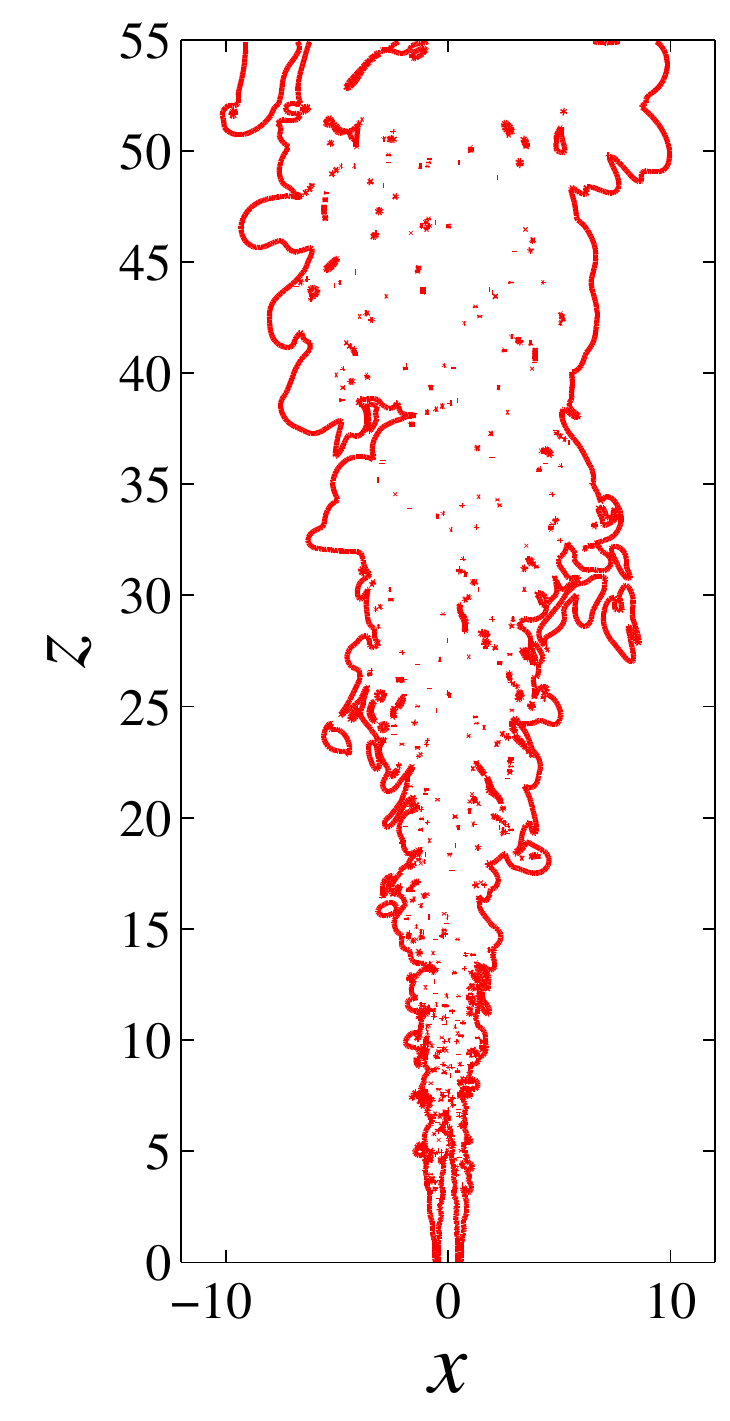}}
\put(105,3){\includegraphics[width = 3.25 cm]{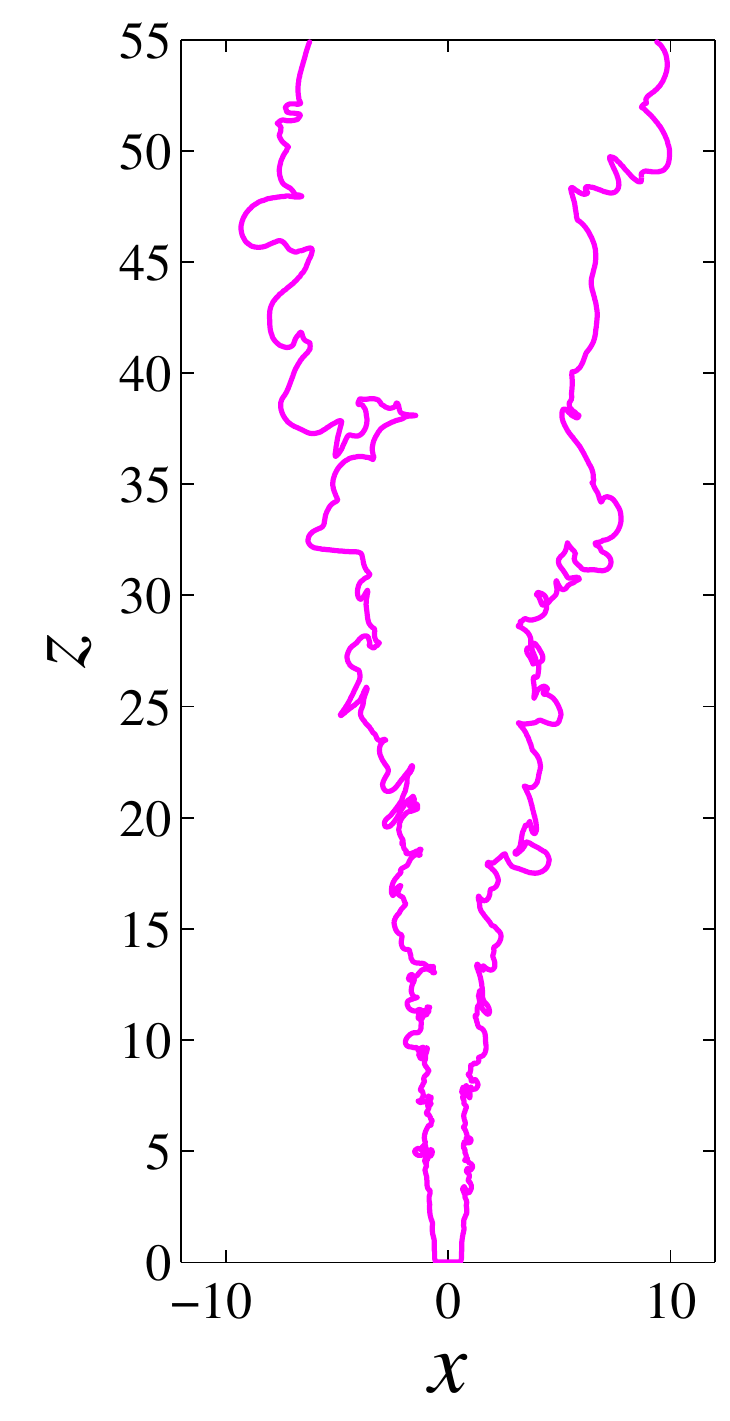}}
\put(15,0){(\textit{a1})}
\put(50,0){(\textit{a2})}
\put(85,0){(\textit{b1})}
\put(120,0){(\textit{b2})}
\end{overpic}
\caption{Instantaneous sectional view of the jet in an axial plane at
$\phi = 0^{\circ}$ showing the boundary of the jet at $t = 1995$ for
two thresholds, namely $|{\omega}|_{\mathrm{th} = 0.1}^*$ (\textit{a1,
a2}) and $|{\omega}|_{\mathrm{th} = 0.5}^*$ (\textit{b1,
b2}). (\textit{a1} and \textit{b1}) show jet section showing the
vorticity patches within (lake) and outside (island) of the jet core
along with the jet boundary. (\textit{a2} and \textit{b2}) show the
corresponding jet section marking only the jet edge after removal of
the islands and lakes. Comparison of (\textit{a1}) and (\textit{b1})
clearly shows the noticeable increase in the number of islands and
lakes at a higher threshold.}
\label{fig:islands and lakes - axial plane}
\end{figure}
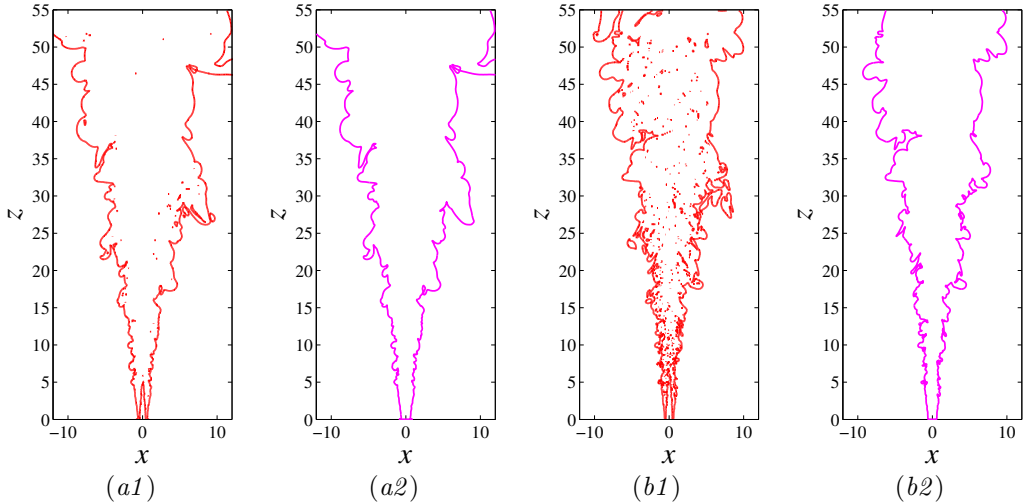

The first step towards the boundary-tracking is to get rid of the
small debris of islands and lakes, which actually do not form a part
of the jet boundary. For instance, Figures \ref{fig:islands and lakes
- axial plane}(a1 and b1) show the jet boundary along with the Islands
and Lakes in a 2D axial plane for two different vorticity
thresholds. Figures \ref{fig:islands and lakes - axial plane}(a2 and
b2) show a cleaned-up jet boundary for a 2D case. In 3D, removing
Islands and Lakes is quite challenging. However, it is possible to
generate the jet surface free from Islands and Lakes.

\begin{figure}[!h]
\centering
\begin{overpic}
[width = 16.0 cm, height = 6.6 cm, unit=1mm]
{Fig_box-eps-converted-to.pdf}
\put(3,0){\includegraphics[width = 7.25 cm]{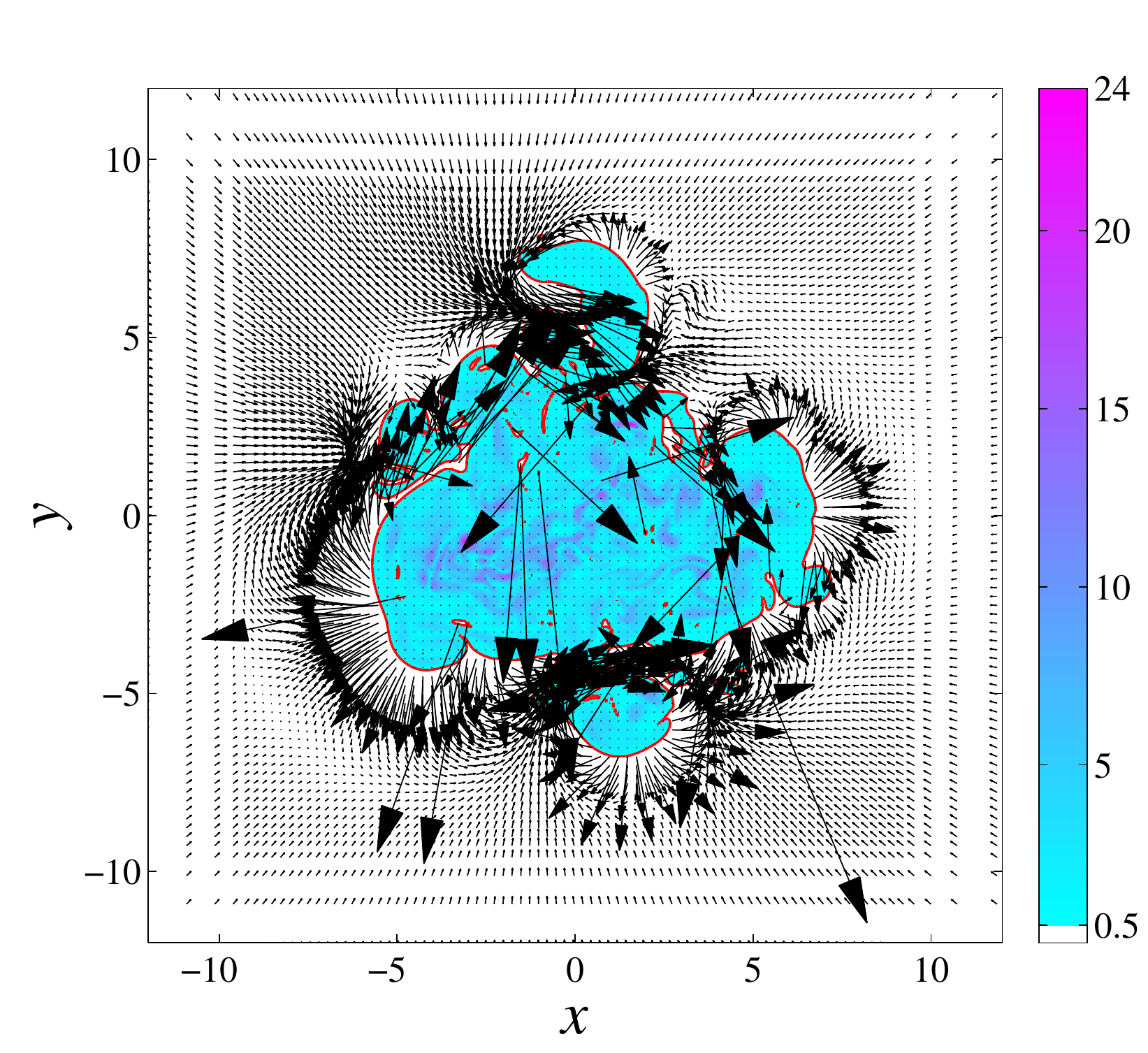}}
\put(82,0){\includegraphics[width = 7.25 cm]{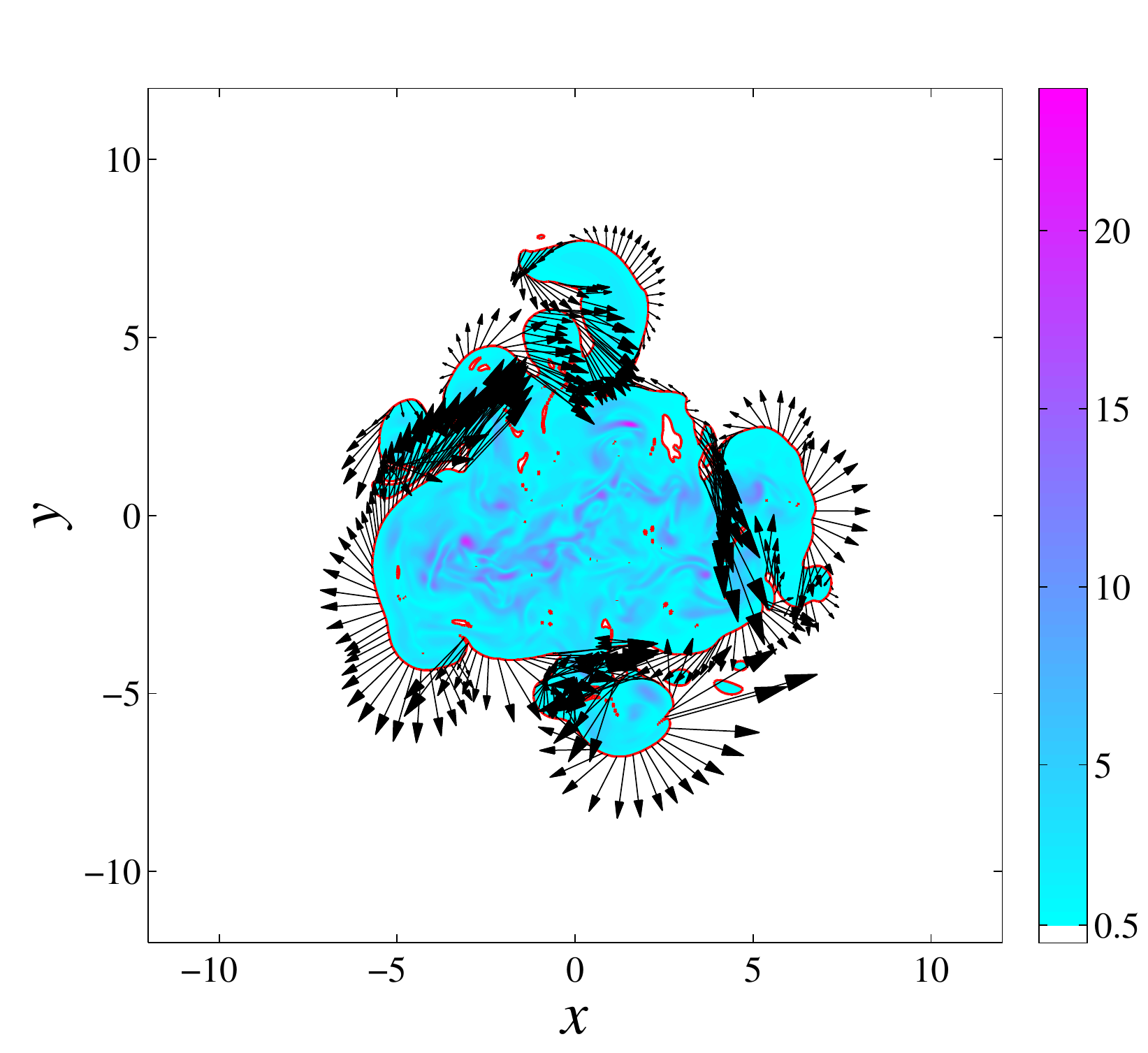}}
\put(0,62){(a)}
\put(80,62){(b)}
\end{overpic}
\caption{Instantaneous velocity and vorticity fields in a diametral
plane at $z = 35$ at $t = 2000$. The jet boundary is at $|\omega|_{th} =
0.5$. (a) Note that the velocity vectors are plotted only outside the
jet boundary. (b) shows the flow velocity only at the jet boundary.}
\label{fig:inst vel vort - diametral}
\end{figure}

In order to understand the entrainment mechanism of the ambient
irrotational fluid into the turbulent core, we are analyzing the flow
field both within and outside the turbulent core. Figure \ref{fig:inst
vel vort - axial} velocity and vorticity fields at one instant in an
axial plane. It appears that the fluid is entrained in to the jet
largely through the deep incursions, which we term as
``Wells''. Figure \ref{fig:inst vel vort - diametral} shows the
velocity and vorticity fields in a diametral plane, for the same time
instant. Further analysis to gain understanding on entrainment process
is in progress.

\begin{figure}[!h]
\centering
\begin{overpic}
[width = 16.0 cm, height = 6.0 cm, unit=1mm]
{Fig_box-eps-converted-to.pdf}
\put(3,0){\includegraphics[width = 7.25 cm]{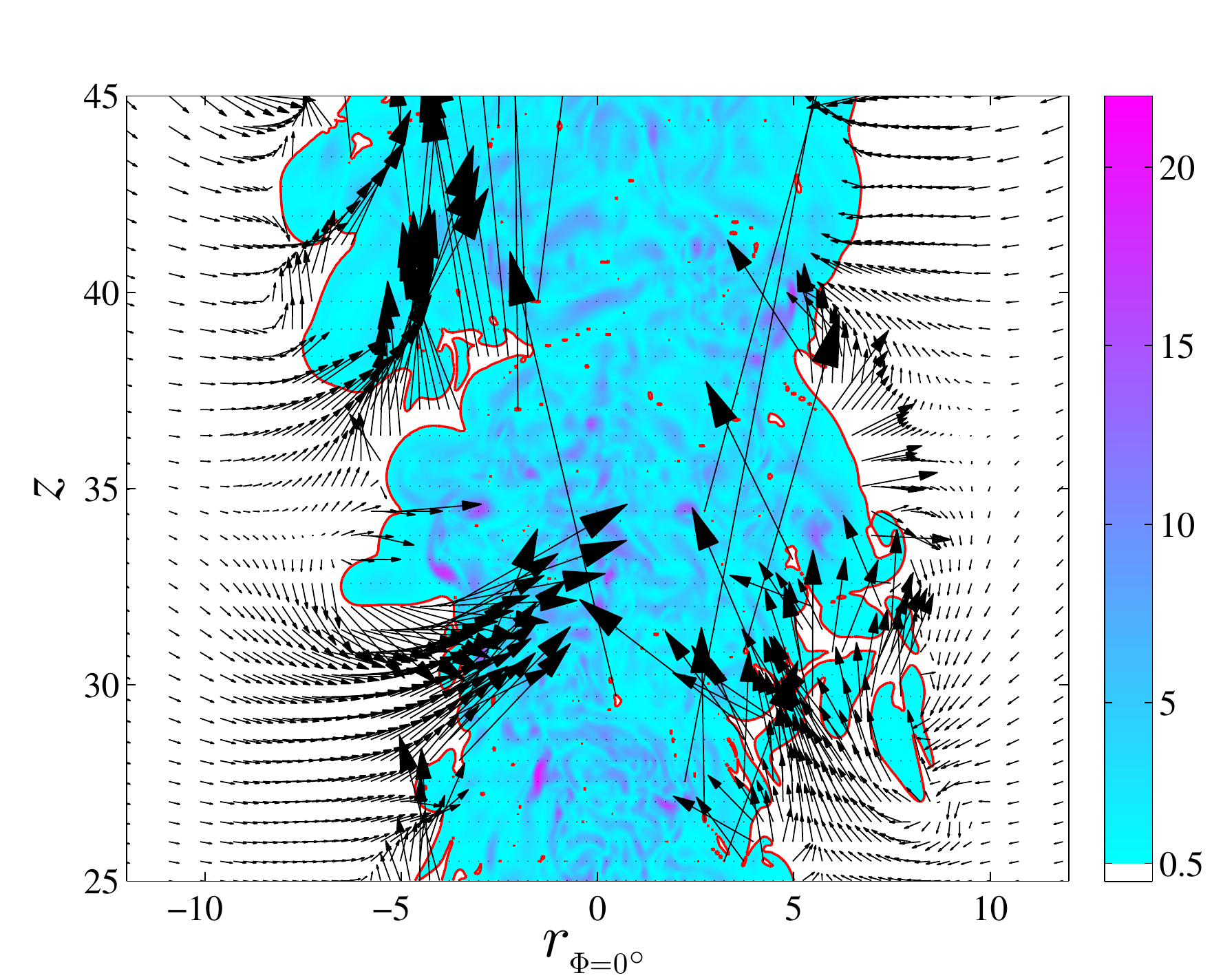}}
\put(82,0){\includegraphics[width = 7.25 cm]{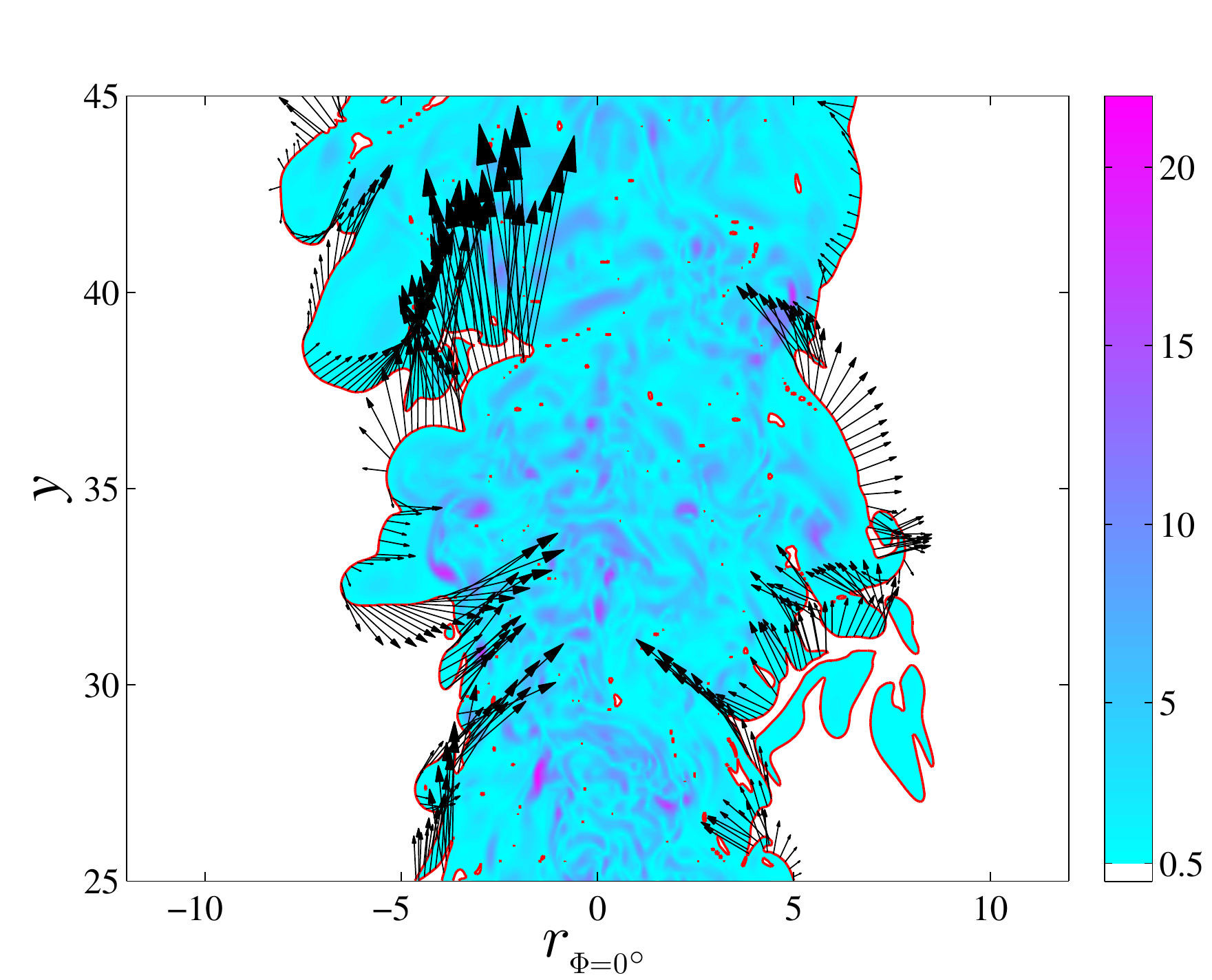}}
\put(0,55){(a)}
\put(80,55){(b)}
\end{overpic}
\caption{Instantaneous velocity and vorticity fields in an axial plane
$(\Phi = 0^{\circ})$ at $t = 2000$. The jet boundary is at
$|\omega|_{th} = 0.5$. (a) Note that the velocity vectors are plotted
only outside the jet boundary. (b) shows the flow velocity only at the
jet boundary.}
\label{fig:inst vel vort - axial}
\end{figure}

\section{Summary}

Direct Numerical Simulation (DNS) has been carried out on an
incompressible turbulent round jet at a Reynolds number of $2400$
(based on orifice diameter $d_0$ and mean exit velocity
$\overline{w}_0$ at floor level $z~=~0$ inlet conditions). The
validation showed that the simulations are of high quality in terms of
self-similarity and conservation of momentum. We identified an
accurately self-preserving region in the jet that spans over 4 orifice
diameters $(32 \leq z \leq 36)$. We also demonstrated that we need to
define two boundaries for the jet: the \enquote{inner} boundary known as T/NT
interface that isolated turbulent fluid from non-turbulent but
rotational fluid; and, the \enquote{outer} boundary known as R/IR interface
that demarcates between rotational and irrotational fluid. We showed
that the appropriate thresholds for defining these two boundaries
based on vorticity modulus are 0.1 for the outer boundary and 0.5 for
the inner boundary. We showed that the fluid confined between these
two boundaries is rotational but non-turbulent as it does not show any
vorticity fluctuations for very long duration. We show that this fluid
is often a long-lasting relic of a turbulent \textquoteleft{tongue}'
or island separated from the turbulent core at some much earlier time
at the previous instants. We also provided the methods for identifying
the boundary in 3D. The analysis on entrainment mechanism is in
progress.


\bibliographystyle{jfm}

\bibliography{main}

\end{document}